# Grid-Based Correlation Analysis to Identify Rare Quantum Transport Behaviors

Nathan D. Bamberger[1], Dylan Dyer[1], Keshaba N. Parida[1], Dominic V. McGrath[1], and Oliver L.A. Monti[1,2,*]

[1]Department of Chemistry and Biochemistry, University of Arizona, 1306 E. University Blvd., Tucson, Arizona 85721, USA

[2]Department of Physics, University of Arizona, 1118 E. Fourth Street, Tucson, Arizona 85721, USA

ABSTRACT: Most single-molecule transport experiments produce large and stochastic datasets containing a wide range of behaviors, presenting both a challenge to their analysis, but also an opportunity for discovering new physical insights. Recently, several unsupervised clustering algorithms have been developed to help extract and separate distinct features from single-molecule transport data. However, these clustering approaches have been primarily designed and used to extract major dataset components, and are consequently likely to struggle with identifying very rare features and behaviors which may nonetheless contain physically meaningful information. In this work, we thus introduce a completely new analysis framework specifically designed for rare event detection in single-molecule break junction data to help unlock such information and provide a new perspective with different implicit assumptions than clustering. Our approach leverages the concept of correlations of breaking traces with their own history to robustly identify paths through distance-conductance space that correspond to reproducible rare behaviors. As both a demonstrative and important example, we focus on rare conductance plateaus for short molecules, which can be essentially invisible when examining raw data. We show that our grid-based correlation tools successfully and reproducibly locate these rare plateaus in real experimental datasets, including in situations that traditional clustering approaches find challenging. This result enables a broader variety of molecules to be considered in the future, and suggests that our new approach is a useful tool for detecting rare yet meaningful behaviors in single molecule transport data more generally.

## 1. INTRODUCTION

Single molecule electronics have the potential to enable cheap and efficient circuit fabrication at the ultimate size limit,[1] and also provide an appealing test-bed for exploring intriguing physical phenomena at the nanoscale such as quantum interference,[2–4] spin filtering,[5,6] and interfacial coupling.[7–9] A significant and ongoing challenge in the investigation of transport through single molecule systems, however, is extracting meaning from the large and stochastic datasets typically produced by experimental techniques such as the scanning tunneling microscope break junction (STM-BJ)[10–16] and mechanically controlled break junction (MCBJ).[17–24] Both of these methods involve forming and then breaking a thin metal constriction to create a single-molecule junction in the nano-gap between two metal electrodes. The primary data collected is the conductance ($G = I / V$) through the junction during the breaking process as a function of how much the two sides have been pulled apart, known as a "breaking trace". Because of the inherently stochastic nature of both the breaking process as well as how/whether a molecule diffuses into and binds in the nano-gap, a wide range of breaking trace behaviors are observed for each single molecule system.[25] The process is thus repeated to collect thousands of breaking traces, with the results commonly summarized in 1D[10] and 2D[26,27] histograms. Conductance in these histograms is near-universally displayed on a logarithmic scale due to the large dynamic range of conductance values exhibited by most molecules.

These 1D and 2D breaking trace histograms are a powerful tool to reveal the average and/or most common behaviors in a dataset, such as an exponentially decaying conductance in the absence of molecules ("tunneling behavior"), or a relatively constant conductance over the length of a bound molecule ("molecular plateaus"). However, creating histograms inherently excludes all "trace history" information—i.e., the specific paths through distance/log(conductance) space followed by different traces. This makes it difficult to distinguish qualitatively different behaviors that may be present in the same dataset, since histograms will effectively average these behaviors together. A particular challenge is presented by behaviors that occur in only a minority of traces, which can become effectively invisible due to histogram averaging. However, previous work has demonstrated that such rare events may nonetheless correspond to physically important behaviors, such as switching between different spin states[28] or different sequences of binding behaviors.[29] Critically, the molecular signature itself—i.e., molecular plateaus—can in some circumstances become an easily lost rare event. Due to the stochastic nature of molecular binding, such plateaus are typically only observed in a fraction of all breaking traces, with the remaining traces displaying tunneling behavior. The magnitude of this so-called "molecular yield" varies depending on binding group strength,[30–32] molecular concentration,[33] and other unknown or uncontrolled variables.[34,35] Figure 1 uses simulated traces to illustrate that, for short molecules whose molecular plateaus mostly overlap with the tunneling background, low molecular yield can make the molecular signature functionally impossible to identify in both the 1D and 2D histograms. Partially for this reason, most break junction experiments focus on systems with molecular yields >10%,[8,11,27,32,35,36] and often approaching 100%,[11,30,31] because this produces histograms with clear molecular features. However, high molecular yields increase the risk of measuring multi-molecule rather than the desired single-molecule features.[35] Moreover, requiring high molecular yields restricts the binding modalities under consideration, which may end up excluding optimal molecular structures for specific applications. For example, a large fraction of all molecules studied in break junctions employ thiol linker groups,[37] in part because the strong sulfur-gold bond typically produces high molecular yields. However, the gold-sulfur bond also has a tendency for strong Fermi-level pinning,[38–42] which can negatively impact tunability of the quantum transport properties. A major benefit of rare event detection in break junction data, therefore, is that identifying infrequent molecular plateaus in the case of low molecular yields could allow a broader variety of molecular structures and metal-organic interfaces to be studied.

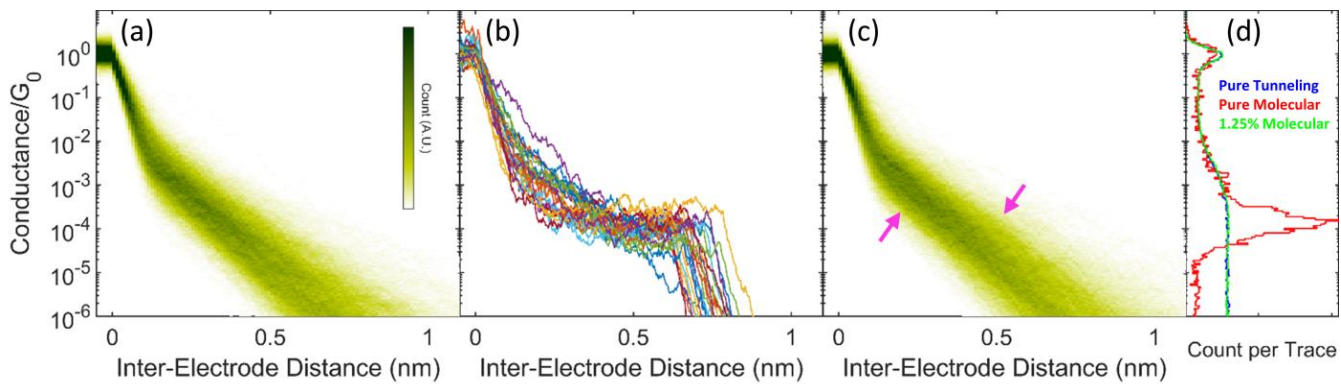

Figure 1. Demonstration, using simulated data, of how a short and rare molecular plateau feature can be effectively invisible when looking at histograms of raw data. (a) 2D histogram of 2000 simulated tunneling traces. (b) 25 overlaid simulated molecular traces. (c) 2D histogram for dataset containing 1975 of the simulated tunneling traces from (a) combined with the 25 molecular traces from (b). Despite looking identical to the 2D histogram in (a), the presence of the 25 molecular traces will produce a positive correlation between different locations along the molecular plateau region (e.g., pink arrows); our new approach thus quantifies such spatial correlations to identify rare-yet-meaningful trace behaviors. (d) Overlaid 1D conductance histograms for the datasets from (a-c), again showing that the plateau feature in the mixed dataset is invisible in the raw data.

One strategy for separating qualitatively different behaviors that may overlap in 1D and 2D histograms is to employ clustering. Indeed, over the past five years several clustering approaches have been designed specifically for breaking traces[34,43–53] and related data,[54,55] and these approaches have had varied success in extracting known and potential features, including "hidden" features, from real and simulated datasets. However, during their design and demonstration these clustering algorithms have been primarily used to investigate features occurring fairly frequently in their respective datasets. Therefore, while such approaches can detect rare behaviors in certain cases, they will in general struggle with this task that falls on the edge of what they were designed for. A few types of filtering or plateau-detection algorithms are able to extract weak molecular features from background behavior,[14,35,56–58] but these tend to be specifically tuned to a single system and/or rely on arbitrary cutoffs. There is therefore need for a new type of robust approach which is *specifically* targeted to the challenge of rare event detection, unlike current clustering approaches, and which unlike filtering requires only minimal *a priori* knowledge of the type of feature to be identified.

Here, we introduce such a new approach by focusing exactly on the very information that is lost when making 1D and 2D histograms: trace history, and in particular, any correlation between the future trajectory of a trace and its past behavior. While previous analysis tools for break junction data have considered correlations between the number of points at each conductance,[59] or between conductance cuts at different distances,[50] these approaches reduce the natural space of breaking traces from two dimensions to one. In contrast, in our approach we calculate correlations between different locations in the full two-dimensional distance/log(conductance) space, which allows us to identify rare paths followed by a subset of traces. For example, due to the 25 molecular traces contained in the dataset shown in Figure 1c, there will be a positive correlation between the two areas indicated with pink arrows—that is, more traces pass through *both* areas than would be expected if traces had no history and simply progressed like random walks through the 2D histogram. This positive correlation will then map out the shape of the rare event, e.g., the molecular plateau. In this work, we therefore define a new framework for rigorously defining two-dimensional spatial correlations, and also present tools based on this framework that can identify rare events like the one contained in Figure 1c. We stress that our approach is designed to be applicable for identifying many types of rare events in break junction data, but here we focus on the particular challenge of recognizing rare molecular plateaus both as a concrete example and because of its importance.

In the remainder of this paper, we first present details of both our collection of experimental breaking traces and our generation of simulated breaking traces. We then use the simulated data to introduce our new approach, which starts by using coarse-gridding to define pairwise correlations between discrete locations in distance/log(conductance) space. Markov-Chain Monte-Carlo simulations are then used to extract particular types of rare features and identify interesting regions for further analysis. We use simulated data for this purpose in order to demonstrate *how* our approach works on datasets containing *known* rare plateau features. Finally, in the last section, we apply our new framework to experimental datasets, thereby validating that we can successfully and reproducibly identify rare plateau features in practice. This application also reveals that molecular signatures may be more common than previously thought, enabling research on systems with low molecular yield and with potential implications for our understanding of the nanoscopic environment of the junction.

## 2. METHODS

**2.1 Generation of Simulated Breaking Traces**. Due to the atomic-scale complexity of single-molecule junctions, creating physical models that faithfully reproduce all features and properties of experimental breaking traces remains a significant challenge. For this work, we therefore instead used an empirical model to generate simulated breaking traces that capture at least the most obvious properties of observed breaking traces. Such traces are sufficient for our purposes, despite obvious shortcomings in terms of capturing the full richness of experimental breaking traces, because we only use them to demonstrate *how* our grid-based correlation tools operate, not as a means of training or validating these tools.

Two types of simulated breaking traces were created: tunneling and molecular. All traces were generated on a logarithmic conductance scale with 500 data points per nm of inter-electrode distance. The tunneling traces consist of three sections: a pre-rupture plateau near 1 $G_0$, followed by a sharp drop-off to represent snap-back,[31,60] and finally a shallower linear drop-off to represent tunneling. The parameters defining each of these sections (e.g., the slope of the tunneling drop-off) were fixed for each individual trace, and normally distributed across the set of all tunneling traces. Low-amplitude random noise was then added on top of each trace to create a more realistic shape. The molecular traces were generated in the same way as the tunneling traces, except that they contained two additional sections—a very gradually sloped plateau and a fairly sharp drop-off from the end of this plateau—which occur after the tunneling drop-off has decreased to the chosen conductance value for the plateau. See SI section S.2 for full details of simulated trace generation.

2000 tunneling traces were generated to produce the 2D histogram in Figure 1b, and then 25 simulated molecular traces were combined with the first 1975 of those tunneling traces to produce the dataset shown in Figure 1c and used throughout section 3 below.

**2.2 MCBJ Fabrication and Experimental Setup.** MCBJ samples were fabricated and run following the methods of Bamberger et al.[50] Each sample was fabricated on a substrate of 0.5 mm thick phosphor bronze coated with a few-micron insulating layer of polyimide. The pattern of a thin metal wire with an ~100 nm constriction in the center was defined using electron beam lithography. The wire itself was created by thermally evaporating a 4 nm titanium adhesion layer followed by 80 nm of gold. Finally, reactive ion etching with an $O_2/CHF_3$ plasma was used to turn the central constriction into an ~1 μm free-standing gold bridge by under-etching the polyimide.

MCBJ experiments were performed in air at room temperature. Each sample was clamped into a custom-built three-point bending apparatus in which a push rod was used to bend the sample and thereby thin and break the gold bridge. To collect each breaking trace, a stepper motor (ThorLabs DRV50) was first used to adjust the push rod until the conductance through the gold bridge was between 5 and 7 $G_0$ (where $G_0$ is the quantum of conductance, equal to 77.48 uS)[61]. When this set-point was reached, the collection of a single breaking trace was triggered by raising the push rod 40 μm at 60 μm/s using a linear piezo actuator (ThorLabs PAS009) while simultaneously recording the conductance through the bridge at 20 kHz using a custom high-bandwidth Wheatstone bridge amplifier.[62] The piezo was then retracted, after which the process can be repeated to collect another breaking trace. Custom LabVIEW software was used to automatically collect thousands of consecutive breaking traces. During trace collection, the bending apparatus was placed on a vibrationally isolated table to reduce mechanical noise and inside a copper Faraday cage to reduce electrical and acoustic noise.

**2.3 Collection of Experimental Datasets.** The molecules studied in this work (OPV2-2SMe, OPV2-2BT, and OPV2-2SAc; see below) were each synthesized on site (SI section S.1.1) and characterized using NMR spectroscopy (SI section S.1.2) and mass spectrometry (SI section S.1.3). Each molecule was dissolved in HPLC grade (>99.7%, Alfa Aesar) dichloromethane (DCM) to form ~1 μM and/or ~10 μM solutions. Note that the acetyl-protected binding groups in OPV2-2SAc are known to spontaneously de-protect on the gold surface for experiments performed in air, forming free thiol binding groups.[63]

Dataset collection also followed the method of Bamberger et al.[50] Each MCBJ sample was cleaned with $O_3$/UV and rinsed with ethanol shortly before use. For each sample, we initially deposited ~10 μL of pure DCM, using a clean glass syringe, on the center of the junction with the aid of a Kalrez gasket. We then collect an "empty" or "tunneling" dataset of a few thousand breaking traces, both as a negative control and in order to calculate an attenuation ratio that was then applied to all subsequent datasets collected on that same MCBJ sample. Next, the LabVIEW program was paused with the gold bridge fully broken, and 10-20 μL of molecular solution deposited on the center of the junction using a clean glass syringe. The LabVIEW program was then allowed to continue collecting breaking traces.

The experimental data in this work were collected using four different MCBJ samples. Due to events such as multiple depositions of molecular solution and/or a full relaxation of the push rod, the set of traces from each sample was broken into multiple chunks of 2000+ sequentially collected traces, and each of these datasets was analyzed independently (see SI section S.3 for details).

## 3. RESULTS AND DISCUSSION

In this section, we begin by developing our new framework for considering two-dimensional spatial correlations, which statistically identifies deviations from random-walk behavior in order to define a quantity we call connection strength. Next, we use connection strength as the basis for a "feature-finder" tool that can extract, with minimal input parameters, different types of rare features from a breaking trace dataset. Finally, we use our new framework and feature-finder tool to extract weak molecular plateau features from experimental datasets for several short molecules and validate the results, demonstrating the power and utility of this new approach.

**3.1 Calculating Correlations and Defining Connection Strength.** Before we can calculate correlations between different locations in distance/log(conductance) space, we first need to formally define such locations. To this end, we superimpose each trace onto a coarse grid and then represent it as a series of lattice points from the grid, which we call "nodes". As shown in Figure 2a, each successive node in a coarsened trace increases by exactly one grid unit in $x$ but can increase or decrease by any amount in $y$. The advantage of using coarsened traces is that nodes represent *finite and discrete* locations in distance/log(conductance) space that each trace unambiguously either does or does not pass through, making it straightforward to consider spatial correlations (see SI sections S.4.1 and S.4.2 for details). It is important to note that the size of the coarse grid can impact the correlations we are subsequently able to find: if the grid is very coarse then subtle correlations can be drowned out by the great number of uncorrelated traces passing through each node; and if the grid is very fine an ever-larger total number of traces is needed to provide the statistical power to identify correlations in the first place. In practice, however, there is a broad range of gridding sizes over which our approach works well and yields reasonable results. Throughout this work, we thus use a grid with 25 nodes per nm and 10 nodes per conductance decade, but our main conclusions are not overly sensitive to modest changes in this grid size (SI section S.7.1).

The main idea behind our strategy for identifying when trace behavior is correlated with past trace history is the realization that if *no* such correlations existed, then traces would simply proceed as (weighted) random walks through distance/log(conductance) space (i.e., whether each trace visits a given node would depend only on the single node visited immediately prior). Therefore, we can find such correlations by looking for the "least random-walk-like" trace behaviors. This idea is illustrated on an extremely simplified case in Figure 2b,c: suppose we have a set of 502 traces (Figure 2b) with 100 traces each sloping downwards through 5 parallel nodes (to represent tunneling traces) and 2 traces which proceed horizontally (to represent molecular plateaus). Based on all of the traces that pass through a given node, we can calculate "exit probabilities" (Figure 2c) for each node to its neighboring nodes (SI section S.4.3). Using the exit probabilities in Figure 2c, it is clear that *if* traces behaved like random walks, then, on average, only $(2\%)^4 = 0.000016\%$ of the traces passing through node X would also pass through node Y. However, for the actual traces (Figure 2b), $2/102 = 2\%$ of the traces from X also pass through Y. Because $2\% \gg 0.000016\%$—that is, *more* traces go from X to Y than expected under random-walk behavior—we conclude that nodes X and Y are *positively* correlated.

To quantify this measure of correlation while also rigorously accounting for the contribution of random chance, we use a pair of one-sided binomial hypothesis tests. As explained in SI section S.4.4, these tests allow us to calculate the probability, under the null hypothesis of random walk behavior, of more or fewer traces than observed passing between two nodes. For the example in Figure 2b,c, these tests compute the probability of seeing 2 or more traces going *from* X *to* Y under random walk conditions as only 5.5 x $10^{-10}$%, and the probability of seeing 2 or fewer such traces under random walk conditions as ~100%. We refer to these two types of probability as $p_{above}(X, Y)$ and $p_{below}(X, Y)$, respectively.

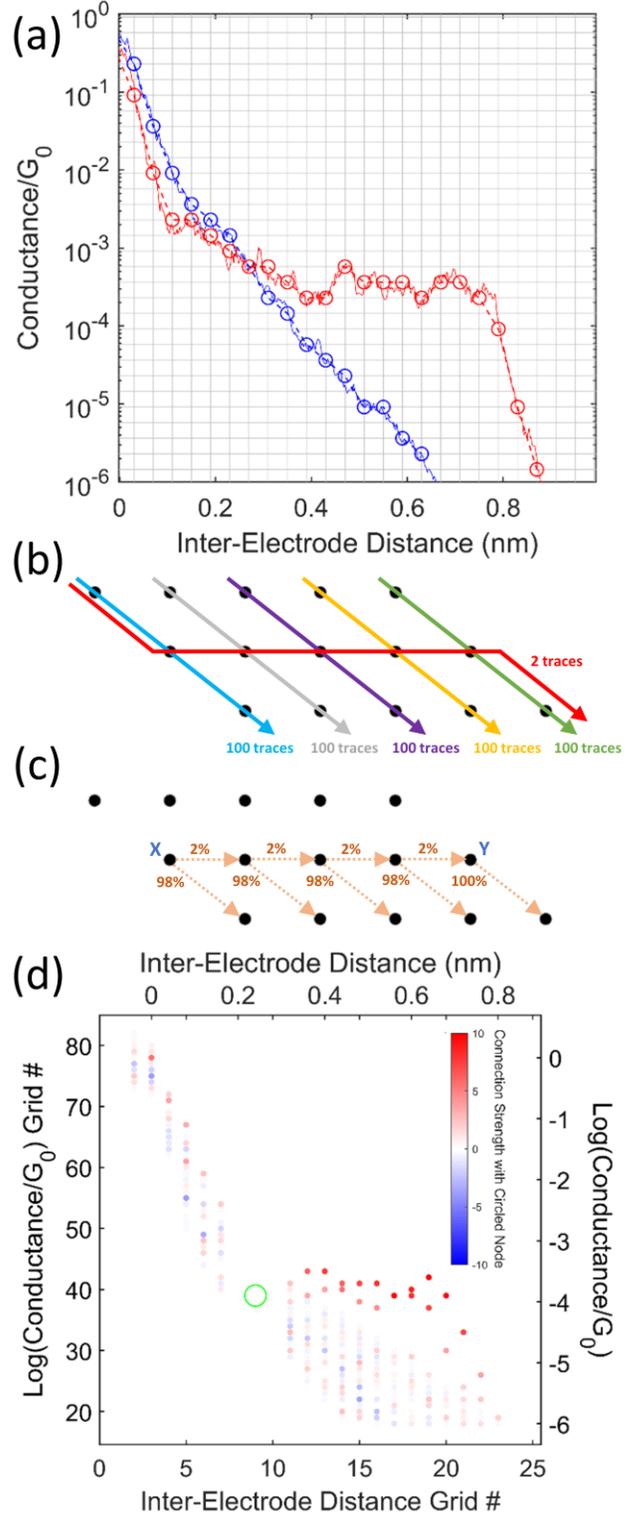

**Figure 2.** (a) Example of coarse-gridding two traces by representing each as a series of lattice points from the same grid. The original traces are represented with solid lines, the coarsened traces are represented with circles connected by dotted lines, and the grid is represented with gray lines. To make the grid easily visible, a coarser gridding is used for this plot than that used elsewhere in this work. **(b) Hypothetical example of a small set of nodes with 502 traces passing through them. (c) Exit probabilities calculated based on the traces in (b). Based solely on these exit probabilities, the probability of going from node *X* to node *Y* is vanishingly small, but in fact two traces followed this path, revealing their correlation with their own history. (d) Connection Strength distribution for all nodes versus the circled green node for the datasets from Figure 1c. The most-positive Connection Strength nodes (red) clearly pick out the molecular plateau feature that was invisible in the raw data.**

Positively correlated node-pairs will have small $p_{above}$ and large $p_{below}$ values, while negatively correlated node-pairs—i.e., cases in which fewer traces pass between both nodes than would be expected under random walk conditions—will have large $p_{above}$ and small $p_{below}$ values, and node-pairs with little-to-no correlation will have relatively large $p_{above}$ and $p_{below}$ values. We therefore combine both values into a single measure of pairwise node correlation that we call "connection strength", or *CS*, using the definition (see SI section S.4.5 for details):

$$CS(X,Y) = -\ln(p_{above}(X,Y)) + \ln(p_{below}(X,Y))$$

Taking logarithms allows us to easily differentiate very small probabilities from one another, and the sign choices ensure that *CS* will be positive (negative) for positively (negatively) correlated node pairs. Moreover, in information theory log-probabilities represent the self-information or "surprisal" of an event,[64] making this a well-motivated definition for our goal of identifying correlations that are surprising relative to the expectation of random-walk behavior. To make *CS* symmetric with respect to nodes X and Y, X is always chosen as the node farther left and Y as the node farther right, since traces are only able to proceed from left to right. For nodes in the same column, *CS* is set to zero, indicating no correlation (since it is impossible by design for a single trace to pass through both such nodes).

This concept of pairwise connection strength between nodes is central to our ability to use spatial correlation to identify rare events in breaking traces. To demonstrate this, Figure 2d shows the connection strength distribution for the simulated dataset in Figure 1c for every node compared to the green-circled node. The positive correlations clearly form a plateau shape, indicating that more traces passing through the circled node follow a plateau-like path than would be expected from pure random walk behavior. This shows that there must exist a subset of traces that do *not* behave like random walks, but rather follow this plateau path instead of the average behavior of the dataset; in other words, pairwise connection strength has exactly picked out the rare event that was purposefully built into this simulated dataset! While "higher-order" correlations—e.g., if a trace passes through nodes X *and* Y, is it more or less likely than a random walk to also pass through node Z?—can be calculated in principle, the number of observations for such events in a given dataset, and hence our statistical power, would drop exponentially. For this reason, we focus exclusively on pairwise connection strength in this work.

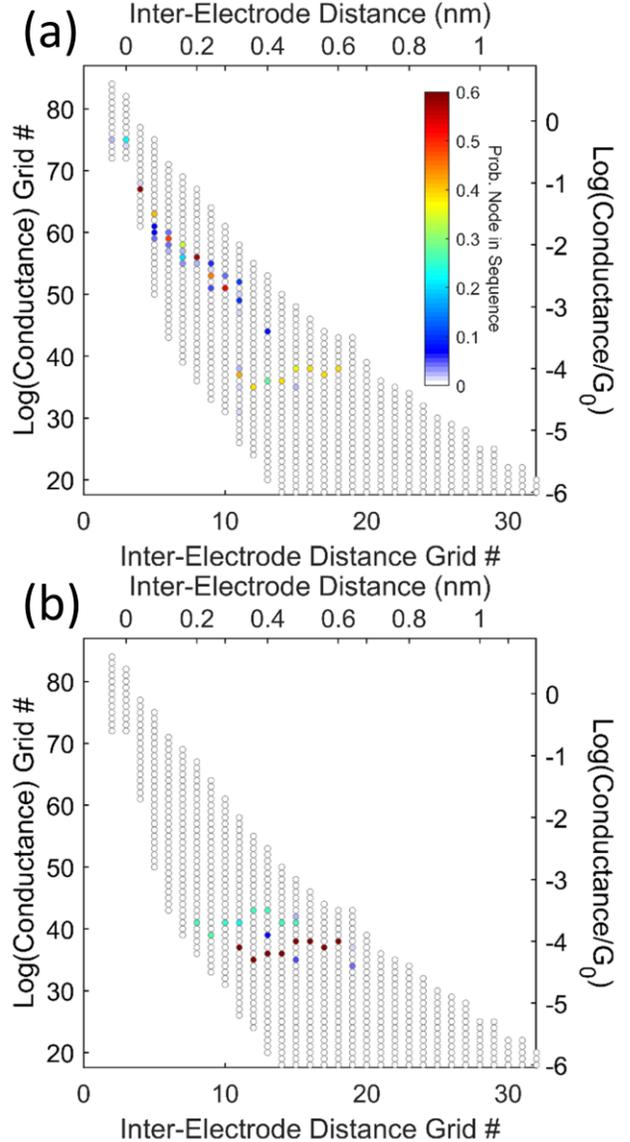

**Figure 3. (a) Final distribution of nodes produced by running the MCMC feature-finder with 8-node sequences and no additional restrictions on the simulated dataset from Figure 1c. The main feature discovered is a tunneling-like feature at high conductance, because tunneling traces are also correlated with their own history. (b) To focus our search on plateau-like features in particular, we add slope criteria to the MCMC feature-finder, leading to the successful recovery of the rare plateau feature hidden in the dataset in Figure 1c.**

**3.2 Identifying Significant Features**. Figure 2d demonstrates that connection strength distributions are a powerful way of visualizing spatial correlations and can reveal rare events. However, for each dataset there will be as many different distributions like the one in Figure 2d as there are nodes, making it unrealistic to examine all of them. The connection strength distribution with a specific node is thus most useful as a tool for investigating locations in distance/log(conductance) space that a researcher is already interested in. To perform a less-directed exploration, we developed a new tool to identify such "interesting nodes" in the first place, with what counts as an "interesting node" naturally depending on what type of feature/rare event is under consideration.

The basis of this new tool is determining where in distance/log(conductance) space the "least-random-walk-like" node sequences meeting certain criteria can be found. To formalize the concept of "least-random-walk-like", we define the "significance" of a sequence of nodes as the average connection strength between all possible node-pairs chosen from that sequence (SI section S.4.6). High significance node sequences intuitively represent paths through distance/log(conductance) space that real traces followed significantly more often than random walk traces would have, and which may thus correspond to physically meaningful rare events. To solve the problem of identifying high-significance node sequences, we make use of Markov-Chain Monte Carlo (MCMC) simulation, which is a way to estimate the distribution of multidimensional objects according to a pre-defined weighting. In our case, the multidimensional objects are node sequences meeting specifiable criteria and the weighting of each sequence is $\exp(\text{significance}/T)$, where $T$ is an "effective temperature" controlling how flattened vs. peaked the distribution will be. This tool, which we refer to as the "MCMC feature-finder", thus produces a distribution of node sequences heavily weighted towards high-significance paths, allowing us to find those very paths and locate the "interesting nodes" introduced above (see SI section S.5 for all MCMC details).

As a first illustration of the MCMC feature-finder, we will consider 8-node sequences for the simulated dataset from Figure 1c, with no additional restrictions placed on the sequences' shape, slope, etc. Because a distribution of node-sequences is difficult to represent visually, we instead examine the feature-finder results by plotting the frequency with which each individual node was included in the sequences produced by the MCMC (Figure 3a). Instead of following the rare molecular plateau feature, these node sequences are concentrated in the high-conductance tunneling region. This is because tunneling is also not a random-walk behavior; as explained in the methods section, each simulated tunneling trace has a fixed average slope, so its future trajectory *is* correlated with its past behavior (and this is likely true of experimental tunneling traces as well). This demonstrates a critically important point: *traces in the same dataset can possess correlation with their histories in multiple different ways*. To identify a specific *type* of correlation in a given dataset using the MCMC feature-finder, therefore, we include the option within this tool to impose extra criteria on the node sequences being generated.

For example, in this paper we are focused on the specific case of plateau-like features. In our second illustration of the MCMC feature-finder (Figure 3b), we thus require the 8-node sequences to have slopes of no more than 2.5 decades/nm (for details on slope calculation see SI section S.5.2). The inclusion of this relatively weak restriction results in the MCMC feature-finder highlighting exactly the rare molecular plateau feature that was built into this simulated dataset. The high-probability nodes in Figure 3b thus represent "interesting nodes" in the context of plateau-like behavior, which could be further explored using connection strength distributions as in Figure 2d. We note that these nodes and the rare plateau feature they constitute were identified without knowing anything *a priori* about its location in distance/log(conductance) space. The only input needed was what *type* of rare event we were looking

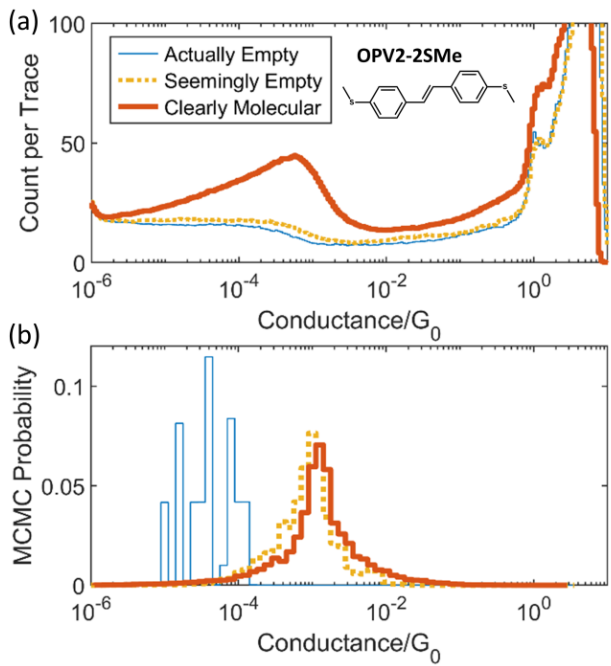

**Figure 4.** (a) Overlaid 1D conductance histograms for three experimental MCBJ datasets collected on the same sample with the molecule OPV2-2SMe (inset). The "actually empty" dataset was collected before any molecules were deposited, the "seemingly empty" dataset was collected after molecular deposition, but its raw data looks indistinguishable from the actually empty dataset, and the "clearly molecular" dataset was collected after molecular deposition and shows a clear molecular peak. (b) Node probabilities, projected onto the conductance axis, from the MCMC feature-finder when it was applied to the three datasets from (a) in order to search for plateau-like features. The same plateau feature is discovered in both the "seemingly empty" and "clearly molecular" datasets, and not discovered in the "actually empty" dataset.

for—i.e., relatively flat sequences that are 8 nodes (~0.32 nm) in length.

While specifying the type of feature ahead of time does provide an avenue of influence for the user's biases, we note that user choice and biases are necessarily involved in *any* type of analysis, for example through the selection of which unsupervised machine-learning algorithm to use. We therefore believe it is advantageous to have the user make certain decisions explicitly and consciously, rather than making them implicitly and perhaps unknowingly via algorithm design. This is especially true for the case of choosing a feature type, because constructing a single universal and automated algorithm for identifying *every* type of rare event in *every* possible dataset is likely unfeasible. Our MCMC feature-finder should therefore be thought of as a guided exploration tool used to locate rare events of a type loosely defined by the user based on their physical intuition and/or the context of their particular application. Under this view, the ability of our MCMC feature-finder to be targeted at different types of rare behavior in the same dataset, depending on the user's focus, is an advantage.

**3.3 Using Grid-Based Correlation Tools to Find Rare Molecular Features in Experimental Data.** In the previous section we used a simulated dataset containing a known rare molecular plateau feature to demonstrate that our new grid-based correlation tools can detect such rare behaviors in principle, and to explain how and why this is the case. Simulated breaking traces, however, can differ in critical ways from experimental breaking traces. Validating that these new tools achieve their goals in practice thus requires turning to such experimental data.

Figure 4a shows overlaid 1D conductance histograms for three datasets collected on the same MCBJ sample with the short molecule OPV2-2SMe (see SI section S.3 for 2D histograms). These datasets collectively form an ideal test case for our ability to detect rare molecular plateau features. The "actually empty" dataset (blue in Figure 4a) is composed of breaking traces collected before any molecule was introduced to the system, and so serves as a control in which no molecular plateaus should exist. The "clearly molecular" dataset (red in Figure 4a) contains breaking traces collected after a solution of the molecule OPV2-2SMe was deposited on the sample. As the name suggests, this dataset contains a clear molecular feature, and thus serves as a positive control for where the plateaus for this particular molecule are expected to appear. Crucially, the "seemingly empty" dataset (yellow in Figure 4a) contains breaking traces that were also collected *after* molecular deposition, and yet appears nearly identical to the "actually empty" dataset when examining histograms. This "seemingly empty" dataset thus serves as our test case, because it could plausibly contain OPV2-2SMe plateaus—since OPV2-2SMe was physically present on the MCBJ sample during data collection—but if so they must be rare, which would inform our understanding of the junction environment (see below).

We applied our MCMC feature-finder to all three datasets from Figure 4a, with the MCMC simulation set to generate 12-node sequences restricted to have a slope of no more than 2.5 decades/nm (see SI section S.5.2). As shown in Figure 4b and SI section S.6, the MCMC discovered a plateau feature in in the "seemingly empty" dataset that is nearly identical to the plateau feature discovered in the "clearly molecular" dataset, and distinct from the feature found in the "actually empty" dataset. Due to the presence of both positive and negative controls, this result provides strong evidence that we have indeed successfully identified rare molecular plateaus for this OPV2-2SMe molecule. To demonstrate the advantages of our approach for such rare plateau detection, in SI section S.10 we have analyzed the "seemingly empty" dataset from Figure 4a using a few representative clustering strategies. The results show that these strategies, while having some success, are generally challenged by this task, especially as the "rareness" of the feature increases. Our approach performs better at this task because it was specifically designed with rare events in mind.

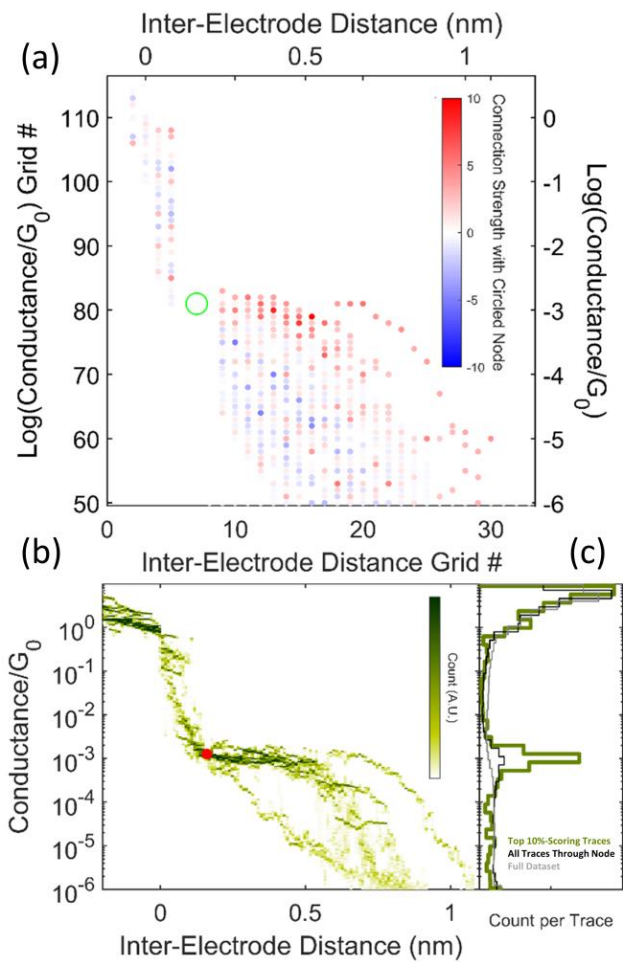

**Figure 5.** (a) Connection strength distribution for the "seemingly empty" dataset from Figure 4 versus the node circled in green, which was identified as a high-probability node by the MCMC feature-finder. To connect these results back to the original traces, we use this distribution to score each of the traces passing through the selected node by the average connection strength of the other nodes they visit. (b) A 2D histogram for the top 10% scoring traces through the selected node (now shown in red) demonstrates that plateaus *are* present in the original experimental traces, not just the MCMC results. (c) These top-scoring traces show a very clear molecular peak (green) which is completely invisible in the raw data (gray), and only weakly visible in all of the traces passing through the selected node (black).

In order to validate the results in Figure 4b, two potential concerns must be addressed. First, there is a concern that the plateau-shapes discovered by the MCMC feature-finder are not representative of the original experimental data. This concern arises because the node sequences in the MCMC simulation are not restricted to node sequences that occurred in the actual experimental traces. Moreover, as described above, in order to focus on rare plateaus in particular, we applied a slope restriction to the MCMC simulations. There is thus a potential risk that the final MCMC results were "forced" into plateau-like shapes for the "seemingly empty" dataset.

To address this concern and connect back to the original breaking traces, we start by considering the connection strength distribution with respect to one of the "interesting nodes" identified by the MCMC feature-finder (Figure 5a). Next, we introduce a new grid-based correlation tool by scoring each trace passing through the selected node by the average connection strength vs. the selected node of all the other nodes the trace passes through (SI section S.8). Intuitively, higher scores identify the least-random-walk-like experimental traces passing through the selected node. Finally, we plot 2D and 1D histograms of just the top 10%-scoring of these traces (Figure 5b,c). This reveals that plateau-like features *are* present in the actual experimentally collected traces in the "seemingly empty" dataset. No slope criteria or restrictions are used at any point in making Figure 5; we just selected the least-random-walk-like traces through a particular node, and those traces turn out to have plateau features. This scoring tool is thus a useful way to validate that a feature discovered with the MCMC feature-finder really exists in the experimental data. It is also a great way to extract for further analysis the traces in a dataset that actually correspond to a particular type of rare behavior. The results in Figure 5 are robust to other choices of high-probability nodes from the MCMC output (SI section S.9).

The second potential concern about the MCMC

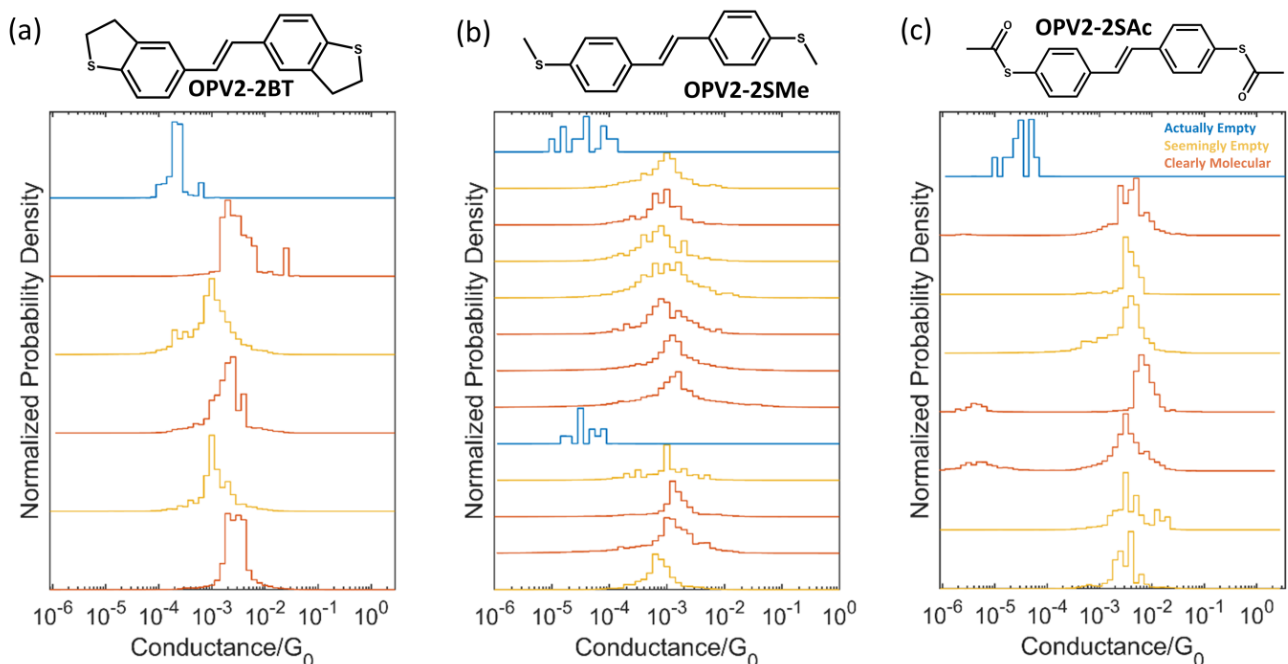

Figure 6. (a-c) Conductance distributions for the plateau-like features identified by the MCMC feature-finder in 27 different experimental datasets for three different short molecules. The distributions for "actually empty" datasets are shown in blue, for "seemingly empty" datasets in yellow, and for "clearly molecular" datasets in red (see SI section S.3 for details). For each molecule, the location of the identified plateau is quite consistent across all "seemingly empty" and "clearly molecular" datasets, providing strong evidence that the rare plateaus discovered by the MCMC feature-finder are in fact signatures of the molecules in question.

feature-finder results in Figure 4b is that the plateau feature discovered in the "seemingly empty" dataset could be located at the same conductance as the plateau feature in the "clearly molecular" dataset simply by chance. After all, the MCMC feature-finder is designed to always find *some* feature matching its input criteria, because some path through distance/log(conductance) space will always be *least* random-walk-like. For example, a "plateau-like" feature (albeit at a different conductance) was still discovered in the "actually empty" dataset, despite our expectation that plateaus will not occur when no molecules are present. This is, in fact, an inherent and unavoidable challenge for any attempt at rare event detection in a highly stochastic and/or noisy system: in any single dataset, it can be difficult to distinguish whether an identified rare event is "real" or just due to random noise-like behavior. The solution is to consider multiple datasets, since "real" rare events should be detected consistently in the same place.

We thus applied our MCMC feature-finder, with the same settings used in Figure 4b (12-node sequences, slope ≤ 2.5 decades/nm), to additional examples of "actually empty", "seemingly empty", and "clearly molecular" datasets collected in the presence of three different short molecules (see SI section S.3 for all dataset details and histograms). As shown in Figure 6, for each molecule the MCMC feature-finder consistently identifies essentially the *same* plateau feature in both the "seemingly empty" and the "clearly molecular" datasets, and this result is robust to changes in the MCMC feature-finder parameters (SI section S.7). This constitutes strong evidence that the rare plateaus detected in the "seemingly empty" datasets are in fact molecular signatures and not just random/noisy behavior. That the identified plateau features appear at different conductances for the different molecules provides yet more evidence that these plateaus originate from the experimental data and are not simply an artifact of the MCMC simulation. In contrast, there is significant variation in the conductance of the features found in the "actually empty" datasets, and these features are mostly not robust to changes in the MCMC feature-finder parameters (SI section S.7). This is in agreement with our hypothesis that these datasets do not contain consistent plateau features, but rather just random behaviors that sometimes approximate plateau-like shapes by chance.

Besides helping to validate our ability to detect rare events in practice, the data in Figure 6 also provide potential insights into the nanoscopic environment of single-molecule junctions. We note that most of the "seemingly empty" datasets were collected under essentially identical conditions to their "clearly molecular" counterparts (SI section S.3). While the reasons for such variation in event frequency are not well studied in the field, one hypothesis is that the concentration of molecules in the immediate vicinity of the junction (and hence the frequency of molecular plateaus) is controlled in large part by random chance, and is only loosely dependent on the concentration of the deposited molecular solution. Therefore, our identification of molecular behavior in so many "seemingly empty" datasets may suggest that, at room temperature, these molecules are mobile enough that they can still find their way into the junction even when their concentration in the immediate vicinity is quite low. Additionally, the fact that there is little-to-no systematic difference between the conductances of the plateaus identified in the "seemingly empty" and "clearly molecular" datasets is consistent with the widely held assumption that MCBJ experiments are truly measuring single-molecule junctions, with little-to-no contribution from multi-molecule junctions.

## 4. CONCLUSIONS

Extracting as much meaningful information as possible from the large, stochastic datasets created by single molecule break junction experiments continues to be a challenging multidimensional problem that will consequently require multiple different approaches to solve. Much effort has recently gone into developing different types of filtering, and, especially, clustering algorithms for break junction data. While these approaches have many advantages and documented successes, to date they have mostly been designed for and applied to the problem of extracting prominent dataset features, rather than the related but distinct challenge of identifying rare behaviors in huge samples of breaking traces. In this work we specifically addressed this latter challenge by introducing a novel approach that uses the trace history information which is ignored by traditional histogram-based analysis to calculate pairwise correlations between discretized locations in distance/log(conductance) space. This framework is quite distinct from other published analysis tools for single-molecule transport data, which we believe has value; all analysis tools make implicit assumptions about dataset structure through their design, and so increasing the variety of these designs and assumptions is important for gaining new perspectives and generating new hypotheses. Using simulated breaking traces, we demonstrated how a suite of tools based on our framework can be used to detect different types of rare features in the same dataset.

To evaluate the utility of these new rare event identification tools, we chose to focus on the specific challenge of detecting rare molecular plateaus in the case of short molecules for which those plateaus overlap with a strong tunneling background signal. Using experimental MCBJ datasets collected for three separate short molecules, we demonstrated a consistent ability to detect molecular plateau features corresponding to the molecular species that was known to be physically present, but whose signature was invisible in the raw data. These rare and hidden molecular plateaus were identified without any *a priori* knowledge or assumptions about their conductance values, and multiple controls and validation tests supported our inference that they are in fact signatures of the molecules in question.

The successful detection of rare molecular plateaus in several experimental datasets is important for multiple reasons. First, it provides us with important insight into the junction environment, suggesting for example that molecules can find their way to the very middle of the junction even when the overall frequency of molecular bridging events is quite low. In future studies our tools may help yield insight into the causes of variable bridging frequency, such as varying local concentration. Second, we believe that addressing the challenge of identifying rare molecular plateaus is of particular importance, because if very weak molecular signatures can be reliably detected, then single-molecule researchers are empowered to explore a greater variety of molecular binding groups, concentration regimes, or perhaps single molecule chemical reactions. Third, despite the importance of rare plateaus in particular, we stress that our grid-based correlation approach is not limited by design to this one type of rare event. Therefore, rigorously validating that the new tools presented here can successfully detect one form of rare behavior is also important because it suggests that they can be used for rare event detection more generally. This could potentially allow new types of physically meaningful break junction behaviors to be found and understood. We also note that our approach was able to identify molecular plateaus in the "clearly molecular" datasets as well, suggesting that while these tools were designed for *rare* event detection in particular, they may in fact have applicability for extracting more common events as well.

In contemplating the possible extension of our grid-based correlation approach to other types of rare events, however, it is also important to consider its limitations. For example, the MCMC feature-finder is designed to focus on the *single* most-correlated feature of a given type. If multiple features match the user-specified criteria—e.g., a dataset with two rare plateau features—then the MCMC feature-finder is likely to be heavily weighted towards the one with even just slightly higher internal correlation (though the second could then be found by modifying the MCMC criteria). An example of this situation is shown in SI section S.11.1 using simulated data. Another limitation of our approach is that it is designed to identify rare events that are localized in distance/log(conductance) space. This is relevant, for example, for the consideration of rare switching events between multiple molecular conductance states. As illustrated in SI section S.11.2 using simulated data, our approach is capable of identifying such switching events if they occur at a preferred distance (e.g., if the switching is caused by stretching-induced conformational change), but is not suitable for detecting switching that stochastically occurs across a broad range of distances (e.g., if the switching mechanism is light-induced).

To enable these and other extensions of our work, we have made all of the MATLAB code needed to implement our new approach freely and publicly available as part of our SMAUG Toolbox at github.com/LabMonti/SMAUG-Toolbox.

## ASSOCIATED CONTENT

**Supporting information:** This material is available free of charge *via* the Internet at http://pubs.acs.org: Molecular Synthesis; Details of Simulated Trace Generation; Experimental Datasets Used in this Work; Further Details on Grid-Based Correlation Framework; Details of MCMC Feature-Finder; Additional Data from MCMC Results; Robustness of MCMC Results; Details on Trace Scoring; Robustness of Conclusions Drawn from Trace Scoring.

The MATLAB code used for this work is available free of charge at github.com/LabMonti/SMAUG-Toolbox.

## AUTHOR INFORMATION

### Corresponding Author

*Oliver L.A. Monti – Department of Chemistry and Biochemistry, University of Arizona, Tucson, Arizona 85721, United States; Department of Physics, University of Arizona, Tucson, Arizona 85721, United States; Email: monti@u.arizona.edu; Phone: ++ 520 626 1177; orcid.org/0000-0002-0974-7253.

### Authors

Nathan D. Bamberger – Department of Chemistry and Biochemistry, University of Arizona, Tucson, Arizona 85721, United States; orcid.org/0000-0001-5348-5695.

Dylan Dyer – Department of Chemistry and Biochemistry, University of Arizona, Tucson, Arizona 85721, United States; orcid.org/0000-0002-2747-9301.

Keshaba N. Parida – Department of Chemistry and Biochemistry, University of Arizona, Tucson, Arizona 85721, United States; orcid.org/0000-0003-3454-0868.

Dominic V. McGrath – Department of Chemistry and Biochemistry, University of Arizona, Tucson, Arizona 85721, United States; orcid.org/0000-0001-9605-2224.

### Author Contributions

N.B. and O.M. conceived the research ideas. K.P. synthesized the molecules, directed by D.M. N.B. and D.D. fabricated the MCBJ samples and collected the break junction data. N.B. designed, developed, and implemented the grid-based correlation framework, with advice and input from O.M. N.B. wrote the manuscript with advice and input from all authors.

### Notes

The authors declare no competing financial interests.

## ACKNOWLEDGMENTS

Financial support from the National Science Foundation, Award No. DMR-1708443, is gratefully acknowledged. Plasma etching of MCBJ samples was performed using a Plasmatherm reactive ion etcher acquired through an NSF MRI grant, Award No. ECCS-1725571. MCMC simulations were performed using High Performance Computing (HPC) resources supported by the University of Arizona TRIF, UITS, and RDI and maintained by the UA Research Technologies department. Quality control was performed using a scanning electron microscope in the W. M. Keck Center for Nano-Scale Imaging in the Department of Chemistry and Biochemistry at the University of Arizona with funding from the W. M. Keck Foundation Grant.

## REFERENCES

(1) Forrest, S. R. The Path to Ubiquitous and Low-Cost Organic Electronic Appliances on Plastic. *Nature* **2004**, *428* (6986), 911–918. https://doi.org/10.1038/nature02498.

(2) Liu, J.; Huang, X.; Wang, F.; Hong, W. Quantum Interference Effects in Charge Transport through Single-Molecule Junctions: Detection, Manipulation, and Application. *Acc. Chem. Res.* **2019**, *52* (1), 151–160. https://doi.org/10.1021/acs.accounts.8b00429.

(3) Chen, Z.; Chen, L.; Li, G.; Chen, Y.; Tang, C.; Zhang, L.; Liu, J.; Chen, L.; Yang, Y.; Shi, J.; Liu, J.; Xia, H.; Hong, W. Control of Quantum Interference in Single-Molecule Junctions via Jahn-Teller Distortion. *Cell Reports Physical Science* **2021**, 100329. https://doi.org/10.1016/j.xcrp.2021.100329.

(4) Camarasa-Gómez, M.; Hernangómez-Pérez, D.; Inkpen, M. S.; Lovat, G.; Fung, E.-D.; Roy, X.; Venkataraman, L.; Evers, F. Mechanically-Tunable Quantum Interference in Ferrocene-Based Single-Molecule Junctions. **2020**. https://doi.org/10.26434/chemrxiv.12252059.v1.

(5) Pal, A. N.; Li, D.; Sarkar, S.; Chakrabarti, S.; Vilan, A.; Kronik, L.; Smogunov, A.; Tal, O. Nonmagnetic Single-Molecule Spin-Filter Based on Quantum Interference. *Nature Communications* **2019**, *10* (1), 5565. https://doi.org/10.1038/s41467-019-13537-z.

(6) Qiu, S.; Miao, Y.-Y.; Zhang, G.-P.; Ren, J.; Wang, C.; Hu, G.-C. Manipulating Current Spin Polarization in Magnetic Single-Molecule Junctions via Destructive Quantum Interference. *J. Phys. Chem. C* **2020**. https://doi.org/10.1021/acs.jpcc.0c02828.

(7) Leary, E.; Rosa, A. L.; González, M. T.; Rubio-Bollinger, G.; Agraït, N.; Martín, N. Incorporating Single Molecules into Electrical Circuits. The Role of the Chemical Anchoring Group. *Chem. Soc. Rev.* **2015**, *44* (4), 920–942. https://doi.org/10.1039/C4CS00264D.

(8) Zeng, B.-F.; Wang, G.; Qian, Q.-Z.; Chen, Z.-X.; Zhang, X.-G.; Lu, Z.-X.; Zhao, S.-Q.; Feng, A.-N.; Shi, J.; Yang, Y.; Hong, W. Selective Fabrication of Single-Molecule Junctions by Interface Engineering. *Small* 2004720. https://doi.org/10.1002/smll.202004720.

(9) Isshiki, Y.; Fujii, S.; Nishino, T.; Kiguchi, M. Fluctuation in Interface and Electronic Structure of Single-Molecule Junctions Investigated by Current versus Bias Voltage Characteristics. *J. Am. Chem. Soc.* **2018**, *140* (10), 3760–3767. https://doi.org/10.1021/jacs.7b13694.

(10) Xu, B.; Tao, N. J. Measurement of Single-Molecule Resistance by Repeated Formation of Molecular Junctions. *Science* **2003**, *301* (5637), 1221–1223. https://doi.org/10.1126/science.1087481.

(11) McNeely, J.; Miller, N.; Pan, X.; Lawson, B.; Kamenetska, M. Angstrom-Scale Ruler Using Single Molecule Conductance Signatures. *J. Phys. Chem. C* **2020**. https://doi.org/10.1021/acs.jpcc.0c02063.

(12) Medina, S.; García-Arroyo, P.; Li, L.; Gunasekaran, S.; Stuyver, T.; José Mancheño, M.; Alonso, M.; Venkataraman, L.; L. Segura, J.; Cordón, J. C. Single-Molecule Conductance in a Unique Cross-Conjugated Tetra(Aminoaryl)-Ethene. *Chemical Communications* **2020**. https://doi.org/10.1039/D0CC07124B.

(13) Schmidt, M.; Wassy, D.; Hermann, M.; Teresa Gonzalez, M.; Agrait, N.; Angela Zotti, L.; Esser, B.; Leary, E. Single-Molecule Conductance of Dibenzopentalenes: Antiaromaticity and Quantum Interference. *Chemical Communications* **2020**. https://doi.org/10.1039/D0CC06810A.

(14) Inkpen, M. S.; Lemmer, M.; Fitzpatrick, N.; Milan, D. C.; Nichols, R. J.; Long, N. J.; Albrecht, T. New Insights into Single-Molecule Junctions Using a Robust, Unsupervised Approach to Data Collection and Analysis. *J. Am. Chem. Soc.* **2015**, *137* (31), 9971–9981. https://doi.org/10.1021/jacs.5b05693.

(15) Li, H. B.; Xi, Y.-F.; Hong, Z.-W.; Yu, J.; Li, X.-X.; Liu, W.-X.; Domulevicz, L.; Jin, S.; Zhou, X.-S.; Hihath, J. Temperature-Dependent Tunneling in Furan Oligomer Single-Molecule Junctions. *ACS Sens.* **2021**. https://doi.org/10.1021/acssensors.0c02278.

(16) Wang, Y.-H.; Yan, F.; Li, D.-F.; Xi, Y.-F.; Cao, R.; Zheng, J.-F.; Shao, Y.; Jin, S.; Chen, J.-Z.; Zhou, X.-S. Enhanced Gating Performance of Single-Molecule Conductance by Heterocyclic Molecules. *J. Phys. Chem. Lett.* **2021**, *12* (2), 758–763. https://doi.org/10.1021/acs.jpclett.0c03430.

(17) Martin, C. A.; Ding, D.; van der Zant, H. S. J.; van Ruitenbeek, J. M. Lithographic Mechanical Break Junctions for Single-Molecule Measurements in Vacuum: Possibilities and Limitations. *New J. Phys.* **2008**, *10* (6), 065008. https://doi.org/10.1088/1367-2630/10/6/065008.

(18) Huisman, E. H.; Trouwborst, M. L.; Bakker, F. L.; van Wees, B. J.; van der Molen, S. J. The Mechanical Response of Lithographically Defined Break Junctions. *Journal of Applied Physics* **2011**, *109* (10), 104305. https://doi.org/10.1063/1.3587192.

(19) Vrouwe, S. A. G.; van der Giessen, E.; van der Molen, S. J.; Dulic, D.; Trouwborst, M. L.; van Wees, B. J. Mechanics of Lithographically Defined Break Junctions. *Phys. Rev. B* **2005**, *71* (3), 035313. https://doi.org/10.1103/PhysRevB.71.035313.

(20) Muller, C. J.; van Ruitenbeek, J. M.; de Jongh, L. J. Conductance and Supercurrent Discontinuities in Atomic-Scale Metallic Constrictions of Variable Width. *Phys. Rev. Lett.* **1992**, *69* (1), 140–143. https://doi.org/10.1103/PhysRevLett.69.140.

(21) Liu, Y.; Ornago, L.; Carlotti, M.; Ai, Y.; El Abbassi, M.; Soni, S.; Asyuda, A.; Zharnikov, M.; van der Zant, H. S. J.; Chiechi, R. C. Intermolecular Effects on Tunneling through Acenes in Large-Area and Single-Molecule Junctions. *J. Phys. Chem. C* **2020**, *124* (41), 22776–22783. https://doi.org/10.1021/acs.jpcc.0c05781.

(22) Kobayashi, S.; Kaneko, S.; Kiguchi, M.; Tsukagoshi, K.; Nishino, T. Tolerance to Stretching in Thiol-Terminated Single-Molecule Junctions Characterized by Surface-Enhanced Raman Scattering. *J. Phys. Chem. Lett.* **2020**, *11* (16), 6712–6717. https://doi.org/10.1021/acs.jpclett.0c01526.

(23) Huang, C.; Jevric, M.; Borges, A.; Olsen, S. T.; Hamill, J. M.; Zheng, J.-T.; Yang, Y.; Rudnev, A.; Baghernejad, M.; Broekmann, P.; Petersen, A. U.; Wandlowski, T.; Mikkelsen, K. V.; Solomon, G. C.; Brøndsted Nielsen, M.; Hong, W. Single-Molecule Detection of Dihydroazulene Photo-Thermal Reaction Using Break Junction Technique. *Nature Communications* **2017**, *8* (1), 15436. https://doi.org/10.1038/ncomms15436.

(24) Chen, L.-C.; Zheng, J.; Liu, J.; Gong, X.-T.; Chen, Z.-Z.; Guo, R.-X.; Huang, X.; Zhang, Y.-P.; Zhang, L.; Li, R.; Shao, X.; Hong, W.; Zhang, H.-L. Nonadditive Transport in Multi-Channel Single-Molecule Circuits. *Small* *16* (39), 2002808. https://doi.org/10.1002/smll.202002808.

(25) Li, Z.; Mejía, L.; Marrs, J.; Jeong, H.; Hihath, J.; Franco, I. Understanding the Conductance Dispersion of Single-Molecule Junctions. *J. Phys. Chem. C* **2020**. https://doi.org/10.1021/acs.jpcc.0c08428.

(26) Martin, C. A.; Ding, D.; Sørensen, J. K.; Bjørnholm, T.; van Ruitenbeek, J. M.; van der Zant, H. S. J. Fullerene-Based Anchoring Groups for Molecular Electronics. *J. Am. Chem. Soc.* **2008**, *130* (40), 13198–13199. https://doi.org/10.1021/ja804699a.

(27) Kamenetska, M.; Koentopp, M.; Whalley, A. C.; Park, Y. S.; Steigerwald, M. L.; Nuckolls, C.; Hybertsen, M. S.; Venkataraman, L. Formation and Evolution of Single-Molecule Junctions. *Phys. Rev. Lett.* **2009**, *102* (12), 126803. https://doi.org/10.1103/PhysRevLett.102.126803.

(28) Harzmann, G. D.; Frisenda, R.; Zant, H. S. J. van der; Mayor, M. Single-Molecule Spin Switch Based on Voltage-Triggered Distortion of the Coordination Sphere. *Angewandte Chemie International Edition* **2015**, *54* (45), 13425–13430. https://doi.org/10.1002/anie.201505447.

(29) Li, Z.; Smeu, M.; Park, T.-H.; Rawson, J.; Xing, Y.; Therien, M. J.; Ratner, M. A.; Borguet, E. Hapticity-Dependent Charge Transport through Carbodithioate-Terminated [5,15-Bis(Phenylethynyl)Porphinato]Zinc(II) Complexes in Metal–Molecule–Metal Junctions. *Nano Lett.* **2014**, *14* (10), 5493–5499. https://doi.org/10.1021/nl502466a.

(30) Moreno-García, P.; Gulcur, M.; Manrique, D. Z.; Pope, T.; Hong, W.; Kaliginedi, V.; Huang, C.; Batsanov, A. S.; Bryce, M. R.; Lambert, C.; Wandlowski, T. Single-Molecule Conductance of Functionalized Oligoynes: Length Dependence and Junction Evolution. *J. Am. Chem. Soc.* **2013**, *135* (33), 12228–12240. https://doi.org/10.1021/ja4015293.

(31) Hong, W.; Manrique, D. Z.; Moreno-García, P.; Gulcur, M.; Mishchenko, A.; Lambert, C. J.; Bryce, M. R.; Wandlowski, T. Single Molecular Conductance of Tolanes: Experimental and Theoretical Study on the Junction Evolution Dependent on the Anchoring Group. *J. Am. Chem. Soc.* **2012**, *134* (4), 2292–2304. https://doi.org/10.1021/ja209844r.

(32) Yoo, P. S.; Kim, T. Linker-Dependent Junction Formation Probability in Single-Molecule Junctions. *Bulletin of the Korean Chemical Society* **2015**, *36* (1), 265–268. https://doi.org/10.1002/bkcs.10061.

(33) Vladyka, A.; Perrin, M. L.; Overbeck, J.; Ferradás, R. R.; García-Suárez, V.; Gantenbein, M.; Brunner, J.; Mayor, M.; Ferrer, J.; Calame, M. In-Situ Formation of One-Dimensional Coordination Polymers in Molecular Junctions. *Nature Communications* **2019**, *10* (1), 262. https://doi.org/10.1038/s41467-018-08025-9.

(34) Cabosart, D.; El Abbassi, M.; Stefani, D.; Frisenda, R.; Calame, M.; van der Zant, H. S. J.; Perrin, M. L. A Reference-Free Clustering Method for the Analysis of Molecular Break-Junction Measurements. *Appl. Phys. Lett.* **2019**, *114* (14), 143102. https://doi.org/10.1063/1.5089198.

(35) Frisenda, R.; Stefani, D.; van der Zant, H. S. J. Quantum Transport through a Single Conjugated Rigid Molecule, a Mechanical Break Junction Study. *Acc. Chem. Res.* **2018**, *51* (6), 1359–1367. https://doi.org/10.1021/acs.accounts.7b00493.

(36) Nováková Lachmanová, Š.; Kolivoška, V.; Šebera, J.; Gasior, J.; Mészáros, G.; Dupeyre, G.; Lainé, P. P.; Hromadová, M. Environmental Control of Single-Molecule Junction Evolution and Conductance: A Case Study of Expanded Pyridinium Wiring. *Angew Chem Int Ed Engl* **2021**, *60* (9), 4732–4739. https://doi.org/10.1002/anie.202013882.

(37) Metzger, R. M. Unimolecular Electronics. *Chem. Rev.* **2015**, *115* (11), 5056–5115. https://doi.org/10.1021/cr500459d.

(38) Kim, B.; Choi, S. H.; Zhu, X.-Y.; Frisbie, C. D. Molecular Tunnel Junctions Based on π-Conjugated Oligoacene Thiols and Dithiols between Ag, Au, and Pt Contacts: Effect of Surface Linking Group and Metal Work Function. *J. Am. Chem. Soc.* **2011**, *133* (49), 19864–19877. https://doi.org/10.1021/ja207751w.

(39) Rodriguez-Gonzalez, S.; Xie, Z.; Galangau, O.; Selvanathan, P.; Norel, L.; Van Dyck, C.; Costuas, K.; Frisbie, C. D.; Rigaut, S.; Cornil, J. HOMO Level Pinning in Molecular Junctions: Joint Theoretical and Experimental Evidence. *J. Phys. Chem. Lett.* **2018**, *9* (9), 2394–2403. https://doi.org/10.1021/acs.jpclett.8b00575.

(40) Dyck, C. V.; Geskin, V.; Cornil, J. Fermi Level Pinning and Orbital Polarization Effects in Molecular Junctions: The Role of Metal Induced Gap States.

*Advanced Functional Materials* **2014**, *24* (39), 6154–6165. https://doi.org/10.1002/adfm.201400809.

(41) Xie, Z.; Bâldea, I.; Frisbie, C. D. Determination of Energy-Level Alignment in Molecular Tunnel Junctions by Transport and Spectroscopy: Self-Consistency for the Case of Oligophenylene Thiols and Dithiols on Ag, Au, and Pt Electrodes. *J. Am. Chem. Soc.* **2019**, *141* (8), 3670–3681. https://doi.org/10.1021/jacs.8b13370.

(42) Heimel, G.; Romaner, L.; Brédas, J.-L.; Zojer, E. Interface Energetics and Level Alignment at Covalent Metal-Molecule Junctions: π-Conjugated Thiols on Gold. *Phys. Rev. Lett.* **2006**, *96* (19), 196806. https://doi.org/10.1103/PhysRevLett.96.196806.

(43) Wu, B. H.; Ivie, J. A.; Johnson, T. K.; Monti, O. L. A. Uncovering Hierarchical Data Structure in Single Molecule Transport. *The Journal of Chemical Physics* **2017**, *146* (9), 092321. https://doi.org/10.1063/1.4974937.

(44) Korshoj, L. E.; Afsari, S.; Chatterjee, A.; Nagpal, P. Conformational Smear Characterization and Binning of Single-Molecule Conductance Measurements for Enhanced Molecular Recognition. *J. Am. Chem. Soc.* **2017**, *139* (43), 15420–15428. https://doi.org/10.1021/jacs.7b08246.

(45) Hamill, J. M.; Zhao, X. T.; Mészáros, G.; Bryce, M. R.; Arenz, M. Fast Data Sorting with Modified Principal Component Analysis to Distinguish Unique Single Molecular Break Junction Trajectories. *Phys. Rev. Lett.* **2018**, *120* (1), 016601. https://doi.org/10.1103/PhysRevLett.120.016601.

(46) Lauritzen, K. P.; Magyarkuti, A.; Balogh, Z.; Halbritter, A.; Solomon, G. C. Classification of Conductance Traces with Recurrent Neural Networks. *The Journal of Chemical Physics* **2018**, *148* (8), 084111. https://doi.org/10.1063/1.5012514.

(47) Huang, F.; Li, R.; Wang, G.; Zheng, J.; Tang, Y.; Liu, J.; Yang, Y.; Yao, Y.; Shi, J.; Hong, W. Automatic Classification of Single-Molecule Charge Transport Data with an Unsupervised Machine-Learning Algorithm. *Physical Chemistry Chemical Physics* **2019**, *22*, 1674–1681. https://doi.org/10.1039/C9CP04496E.

(48) Magyarkuti, A.; Balogh, N.; Balogh, Z.; Venkataraman, L.; Halbritter, A. Unsupervised Feature Recognition in Single Molecule Break Junction Data. *Nanoscale* **2020**, *12*, 8355–8363. https://doi.org/10.1039/D0NR00467G.

(49) Vladyka, A.; Albrecht, T. Unsupervised Classification of Single-Molecule Data with Autoencoders and Transfer Learning. *Mach. Learn.: Sci. Technol.* **2020**, *1* (3), 035013. https://doi.org/10.1088/2632-2153/aba6f2.

(50) Bamberger, N. D.; Ivie, J. A.; Parida, K.; McGrath, D. V.; Monti, O. L. A. Unsupervised Segmentation-Based Machine Learning as an Advanced Analysis Tool for Single Molecule Break Junction Data. *J. Phys. Chem. C* **2020**, *124* (33), 18302–18315. https://doi.org/10.1021/acs.jpcc.0c03612.

(51) Liu, B.; Murayama, S.; Komoto, Y.; Tsutsui, M.; Taniguchi, M. Dissecting Time-Evolved Conductance Behavior of Single Molecule Junctions by Non-Parametric Machine Learning. *J. Phys. Chem. Lett.* **2020**, *11* (16), 6567–6572. https://doi.org/10.1021/acs.jpclett.0c01948.

(52) Lin, L.; Tang, C.; Dong, G.; Chen, Z.; Pan, Z.; Liu, J.; Yang, Y.; Shi, J.; Ji, R.; Hong, W. Spectral Clustering to Analyze the Hidden Events in Single-Molecule Break Junctions. *J. Phys. Chem. C* **2021**, *125* (6), 3623–3630. https://doi.org/10.1021/acs.jpcc.0c11473.

(53) El Abbassi, M.; Overbeck, J.; Braun, O.; Calame, M.; van der Zant, H. S. J.; Perrin, M. L. Benchmark and Application of Unsupervised Classification Approaches for Univariate Data. *Communications Physics* **2021**, *4* (1), 1–9. https://doi.org/10.1038/s42005-021-00549-9.

(54) Lemmer, M.; Inkpen, M. S.; Kornysheva, K.; Long, N. J.; Albrecht, T. Unsupervised Vector-Based Classification of Single-Molecule Charge Transport Data. *Nature Communications* **2016**, *7*, 12922. https://doi.org/10.1038/ncomms12922.

(55) Albrecht, T.; Slabaugh, G.; Alonso, E.; Al-Arif, S. M. R. Deep Learning for Single-Molecule Science. *Nanotechnology* **2017**, *28* (42), 423001. https://doi.org/10.1088/1361-6528/aa8334.

(56) Quek, S. Y.; Venkataraman, L.; Choi, H. J.; Louie, S. G.; Hybertsen, M. S.; Neaton, J. B. Amine−Gold Linked Single-Molecule Circuits: Experiment and Theory. *Nano Lett.* **2007**, *7* (11), 3477–3482. https://doi.org/10.1021/nl072058i.

(57) Jang, S.-Y.; Reddy, P.; Majumdar, A.; Segalman, R. A. Interpretation of Stochastic Events in Single Molecule Conductance Measurements. *Nano Lett.* **2006**, *6* (10), 2362–2367. https://doi.org/10.1021/nl0609495.

(58) Brooke, R. J.; Szumski, D. S.; Vezzoli, A.; Higgins, S. J.; Nichols, R. J.; Schwarzacher, W. Dual Control of Molecular Conductance through PH and Potential in Single-Molecule Devices. *Nano Lett.* **2018**, *18* (2), 1317–1322. https://doi.org/10.1021/acs.nanolett.7b04995.

(59) Halbritter, A.; Makk, P.; Mackowiak, Sz.; Csonka, Sz.; Wawrzyniak, M.; Martinek, J. Regular Atomic Narrowing of Ni, Fe, and V Nanowires Resolved by Two-Dimensional Correlation Analysis. *Phys. Rev. Lett.* **2010**, *105* (26), 266805. https://doi.org/10.1103/PhysRevLett.105.266805.

(60) Yanson, A. I.; Bollinger, G. R.; van den Brom, H. E.; Agraït, N.; van Ruitenbeek, J. M. Formation and Manipulation of a Metallic Wire of Single Gold Atoms. *Nature* **1998**, *395* (6704), 783–785. https://doi.org/10.1038/27405.

(61) Ohnishi, H.; Kondo, Y.; Takayanagi, K. Quantized Conductance through Individual Rows of Suspended Gold Atoms. *Nature* **1998**, *395* (6704), 780–783. https://doi.org/10.1038/27399.

(62) Johnson, T. K.; Ivie, J. A.; Jaruvang, J.; Monti, O. L. A. Fast Sensitive Amplifier for Two-Probe Conductance

Measurements in Single Molecule Break Junctions. *Review of Scientific Instruments* **2017**, *88*. https://doi.org/10.1063/1.4978962.

(63) González, M. T.; Leary, E.; García, R.; Verma, P.; Herranz, M. Á.; Rubio-Bollinger, G.; Martín, N.; Agraït, N. Break-Junction Experiments on Acetyl-Protected Conjugated Dithiols under Different Environmental Conditions. *J. Phys. Chem. C* **2011**, *115* (36), 17973–17978. https://doi.org/10.1021/jp204005v.

(64) Jones, D. S. *Elementary Information Theory*; Oxford University Press, 1979.

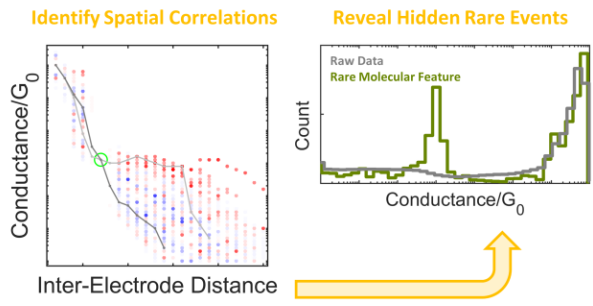

TOC Graphic

# Supplementary Information for:

# Grid-Based Correlation Analysis to Identify Rare Quantum Transport Behaviors

Nathan D. Bamberger[1], Dylan Dyer[1], Keshaba N. Parida[1], Dominic V. McGrath[1], and Oliver L.A. Monti[1,2,*]

[1]Department of Chemistry and Biochemistry, University of Arizona, 1306 E. University Blvd., Tucson, Arizona 85721, USA
[2]Department of Physics, University of Arizona, 1118 E. Fourth Street, Tucson, Arizona 85721, USA

## Contents

S.1 Molecular Synthesis

*S.1.1 Synthetic Procedures*

All reactions were carried out using oven-dried (120 °C) glassware. All commercially available reagents and common solvents were used without further purification. Reaction mixtures were magnetically stirred and progress was monitored by thin layer chromatography using Merck Silica Gel 60 F254 plates visualized by fluorescence quenching under UV light. Flash chromatographic purification was performed on silica 32-63, 60 Å using flash chromatography under positive nitrogen pressure. Concentration under reduced pressure was performed by rotary evaporation at <40 °C at the appropriate pressure. Structural characterization data was obtained on commercially available instrumentation with specifications as indicated in the experimental descriptions.

**(*E*)-1,2-*bis*(2,3-dihydrobenzo[*b*]thiophen-5-yl)ethene (a.k.a. OPV2-2BT):** McMurry coupling was 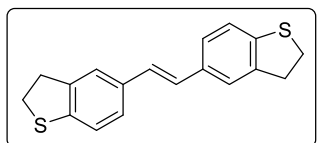 performed using a procedure reported in the literature with slight modification.[1] To a chilled suspension of Zn (1.98 g, 30.4 mmol) in THF-$CH_2Cl_2$ (40 mL, 3:1 *v/v*) was slowly added $TiCl_4$ (1.73 mL, 14.7 mmol) at 0 °C under $N_2$ atmosphere. The resulting mixture was maintained at reflux for 2 h during which time the color of the mixture became dark brown. The mixture was cooled to 0 °C using an ice bath, followed by slow addition of a solution of 2,3-dihydrobenzo[*b*]thiophene-5-carbaldehyde[2] (1.3 g, 7.9 mmol) and $CH_2Cl_2$ (3 mL). The mixture was stirred while warming to rt for 10 mins, then maintained at reflux overnight. The mixture was cooled to 0 °C and subjected to dropwise addition of aqueous saturated $Na_2CO_3$ (34 mL). The resulting solid was filtered, washed with acetone (100 mL), and dried under reduced pressure. The solid was triturated with $CHCl_3$ (20 mL*4) and the remaining solid residue was transferred to a beaker and extracted with boiling $CHCl_3$ (100 mL) (repeated thrice). *Caution: Product has poor solubility!* The organic layers were combined, evaporated, and subjected to a silica gel

column chromatography purification to afford the desired product as a light-yellow powder (0.81 g, 69%): **R**$_f$: 0.4 (1:19 *v/v*, EtOAc/hexane); **¹H NMR** (400 MHz, CDCl$_3$): δ 7.34 (d, *J* = 1.7 Hz, 2H), 7.23 (dd, *J* = 8.1, 1.7 Hz, 2H), 7.18 (d, *J* = 8.0 Hz, 2H), 6.96 (s, 2H), 3.34–3.46 (m, 4H), 3.26-3.33 (m, 4H) ppm; **¹³C NMR** (101 MHz, CDCl$_3$): δ 141.1, 140.8, 134.1, 127.2, 126.1, 122.3, 122.1, 36.2, 33.7 ppm; **LRMS (MS+)**: *m/z* calculated for C$_{18}$H$_{16}$S$_2$ 296.0693 [M]$^+$; found 296.95.

*Synthetic Procedure of (E)-S,S'-(Ethene-1,2-diylbis(4,1-phenylene)) diethanethioate:*

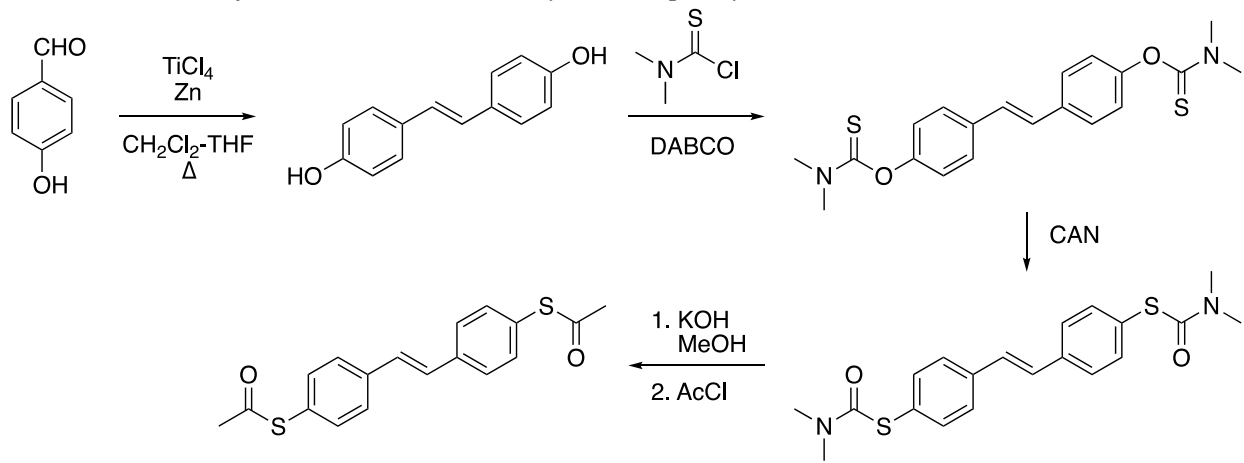

**(*E*)-*O*,*O*'-(Ethene-1,2-diyl*bis*(4,1-phenylene))*bis*(dimethylcarbamothioate):**[3] (*E*)-*O*,*O*'-(Ethene-1,2-diyl*bis*(4,1-phenylene))*bis*(dimethylcarbamothioate) was prepared following a literature procedure.[3,4] McMurry coupling of 4-hydroxybenzaldehyde gave (*E*)-4,4'-(ethene-1,2-diyl)diphenol[4] which was further subjected to *bis*-thiocarbamoylation to afford (*E*)-*O*,*O*'-(ethene-1,2-diyl*bis*(4,1-phenylene))*bis*(dimethylcarbamothioate):[3] **¹H NMR** (400 MHz, CDCl$_3$): δ 7.52 (d, *J* = 8.0 Hz, 4H), 7.04-7.08 (m, 6H), 3.47 (s, 6H), 3.36 (s, 6H) ppm; **¹³C NMR** (101 MHz, CDCl$_3$): δ 187.8, 153.5, 135.2, 128.2, 127.3, 123.1, 43.4, 38.9 ppm.

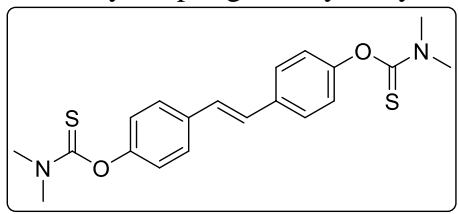

**(*E*)-*S*,*S*'-(Ethene-1,2-diyl*bis*(4,1-phenylene))*bis*(dimethylcarbamothioate):**[3] A mixture of (*E*)-*O*,*O*'-(Ethene-1,2-diyl*bis*(4,1-phenylene))*bis*(dimethylcarbamothioate)

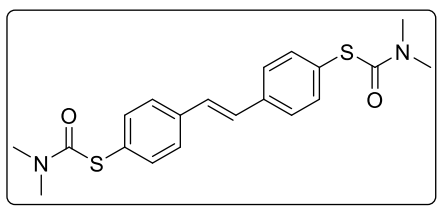

(100 mg, 0.258 mmol), ceric ammonium nitrate (283 mg, 0.517 mmol), and DMSO (5 mL) was stirred at rt for 36 h.[5] The reaction was quenched with H$_2$O (20 mL) and extracted with CH$_2$Cl$_2$ (3 x 20 mL). The combined organic layer was dried (MgSO$_4$), filtered, and concentrated under reduced pressure. The crude residue was triturated with EtOAc and filtered to afford the desired product as a pale-yellow powder (48 mg, 48%): **R**$_f$: 0.2 (CHCl$_3$); **¹H NMR** (400 MHz, CDCl$_3$): δ 7.46-7.53 (m, 8H), 7.12 (s, 2H), 3.05-3.10 (m, 12H) ppm; **¹³C NMR** (101 MHz, CDCl$_3$): δ 166.9, 138.1, 136.0, 129.2, 128.1, 127.2, 37.1.

**(*E*)-*S,S'*-(Ethene-1,2-diyl*bis*(4,1-phenylene)) diethanethioate (a.k.a. OPV2-2SAc):**[6] A mixture of (*E*)-*S,S'*-(Ethene-1,2-diyl*bis*(4,1-phenylene))*bis*(dimethylcarbamothioate) (0.15 g, 0.39 mmol), KOH (0.49 g, 0.87 mmol), and THF-MeOH (4.5 mL, 7:2 *v/v*)[7] was purged with $N_2$[8] and maintained at 50 °C for overnight. The mixture was cooled to rt, ice was added and the resulting mixture was acidified with 6N HCl to a pH ~ 2. A solid emulsion was observed which was evaporated to dryness. To this residue was added 1 mL AcCl and 0.2 mL Et$_3$N at 0 °C. *Caution: the addition is very exothermic!* After addition, the mixture was maintained at 50 °C for overnight. The mixture was cooled to rt, ice was added and the resulting mixture was extracted with CH$_2$Cl$_2$ (2 x 5 mL) and EtOAc (5 mL). The combined organic layers were dried (MgSO$_4$), filtered, and concentrated under reduced pressure. The crude residue was subjected to a silica gel column chromatography to afford the desired product as an off-white solid (23 mg, 18%): **R$_f$**: 0.2 (1:9 *v/v*, EtOAc/hexane); **$^1$H NMR** (400 MHz, CDCl$_3$): δ 7.55 (d, *J* = 8.0 Hz, 4H), 7.41 (d, *J* = 8.0, 4H), 7.1s (s, 2H), 2.44 (s, 6H) ppm; **$^{13}$C NMR** (101 MHz, CDCl$_3$): δ 194.1, 138.3, 134.8, 129.3, 127.45, 127.4, 30.4 ppm.

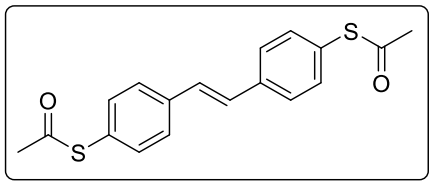

**(*E*)-1,2-*bis*(4-(methylthio)phenyl)ethene (a.k.a. OPV2-2SMe):**[9] McMurry coupling was performed using a literature reported procedure with slight modification.[1] To a chilled suspension of Zn (1.64 g, 25.2 mmol) in THF-CH$_2$Cl$_2$ mixture (40 mL, 3:1 *v/v*) was slowly added TiCl$_4$ (1.33 mL, 12.2 mmol) at 0 °C under N$_2$ atmosphere. The resulting mixture was maintained at reflux for 2 h during which the color became dark brown. The mixture was cooled to 0 °C and a solution of 4-(methylthio)benzaldehyde (1.00 g, 6.57 mmol) and CH$_2$Cl$_2$ (5 mL) was slowly added. It was allowed to stir at rt for 10 mins, then refluxed for overnight. After that, reaction was cooled to 0 °C, and dropwise added 30 mL aqueous saturated Na$_2$CO$_3$ solution. Filtered off the solid, washed with acetone (30 mL) and CHCl$_3$ (50 mL). The filtrate was evaporated (to remove acetone and THF) and extracted with CHCl$_3$ (30 mL*3). The combined organic layer was dried (MgSO$_4$) and concentrated, and the residue was purified by silica gel column chromatography to afford the desired product as a colorless solid (0.6 g, 67%): **R$_f$**: 0.2 (1:9 *v/v*, EtOAc/hexane); **$^1$H NMR** (400 MHz, CDCl$_3$): δ 7.42 (d, *J* = 8.0 Hz, 4H), 7.24 (d, *J* = 8.0, 4H), 7.02 (s, 2H), 2.51 (s, 6H) ppm; **$^{13}$C NMR** (101 MHz, CDCl$_3$): δ 137.9, 134.5, 127.6, 126.96, 126.9, 16.01 ppm.

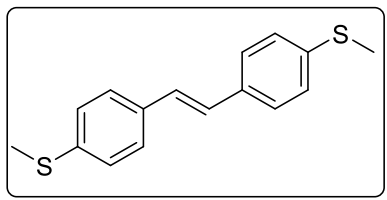

*S.1.2 NMR Characterization*

**The nuclear magnetic resonance** (NMR) spectra were recorded on Bruker Avance III 400 spectrometer operating at 400 MHz ($^1$H) and 101 MHz ($^{13}$C) respectively. Chemical shifts (δ) are reported in parts per million (ppm) with tetramethylsilane (TMS) as the internal standard (at 0 ppm) for both $^1$H and $^{13}$C. TMS and residual NMR solvent peaks are not labelled. All $^{13}$C spectra were measured with complete proton decoupling. Data are reported as follows: s = singlet, d = doublet, t = triplet, q = quartet, m = multiplet, dd = doublet of doublets, coupling constants (*J*) in Hz. MNova software was used to plot NMR spectra.

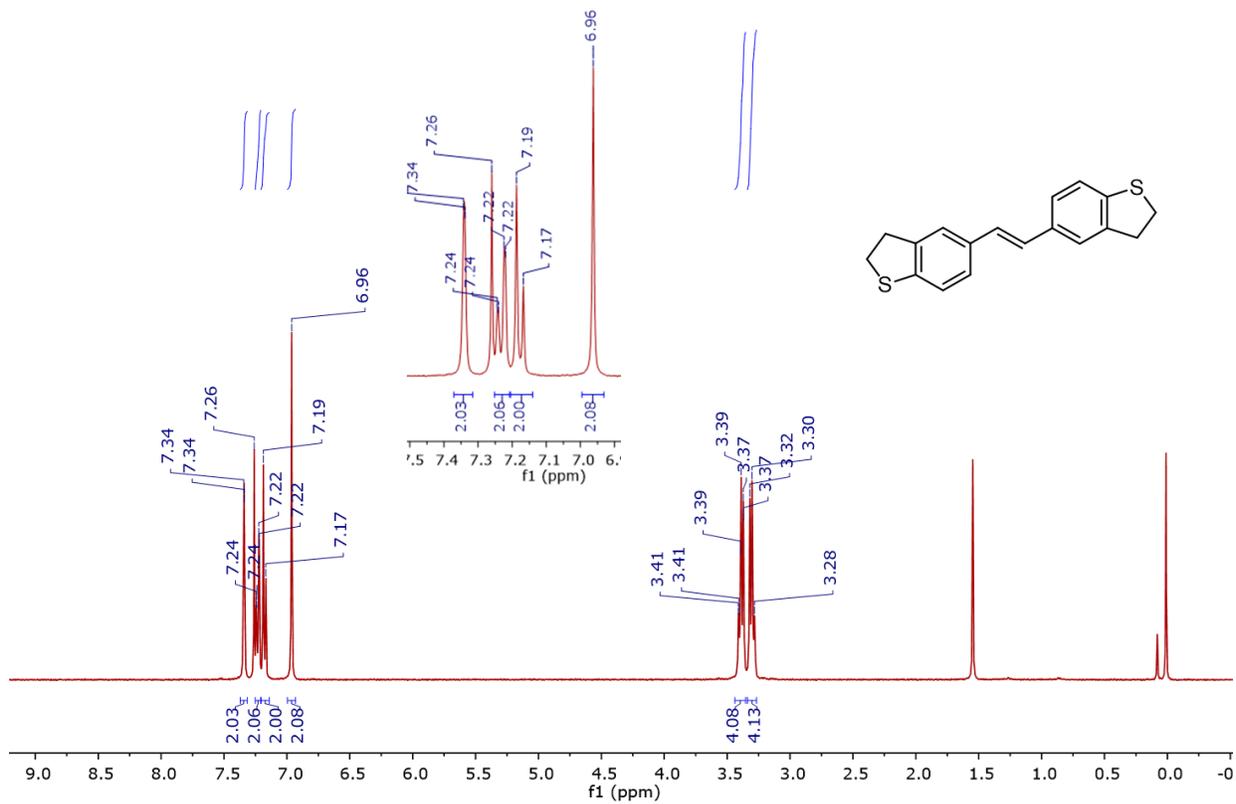

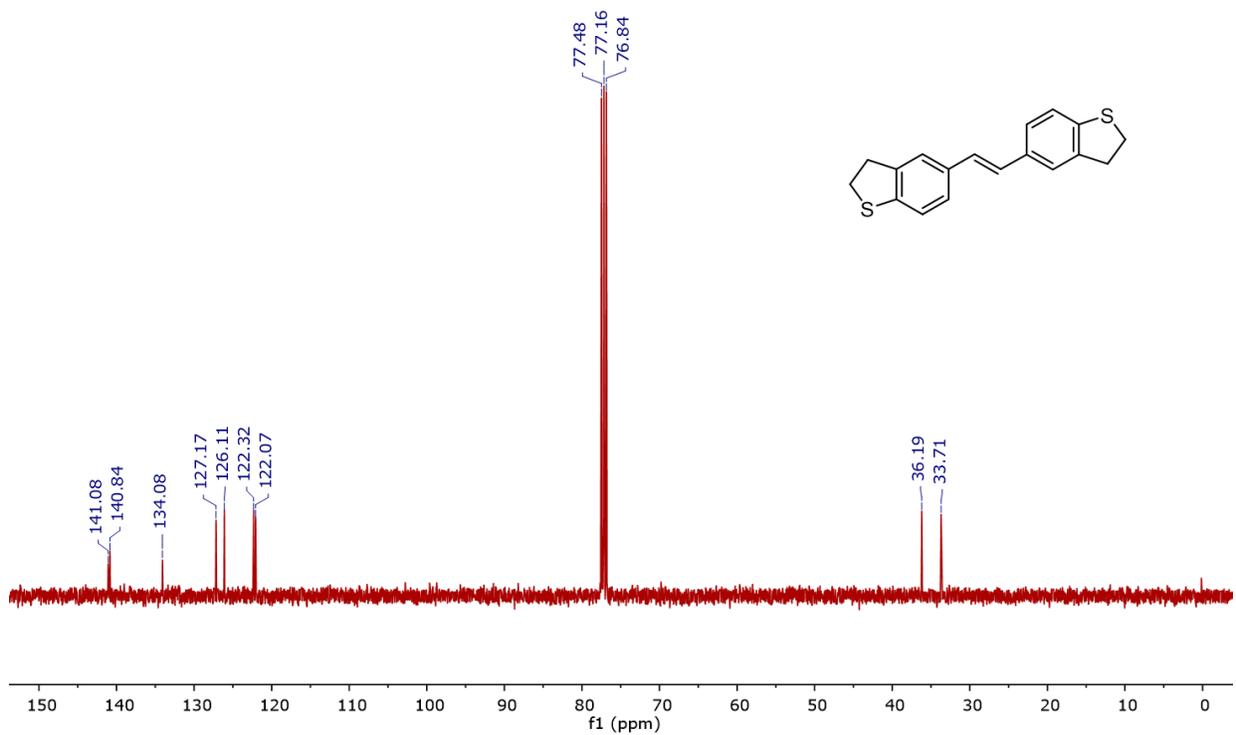

**Figure S1**. ¹H NMR (400 MHz) and ¹³C NMR (101 MHz) spectra of (*E*)-1,2-*bis*(2,3-dihydrobenzo[*b*]thiophen-5-yl)ethene (a.k.a. OPV2-2BT) in CDCl₃.

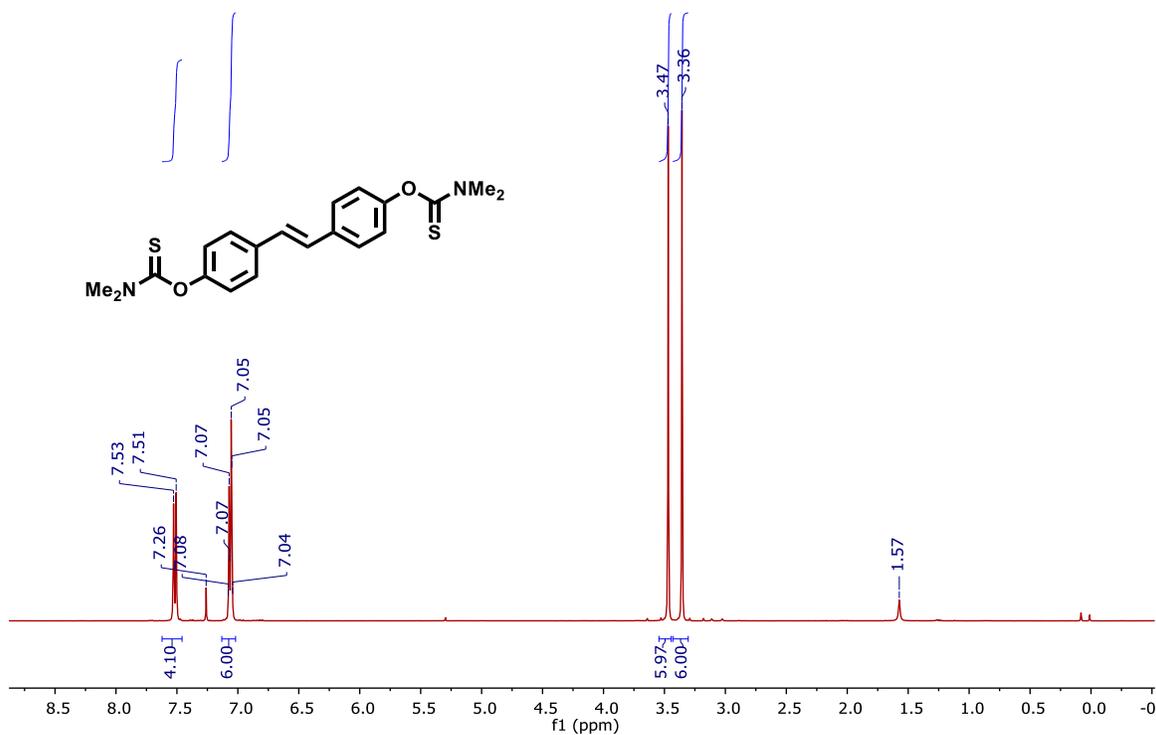

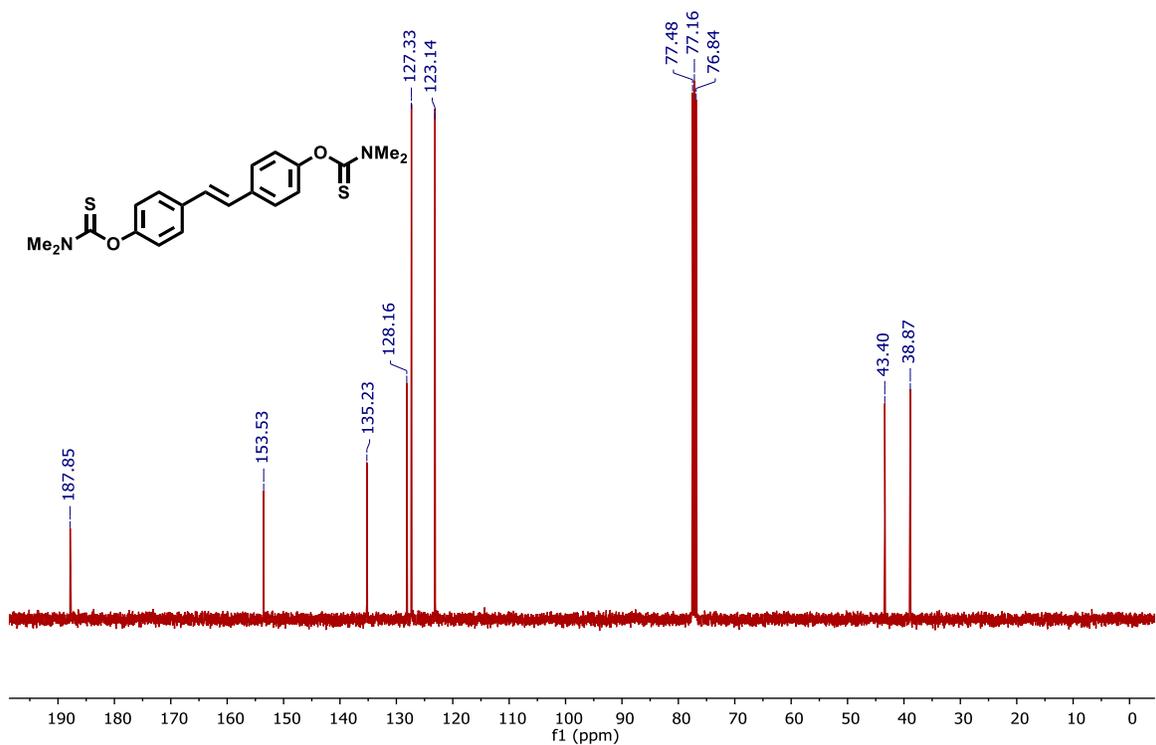

**Figure S2**. ¹H NMR (400 MHz) and ¹³C NMR (101 MHz) spectra of (*E*)-*O,O'*-(Ethene-1,2-diyl*bis*(4,1-phenylene))*bis*(dimethylcarbamothioate) in CDCl₃.

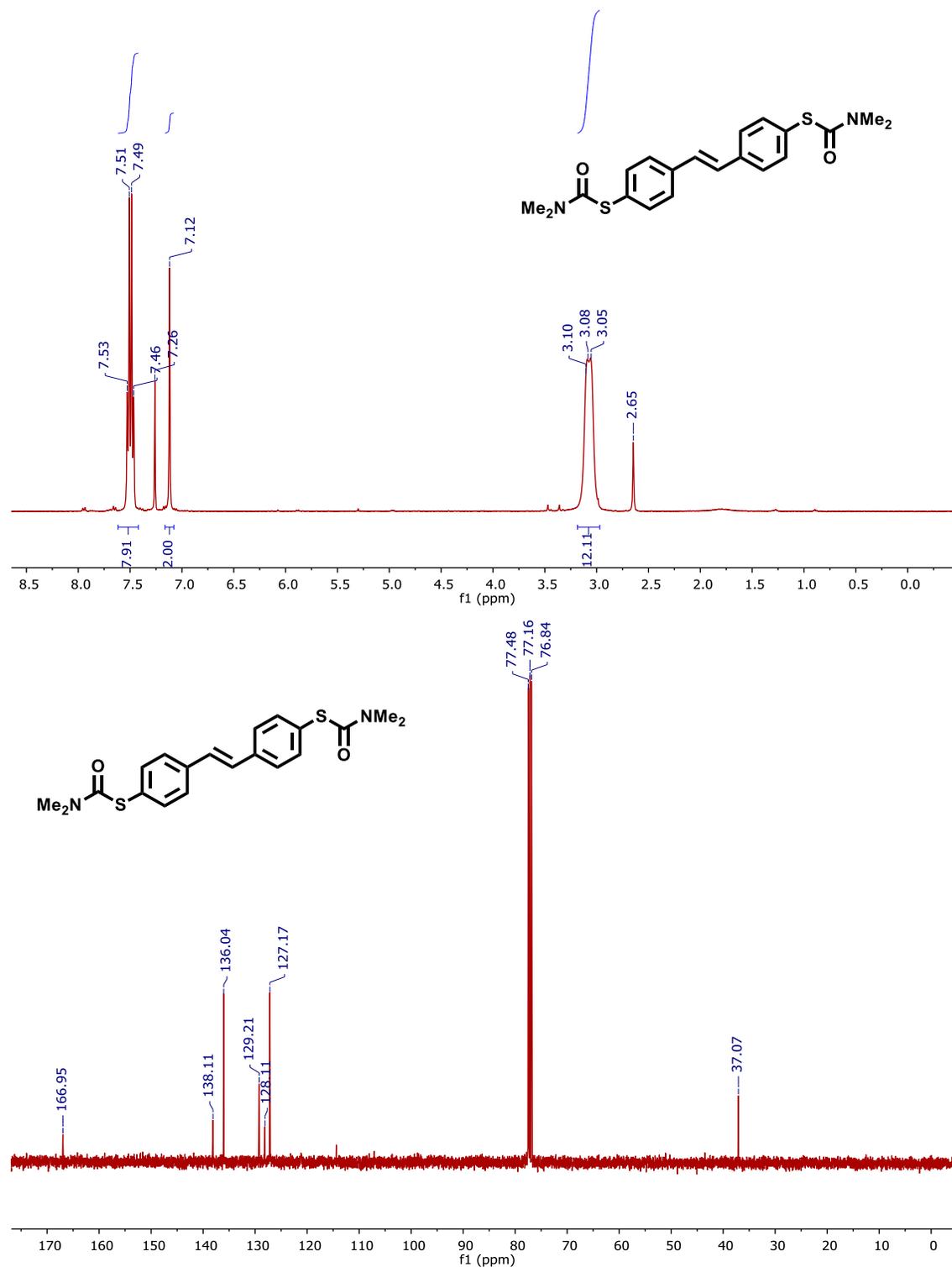

**Figure S3**. ¹H NMR (400 MHz) and ¹³C NMR (101 MHz) spectra of (*E*)-*S*,*S'*-(ethene-1,2-diyl*bis*(4,1-phenylene))*bis*(dimethylcarbamothioate) in CDCl₃.

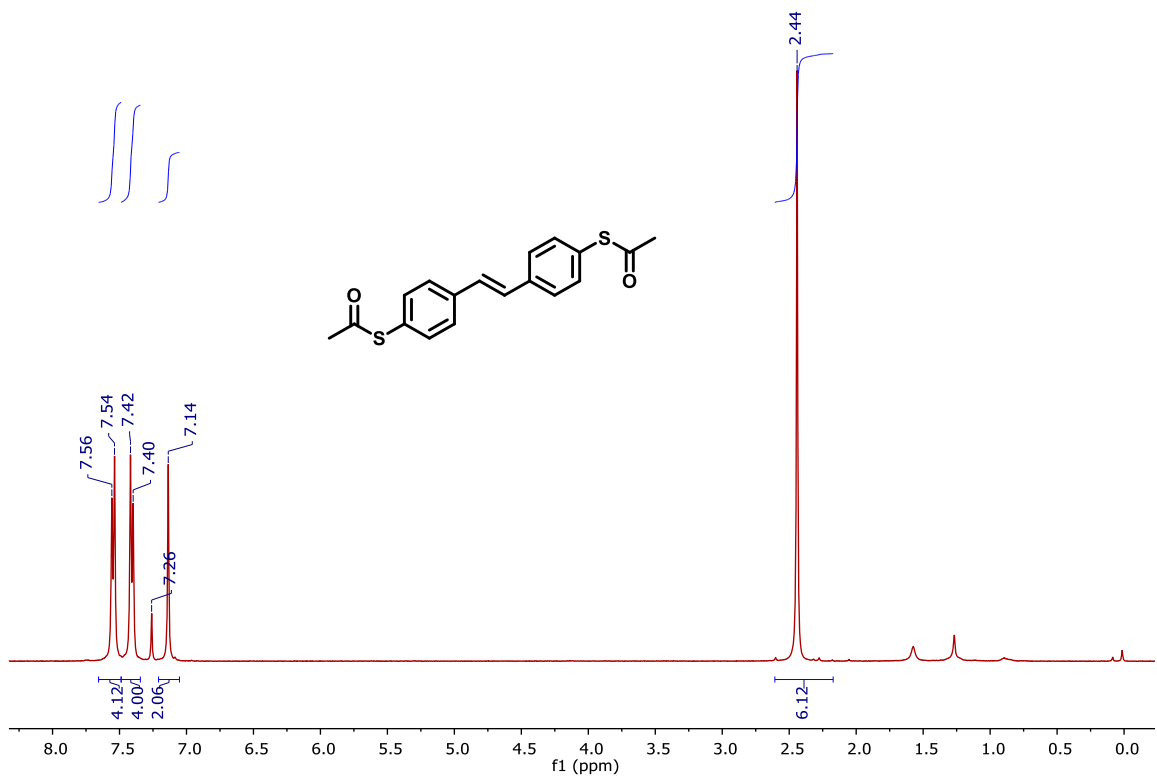

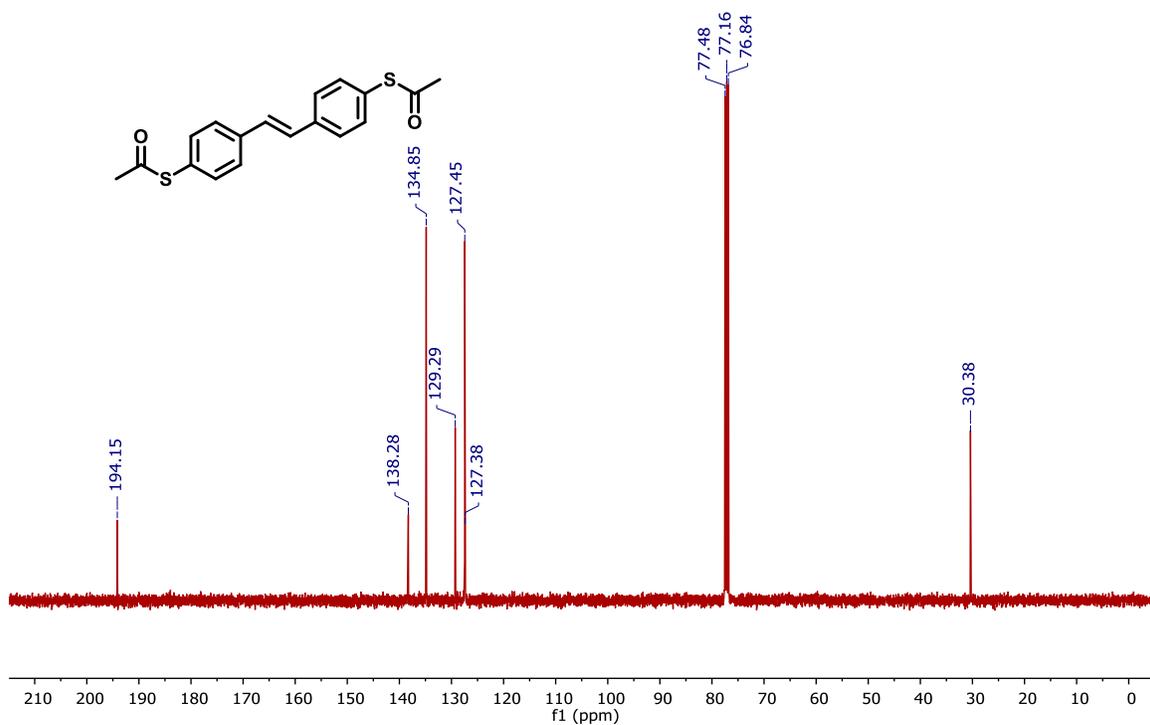

**Figure S4**. $^1$H NMR (400 MHz) and $^{13}$C NMR (101 MHz) spectra of (*E*)-*S*,*S*'-(Ethene-1,2-diyl*bis*(4,1-phenylene)) diethanethioate (a.k.a OPV2-2SAc) in CDCl$_3$.

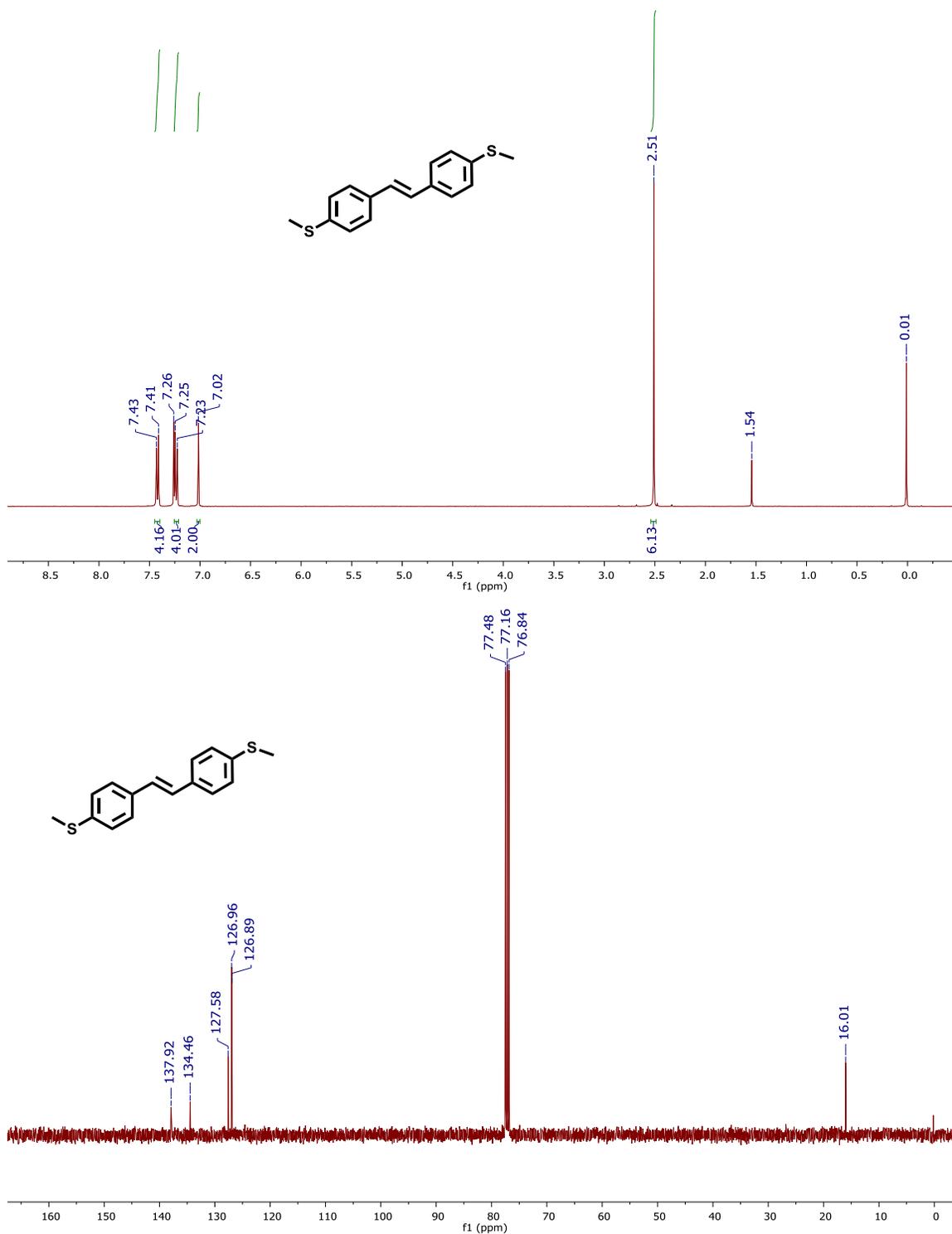

**Figure S5**. ¹H NMR (400 MHz) and ¹³C NMR (101 MHz) spectra of (*E*)-1,2-*bis*(4-(methylthio)phenyl)ethene (a.k.a. OPV2-2SMe) in CDCl₃.

*S.1.3 Mass Spectrometry Characterization*

Low resolution mass spectra were recorded in amaZon SL instrument by the MS service at Analytical & Biological Mass Spectrometry Facility, University of Arizona. LRMS (*m/z*) was recorded in the positive/negative mode using $CH_2Cl_2$ in MeOH, or solvent free probe mode using Tuning Mix.m method.

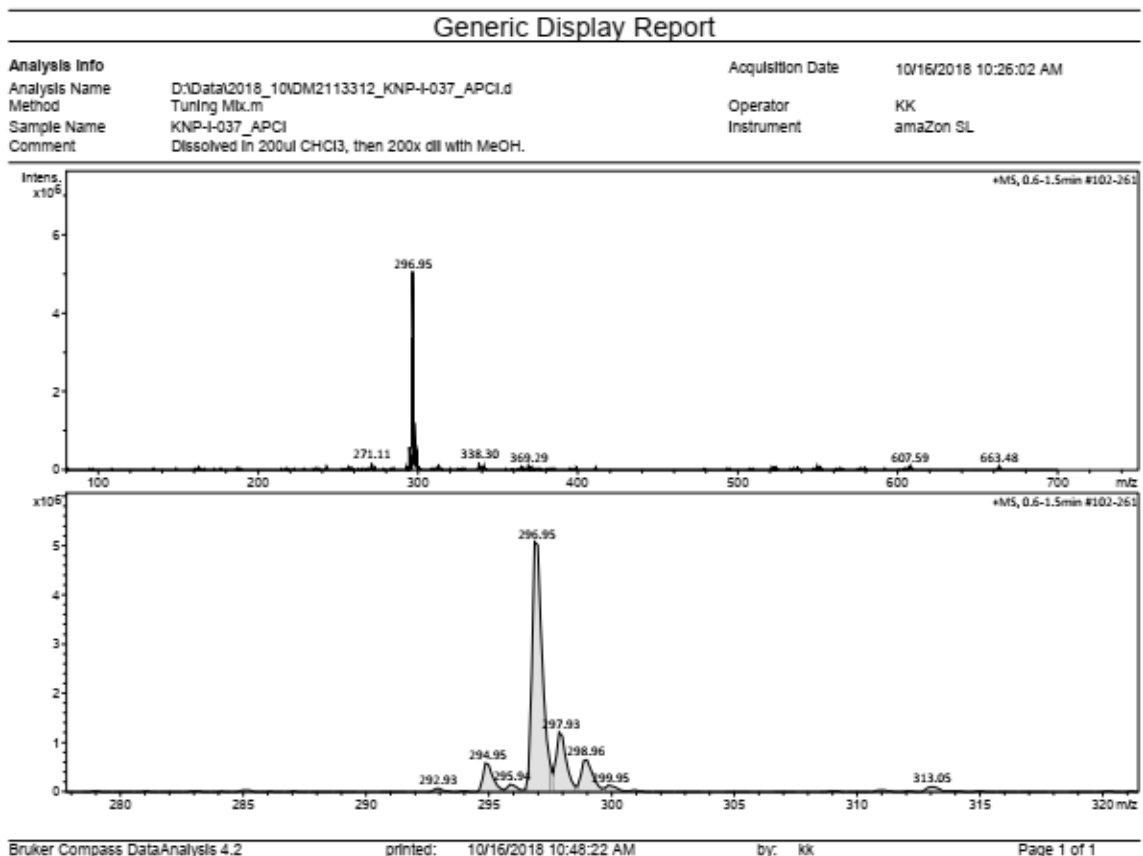

**Figure S6**. LRMS spectrum of (*E*)-1,2-*bis*(2,3-dihydrobenzo[*b*]thiophen-5-yl)ethene (a.k.a. OPV2-2BT).

S.2 Details of Simulated Trace Generation

Each simulated trace used in this work was generated as a list of points ($x_i$, $y_i$) with $x$ representing inter-electrode distance, in nm, and $y$ representing the logarithm of the junction conductance, in $\log(G/G_0)$. Each trace begins at $x = -0.25$ nm and proceeds, with 0.002 nm steps between each successive point, until $y = -7$. Each trace was initially generated as a "model trace", after which low-amplitude random noise was added on to create the final trace.

Each tunneling model trace begins with a perfectly flat pre-rupture plateau at $y = G_{pr}$ which continues until $x = 0$ nm. Next, the model trace drops linearly with a slope of *DropSlope* until a *y*-value of $G_{drop}$ is reached. Finally, the model trace drops linearly with a slope of *TunnSlope* until the trace ends at $y = -7$ $\log(G/G_0)$. For each model trace, these parameters are independently chosen from respective normal distributions (see **Table S1**).

Each molecular model trace begins with the same three sections as the model tunneling traces, but the third section ends when $y$ reaches $G_{plateau}$. The model trace then proceeds linearly with a slope of *PlateauSlope* until $x = PlateauLength$. Finally, the model trace drops linearly with a slope of *BreakOffSlope* until the trace ends at y = -7 log($G/G_0$). Just as with the tunneling model traces, all parameters are independently chosen from their respective normal distributions for each model molecular trace (see **Table S1**).

To each model trace, a small amount of noise $n_i$ was added to the *y*-value for each point in the following manner. We first drew values of *NoiseMag* and *Tether* from their respective normal distributions (see **Table S1**). For each point in the trace, $n_i$ was then randomly drawn from a normal distribution with $\mu = x_i + (1 - Tether)n_{i-1}$ and $\sigma = NoiseMag$. The result is that the amount of noise increases cumulatively along the trace, but this cumulative noise cannot increase too much because the trace gets slightly pulled back towards the location of the master trace by a "tether".

It is important to emphasize that all of the above details were chosen in order to produce simulated traces that visually resemble what is observed in our experiments, and are not based on any physical model of charge transport. Since the only purpose of these simulated traces is to demonstrate *how* our grid-based correlation tools work, traces that resemble observed behaviors, even if only superficially, are appropriate for this work.

**Table S1**. Mean ($\mu$) and standard deviation ($\sigma$) for the normal distributions that each of the parameters used to generate simulated traces is drawn from. $G_{plateau}$, *PlateauSlope*, *PlateauLength*, and *BreakOffSlope* are only used for molecular traces, while all other parameters are used for both tunneling and molecular traces. For every individual trace, each parameter has a single fixed value, but the parameters are chosen independently for each trace using these normal distributions.

| Parameter Name | $\mu$ | $\sigma$ | Units |
|---|---|---|---|
| $G_{pr}$ | 0 | 0.05 | log($G/G_0$) |
| *DropSlope* | -20 | 5 | decades/nm |
| $G_{drop}$ | -2.5 | 0.5 | log($G/G_0$) |
| *TunnSlope* | -6 | 1.2 | decades/nm |
| $G_{plateau}$ | -4 | 0.35 | log($G/G_0$) |
| *PlateauSlope* | -0.05 | 0.01 | decades/nm |
| *PlateauLength* | 0.65 | 0.05 | nm |
| *BreakOffSlope* | -25 | 5 | decades/nm |
| *NoiseMag* | 0.05 | 0.01 | log($G/G_0$) |
| *Tether* | 0.05 | 0.001 | - |

S.3 Experimental Datasets Used in this Work

As previously described in Bamberger et al. 2020,[10] certain events during the collection of breaking traces on a given MCBJ sample often lead to discrete, qualitative changes in trace behavior. Our standard practice is thus to treat the chunk of traces collected between each pair of such events as a separate dataset, to be analyzed independently. The events that we consider to begin a new dataset are: pausing the LabVIEW program to deposit *or* re-deposit molecular solution, or to deposit pure dichloromethane; and stopping the LabVIEW program to fully relax the push rod, then restarting the LABVIEW program (we refer to this as "starting a new trial"). **Table S2** lists all of the experimental datasets considered in this work, each assigned a unique identification number, as well as the trial number and number of molecular depositions for each dataset. We stress that **all usable datasets from these four MCBJ samples are analyzed and reported on in this work**. If a certain dataset appears to be missing—e.g., deposition #3 for sample #108-5—it is because that dataset contained only a few hundred traces, and thus was not suitable for analysis, *not* because it was excluded for producing "undesired" results.

In some of the large datasets collected for this work, the molecular signature seemingly disappeared partway through the block of sequential traces. We chose to break these datasets into multiple subsets and to analyze each subset independently. This decision was taken solely to provide more test cases for our rare event detection tools; unlike the dataset-splitting events discussed above, we do not believe there is any fundamental reason to analyze these blocks of traces separately, and the dividing points for the different subsets were chosen somewhat arbitrarily. These subsets, and what block of traces each one corresponds to, can be found in **Table S2**.

**Table S2**. List of each experimental dataset, or dataset subset, analyzed in this work. Sample # specifies the specific physical MCBJ sample that the breaking traces were collected on, and Dataset ID# specifies the specific block of traces collected for a given trial/deposition combination. In some cases, a single dataset was broken into multiple subsets for independent analysis. All usable datasets from four MCBJ samples run with three different molecules are included. The order of datasets and dataset subsections in this table matches the order of the results shown in Figure 6 in the main text, going first top-to-bottom and then left-to-right.

| Dataset ID# | Sample # | Trial # | Deposition # | Concentration (µM) | # of Traces | Subset? | Molecule Present |
|---|---|---|---|---|---|---|---|
| 130 | 108-5 | 1 | 0 | N/A | 3847 | No | None |
| 131 | 108-5 | 1 | 2 | 1 | 2500 | Yes, 1-2500 | OPV2-2BT |
| 131 | 108-5 | 1 | 2 | 1 | 7592 | Yes, 2501-10092 | OPV2-2BT |
| 133 | 108-5 | 3 | 5 | 1 | 6562 | No | OPV2-2BT |
| 134 | 108-5 | 3 | 5[a] | 1 | 11897 | No | OPV2-2BT |
| 135 | 108-5 | 4 | 1[b] | 10 | 5807 | No | OPV2-2BT |
| 184 | 114-5 | 1 | 0 | N/A | 4093 | No | None |
| 185 | 114-5 | 1 | 2 | 1 | 4650 | No | OPV2-2SMe |
| 186 | 114-5 | 1 | 3 | 1 | 2594 | No | OPV2-2SMe |
| 187 | 114-5 | 1 | 4 | 1 | 4467 | No | OPV2-2SMe |
| 188 | 114-5 | 2 | 7 | 1 | 4212 | No | OPV2-2SMe |
| 189 | 114-5 | 2 | 8 | 1 | 2543 | No | OPV2-2SMe |
| 190 | 114-5 | 2 | 1[c] | 10 | 8605 | No | OPV2-2SMe |
| 191 | 114-5 | 2 | 2[d] | 10 | 9236 | No | OPV2-2SMe |
| 192 | 125-3 | 1 | 0 | N/A | 2861 | No | None |
| 193 | 125-3 | 1 | 2 | 1 | 3832 | No | OPV2-2SMe |
| 194 | 125-3 | 1 | 1[e] | 10 | 6353 | No | OPV2-2SMe |
| 195 | 125-3 | 1 | 2[f] | 10 | 2837 | No | OPV2-2SMe |
| 196 | 125-3 | 1 | 2[g] | 10 | 5812 | No | OPV2-2SMe |
| 231 | 119-1 | 1 | 0 | N/A | 5066 | No | None |
| 232 | 119-1 | 1 | 1 | 1 | 2500 | Yes, 1-2500 | OPV2-2SAc |
| 232 | 119-1 | 1 | 1 | 1 | 2500 | Yes, 2501-5000 | OPV2-2SAc |
| 232 | 119-1 | 1 | 1 | 1 | 3885 | Yes, 5001-8885 | OPV2-2SAc |
| 233 | 119-1 | 1 | 2 | 1 | 2745 | No | OPV2-2SAc |
| 234 | 119-1 | 1 | 3 | 1 | 2500 | Yes, 1-2500 | OPV2-2SAc |
| 234 | 119-1 | 1 | 3 | 1 | 2500 | Yes, 2501-5000 | OPV2-2SAc |
| 234 | 119-1 | 1 | 3 | 1 | 3655 | Yes, 5001-8655 | OPV2-2SAc |

[a]Pure dichloromethane was deposited between datasets 133 and 134, so they are considered distinct despite having the same trial number and number of depositions.
[b]First 10 µM deposition, preceded by 5 1 µM depositions.
[c]First 10 µM deposition, preceded by 8 1 µM depositions.
[d]Second 10 µM deposition, 10th deposition overall.
[e]First 10 µM deposition, preceded by 2 1 µM depositions.
[f]Second 10 µM deposition, 4th deposition overall.
[g]Pure dichloromethane was deposited between datasets 195 and 196, so they are considered distinct despite having the same trial number and number of depositions.

The 1D conductance histograms for each dataset or subset from **Table S2** are shown in **Figure S7**, **Figure S8**, **Figure S9**, and **Figure S10**, grouped by the different MCBJ samples that the datasets were collected from. The 2D distance/conductance histograms for all of these datasets and dataset subsections are shown in **Figure S11**, **Figure S12**, **Figure S13**, and **Figure S14**, again grouped by MCBJ sample. Datasets collected before any molecular solution was deposited are shown in blue in the 1D histogram overlays and labeled as "actually empty" in the 2D histograms.

For datasets collected after molecular solution had been deposited, we considered both the 1D and 2D histograms in order to categorize each dataset as either "seemingly empty" (yellow) or "clearly molecular" (red). Our primary criteria for making these determinations was whether a molecular peak is present in the 1D histogram and whether a plateau-like feature is distinguishable in the 2D histogram. Not all of the datasets that we categorize as "seemingly empty" look identical to the "actually empty" datasets collected on the same MCBJ sample. However, as shown in **Figure S15**, different "actually empty" datasets collected on the same MCBJ sample also do not always look identical. We therefore believe that our "seemingly empty" assignments are justified because in all such cases it is not *a priori* obvious from the 1D and 2D histograms whether or not molecules were truly present. However, we acknowledge that the datasets we label as "seemingly empty" in fact span a range, with datasets which truly appear indistinguishable from the "actually empty" datasets (e.g., #185 or #193) on one end, and datasets that do not have a molecular peak in their 1D histograms, but maybe do show a hint of molecular behavior in their 2D histograms (e.g. #232 Tr. 2501-5000) on the other end. As shown in the main text, our rare event detection tools work well across this range.

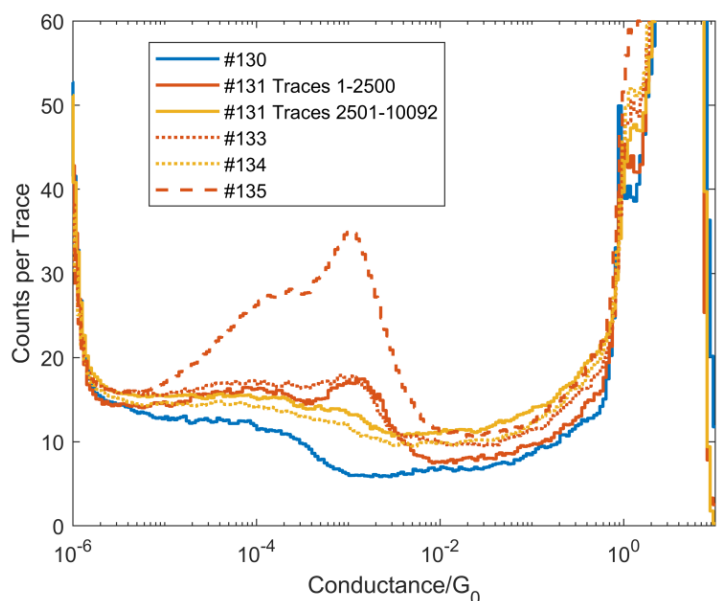

**Figure S7**. Overlaid 1D conductance histograms for the six independently analyzed datasets or dataset subsections collected using MCBJ sample #108-5, with the molecule OPV2-2BT. Legend entries refer to the dataset ID numbers and/or subset ranges in **Table S2**. The "actually empty" dataset is shown in blue, and the datasets we have identified as "seemingly empty" or "clearly molecular" are shown in yellow and red, respectively.

S14

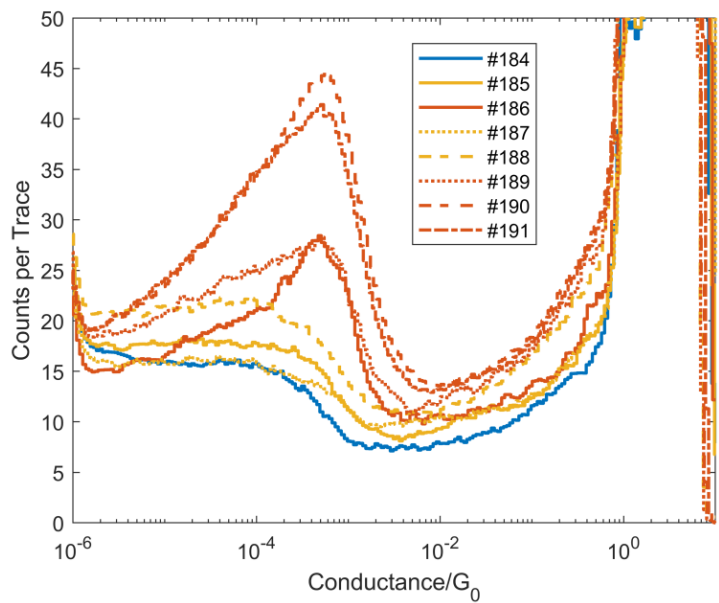

**Figure S8**. Overlaid 1D conductance histograms for the eight independently analyzed datasets collected using MCBJ sample #114-5, with the molecule OPV2-2SMe. Legend entries refer to the dataset ID numbers in **Table S2**. The "actually empty" dataset is shown in blue, and the datasets we have identified as "seemingly empty" or "clearly molecular" are shown in yellow and red, respectively.

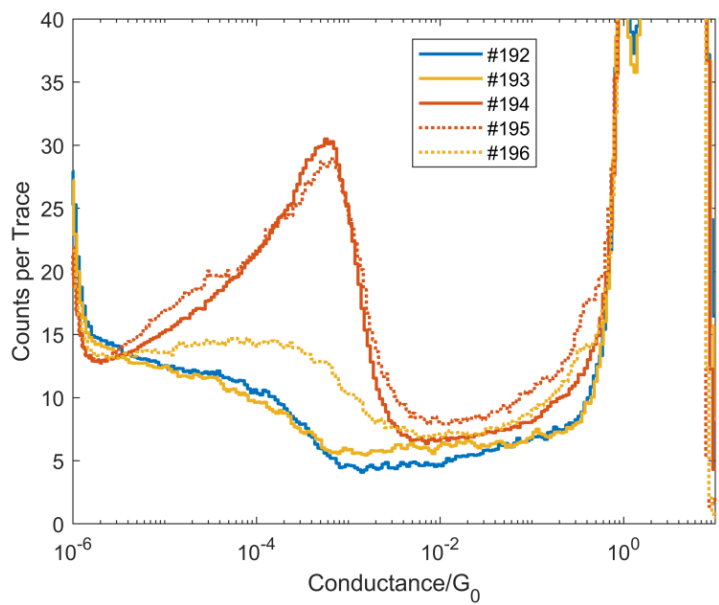

**Figure S9**. Overlaid 1D conductance histograms for the five independently analyzed datasets collected using MCBJ sample #125-3, with the molecule OPV2-2SMe. Legend entries refer to the dataset ID numbers in **Table S2**. The "actually empty" dataset is shown in blue, and the datasets we have identified as "seemingly empty" or "clearly molecular" are shown in yellow and red, respectively.

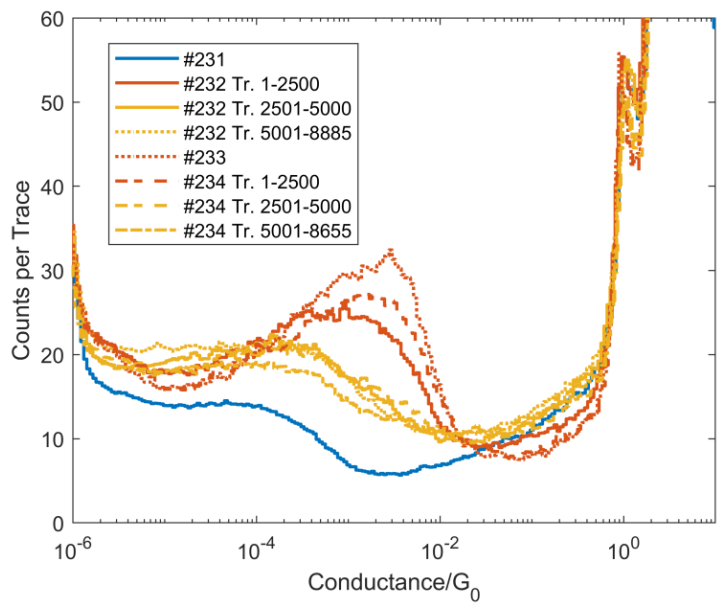

**Figure S10**. Overlaid 1D conductance histograms for the eight independently analyzed datasets or dataset subsections collected using MCBJ sample #119-1, with the molecule OPV2-2SAc. Legend entries refer to the dataset ID numbers and/or subset ranges in **Table S2**. The "actually empty" dataset is shown in blue, and the datasets we have identified as "seemingly empty" or "clearly molecular" are shown in yellow and red, respectively.

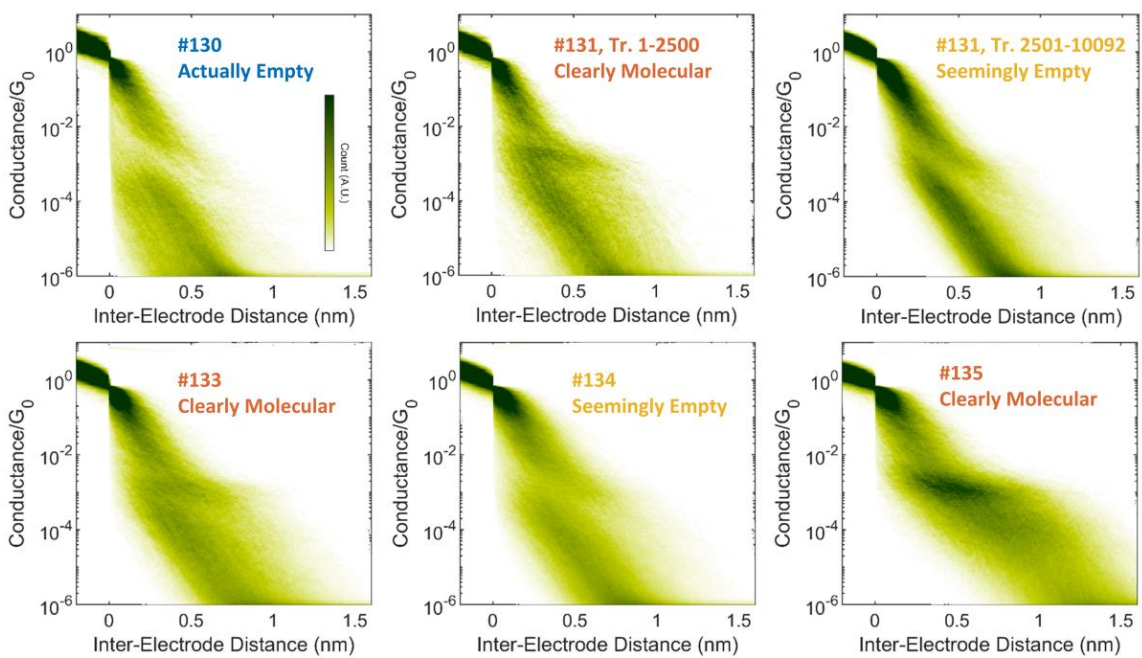

**Figure S11**. 2D conductance/distance histograms for each dataset or dataset subsection collected using MCBJ sample #108-5, with the molecule OPV2-2BT. Labels refer to the dataset ID numbers and/or subset ranges in **Table S2**. We categorized datasets collected after molecular deposition as either "seemingly empty" or "clearly molecular" based on both their 1D and 2D histograms, and these determinations are included in the labels.

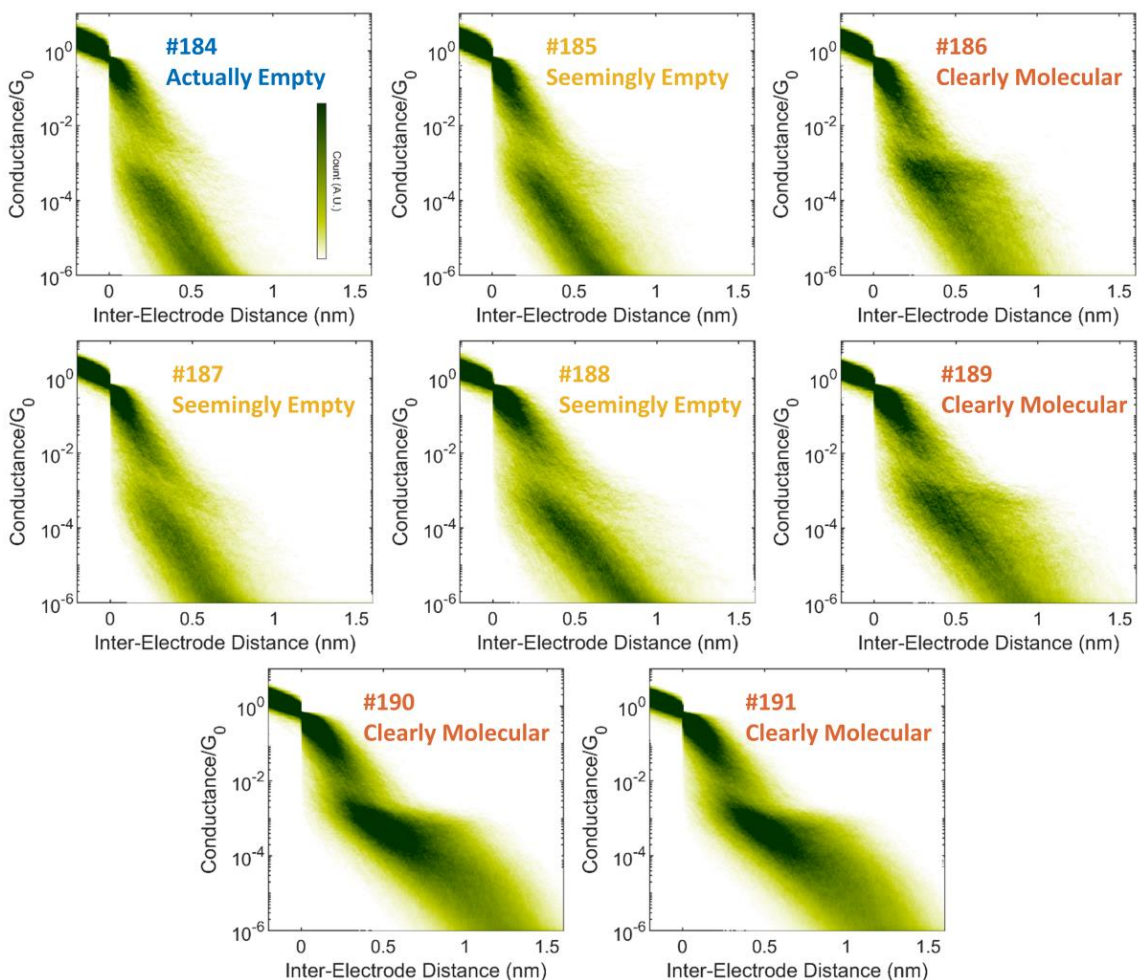

**Figure S12**. 2D conductance/distance histograms for each dataset collected using MCBJ sample #114-5, with the molecule OPV2-2SMe. Labels refer to the dataset ID numbers in **Table S2**. We categorized datasets collected after molecular deposition as either "seemingly empty" or "clearly molecular" based on both their 1D and 2D histograms, and these determinations are included in the labels.

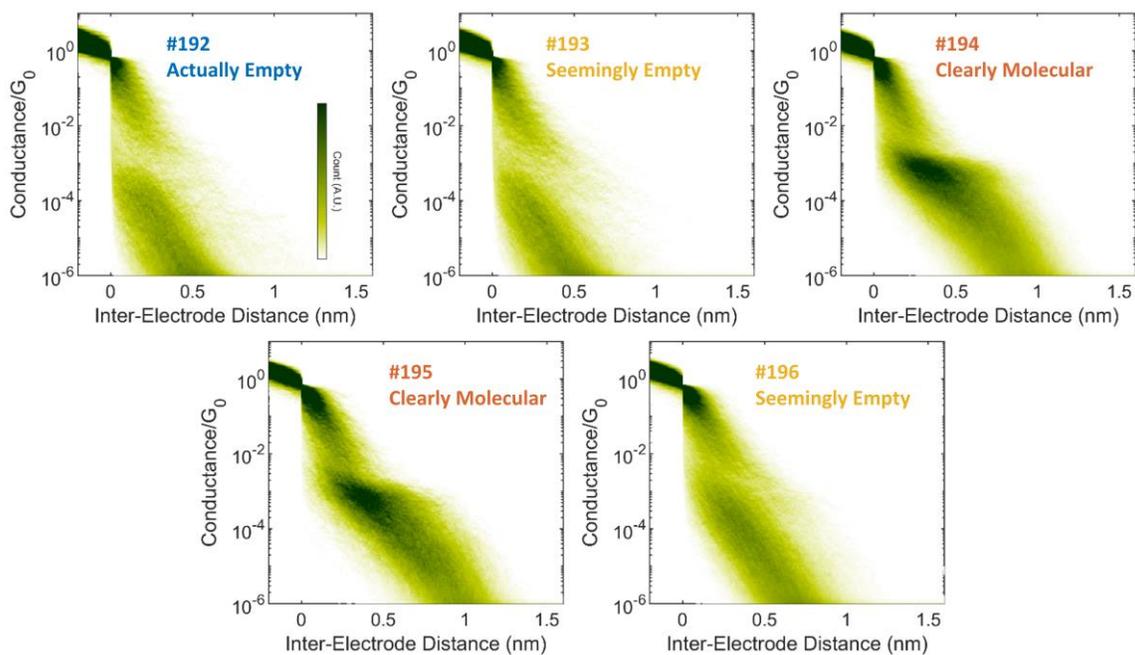

**Figure S13**. 2D conductance/distance histograms for each dataset collected using MCBJ sample #125-3, with the molecule OPV2-2SMe. Labels refer to the dataset ID numbers in **Table S2**. We categorized datasets collected after molecular deposition as either "seemingly empty" or "clearly molecular" based on both their 1D and 2D histograms, and these determinations are included in the labels.

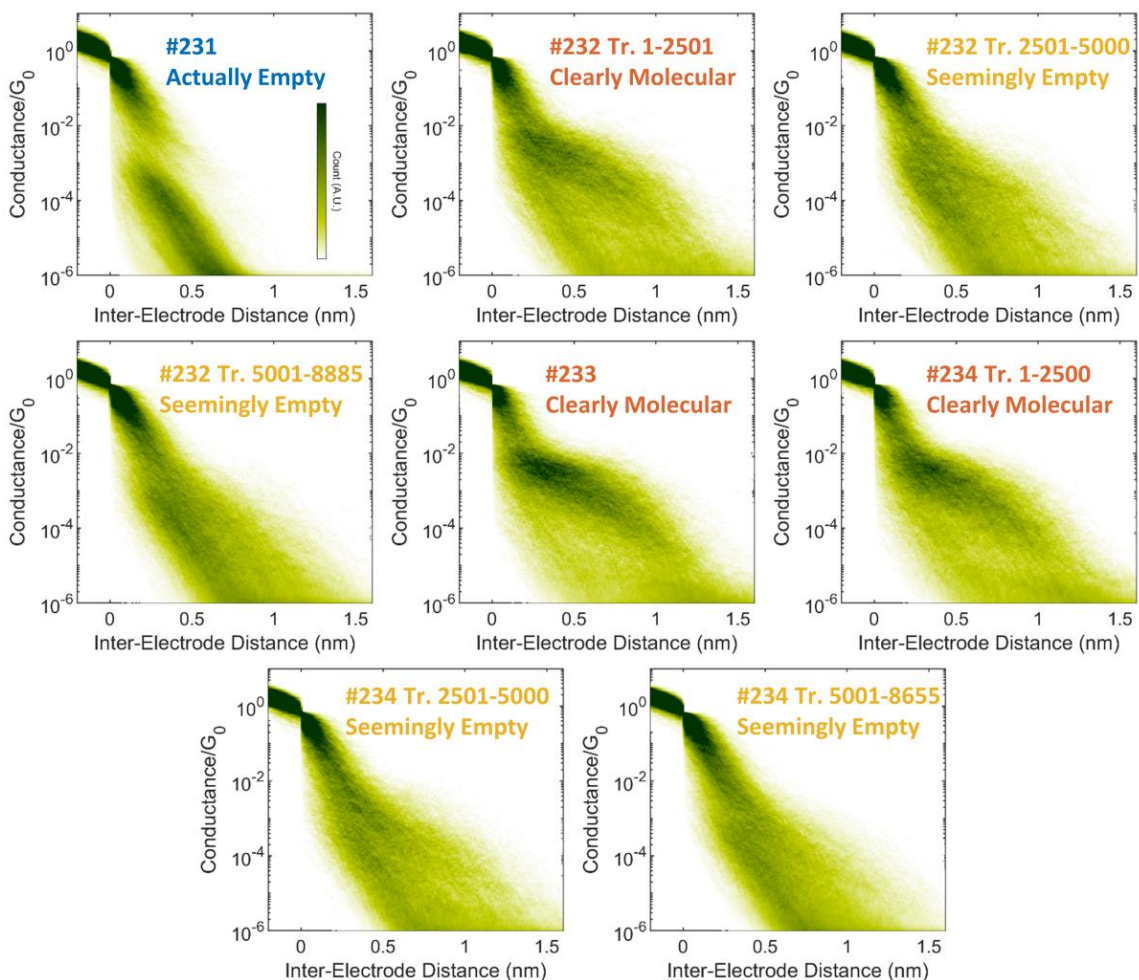

**Figure S14**. 2D conductance/distance histograms for each dataset or dataset subsection collected using MCBJ sample #119-1, with the molecule OPV2-2SAc. Labels refer to the dataset ID numbers and/or subset ranges in **Table S2**. We categorized datasets collected after molecular deposition as either "seemingly empty" or "clearly molecular" based on both their 1D and 2D histograms, and these determinations are included in the labels.

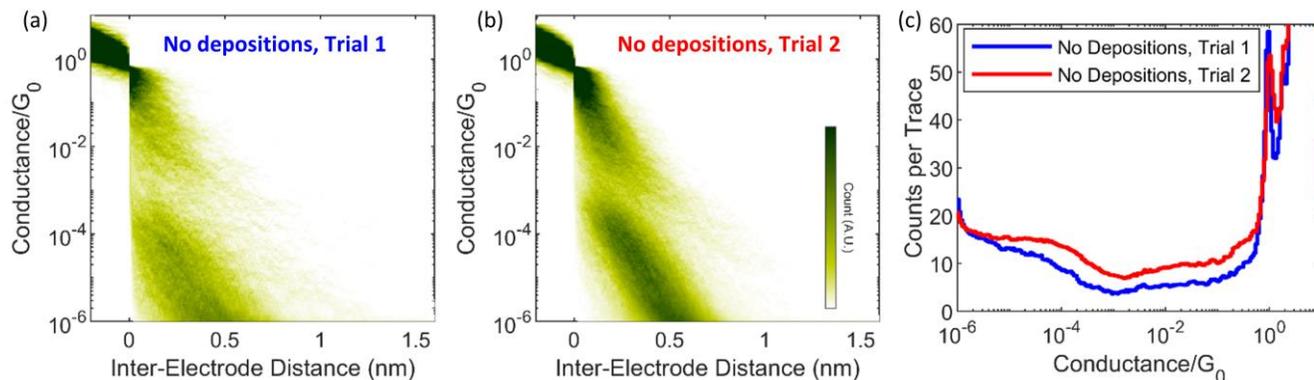

**Figure S15**. Example of a case when multiple "actually empty" datasets (i.e., traces collected before any molecular solution was deposited) were collected on the same MCBJ sample. (a) 2D histogram for the actually empty dataset collected in trial 1. (b) 2D histogram for the actually empty dataset collected in trial 2 (i.e., after the sample had been fully relaxed and then restarted). (c) Overlaid 1D conductance histograms for the two datasets from (a) and (b). This example demonstrates that there can be meaningful quantitative variation between "actually empty" datasets, even those collected on the same sample. This helps explain why above we label some datasets as "seemingly empty" when they do not show any obvious evidence of molecular features, even if they do not appear identical to the actually empty dataset collected on the same sample. Note that the two datasets in this figure are not included in **Table S2** because they were not analyzed for rare events.

S.4 Further Details on Grid-Based Correlation Framework

*S.4.1 Coarse-Gridding Traces*

Before coarse-gridding, the traces in each dataset were chopped to begin at -0.05 nm inter-electrode distance, thus excluding most of the above-$G_0$ region. Potential correlations between trace behavior before and after final gold rupture (above and below $G_0$ in conductance) are certainly an interesting subject for future work. However, while the amplifier used in our MCBJ experiments is very stable for measuring conductances between $10^{-6}$ $G_0$ and 1 $G_0$,[11] it displays a small amount of drift over time for conductances much above 1 $G_0$. This drift would produce non-meaningful correlations within the above-$G_0$ part of our datasets, and so for this work we limit our focus to the sub-$G_0$ conductance regime.

The coarse grid of nodes is always aligned so that a node lies at 0 nm inter-electrode distance and 1 $G_0$ (i.e. 0 log(G/$G_0$)) of conductance. This ensures that the grid alignment is meaningful and consistent across datasets, rather than random. To convert each trace into a coarse trace, each successive node in the coarse trace is required to advance by exactly one grid-step in the *x*-direction. The *y*-grid value at each step is obtained by averaging together all data points from the original trace that are closest to the current *x*-grid value, then rounding to the nearest *y*-grid.

All datasets used in this work have a nominal noise floor of $10^{-6}$ $G_0$, and so data below this conductance value are not included. For traces that dip below the noise floor but then come back above it, multiple independent coarse traces were created for each consecutive section of at least 2 above-the-noise nodes. This choice ensures that meaningless sub-noise-floor measurements do not affect our results while still making use of as much meaningful data as possible.

*S.4.2 Treatment of Nodes*

Within each dataset, we only have information on nodes that were visited by at least one trace. For the purposes of calculating node-pair properties such as exit probabilities, connection strengths, etc., we therefore consider all nodes never visited by any traces to "not exist" in that particular dataset. As described in the main text, the heart of our grid-based correlation approach considers the different paths a trace might take between any two nodes. This concept struggles in cases where only a single path exists between two nodes, e.g. because the nodes in question were only visited by a single trace. We thus define a "dangling node" as any node having neither a direct neighbor above nor below itself (see **Figure S16**), and we remove all such dangling nodes from each dataset. To maintain the appropriate connectivity of all remaining nodes, dangling nodes are removed in practice by removing all of the coarse traces that passed through them. This process results in the removal of only 1-2% of traces from each dataset. We note that removing these dangling nodes is a conservative choice, because, at worst, it could result in the removal of rare events of interest, but should not introduce any bias to rare behaviors that are retained.

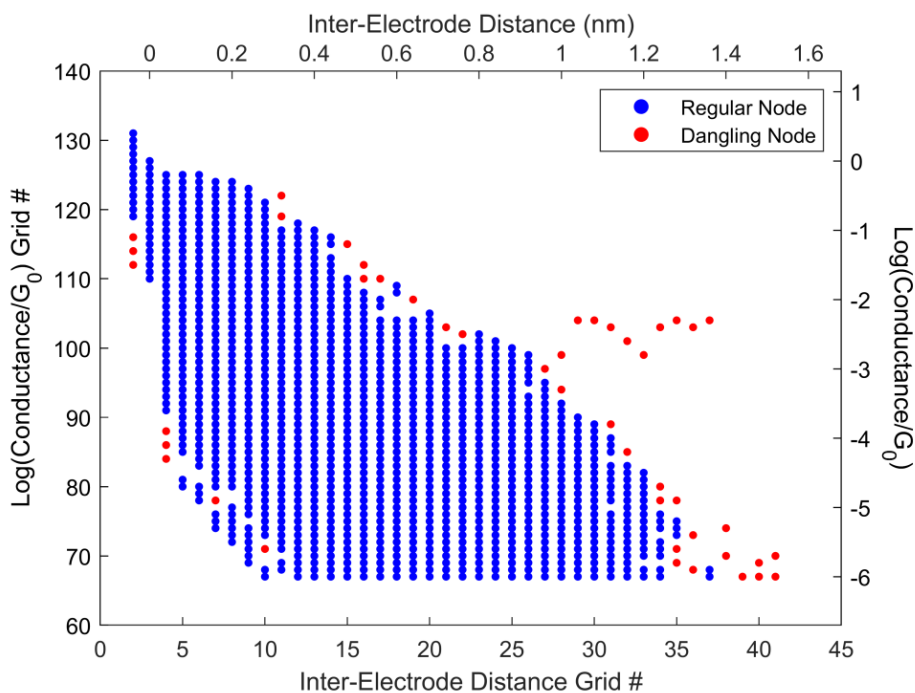

**Figure S16**. Example dataset with all dangling nodes—those missing a neighboring node *both* immediately above and immediately below—highlighted in red. As this example shows, dangling nodes are typically found on the very edge of the distribution and often were visited by only a single trace. These dangling nodes are removed before our grid-based correlation tools are applied because these tools are designed for nodes with multiple potential paths through them, not just one. Nodes are removed by removing the traces that passed through them, so removing dangling nodes is an iterative process.

*S.4.3 Exit Probabilities*

We refer to the empirically derived probability that a trace exiting node *X* will pass immediately to node *Y* as an "exit probability" (also called "transfer probabilities" in our MATLAB code). Exit probabilities are thus defined whenever node *Y* lies in the column immediately to the right of node *X*, and undefined in all other cases. As shown in **Figure S17**, exit probabilities are directly calculated from the observed coarse traces in a given dataset by taking the fraction of those traces that followed each exiting path.

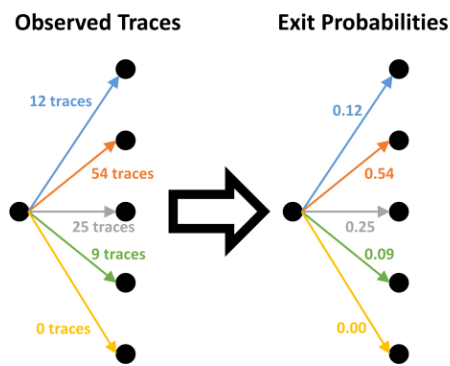

**Figure S17**. Diagram illustrating how exit probabilities are directly calculated from the exit trajectories taken by all of the traces in a given dataset. In this example, 100 traces pass through the node on the left, and only the exit probabilities for that specific node are shown. Exit probabilities are only defined for a pair of nodes in which the second node is in the column immediately to the right of the first node.

*S.4.4 Binomial Hypothesis Tests for Node-Pairs*

For each unique pair of nodes *X* and *Y*, two binomial hypothesis tests are computed, producing values for $p_{above}(X,Y)$ and $p_{below}(X,Y)$. First, we calculate $p_{null}(X,Y)$: the probability that a trace starting from node *X* will eventually pass through node *Y*, under the null hypothesis that the trace behaves like a random weighted walk according to the exit probabilities between nodes, which can be calculated from the set of all exit probabilities using the theory of absorbing Markov chains.[12] If *N* traces pass through node *X*, and *m* of those traces also pass through node *Y*, $p_{above}(X,Y)$ and $p_{below}(X,Y)$ are then calculated as:

$$p_{above}(X,Y) = \sum_{i=m}^{N} \frac{N!}{i!\,(N-i)!} (p_{null})^i (1-p_{null})^{N-i}$$

$$p_{below}(X,Y) = \sum_{i=0}^{m} \frac{N!}{i!\,(N-i)!} (p_{null})^i (1-p_{null})^{N-i}$$

In other words, $p_{above}(X,Y)$ and $p_{below}(X,Y)$ are the upper-tail and lower-tail *p*-values for these two binomial hypothesis tests. They represent how likely it is that we would observe, respectively, at least as many traces going from *X* to *Y* as we did, or no more than the number of traces going from *X* to *Y* as we did, *if* traces really did behave like random walks.

For cases in which $p_{null}$ is zero (e.g., if node *X* is to the right of node *Y*), $p_{above}(X,Y)$ and $p_{below}(X,Y)$ are both undefined. In a few rare cases, $p_{above}$ or $p_{below}$ is calculated as zero, despite $p_{null}$ being nonzero, simply due to limited numerical precision. Such zero values would cause problems when we calculate connection strength by taking logarithms of $p_{above}$ and $p_{below}$ (see below). To deal with this possibility, we impose a minimum value for all $p_{above}$ and $p_{below}$ values of $1/(\# \ of \ nodes \ in \ dataset)^2$. This is a logical choice because $(\# \ of \ nodes \ in \ dataset)^2$ is the number of different hypothesis tests potentially performed for a given dataset, and so we cannot expect reliable measurement of any *p*-values for events that would be observed in the entire dataset less than once, on average, under the null hypothesis.

*S.4.5 Connection Strength*

To calculate the connection strength, or *CS*, between any two nodes, we first define *X* to be the node farther to the left and *Y* to be the node farther to the right. This ensures that $p_{above}(X,Y)$ and $p_{below}(X,Y)$ will both be defined. If the two nodes are in the same column as each other, the connection strength between them is automatically set to zero (to represent no correlation, since a single trace cannot visit multiple nodes in the same column). Connection strength is then defined as:

$$CS(X,Y) = -\ln(p_{above}(X,Y)) + \ln(p_{above}(X,Y))$$

*S.4.6 Node-Sequence Significance*

Our MCMC feature-finder (see below) operates on "node sequences", which are a series of *n* nodes $(m_1, m_2, \cdots, m_n)$ with each node one *x*-grid to the right of the previous—in other words, a node sequence is a section cut out of a potential coarse trace. In order to represent how correlated with each other the nodes in a particular sequence are, we define the "significance" of a particular *n*-node sequence *s* as:

$$SIG(s) = \frac{2}{n(n-1)} \sum_{i=1}^{n} \sum_{j=i+1}^{n} CS(m_i, m_j)$$

In other words, the significance of a node sequence is simply the average connection strength between every possible unique pair of nodes which can be drawn from that sequence. All possible node pairs are included because meaningful spatial correlations might exist between any two points within the node sequence. In our MATLAB code, the significance of a node sequence is also referred to as its "weight".

S.5 Details of MCMC Feature-Finder

*S.5.1 General Operation of MCMC*

We generate Markov Chains of *n*-node sequences using the Metropolis MCMC algorithm. Briefly, Metropolis MCMC creates a list of points—in our case, *n*-node sequences—by iteratively generating a new point, known as a trial step, based on the current last point in the list. The Metropolis criterion (see below) is used to determine whether that trial step is accepted—meaning that the new point is added to the list—or rejected, meaning that the current last point in the list is repeated. After a large number of iterations, the list of points will converge to the desired distribution,[13,14] in our case a distribution of the most significant node-sequences.

In our specific MCMC simulation, six different types of trial steps are used to enable the node sequences to efficiently explore the entire available space (see **Figure S18**): a "single shift", in which one of the sequence's nodes at random is moved up or down by one grid unit; a "rigid shift", in which all of the sequence's nodes are moved up or down by one grid unit together; a "split shift", in which the left half of the sequence is moved up or down by one grid unit while the right half is moved in the opposite direction by one grid unit; a "scramble shift", in which each node in the sequence is randomly moved up or down by one grid unit independently of the others; a "rigid translate", in which all of the sequence's nodes are moved left or right by one grid unit together; and a "snake translate", in which the left-most node is removed and then added to the right end of the sequence at the same vertical location, or vice versa. For each step, the movement direction(s) of any nodes are chosen uniformly at random to ensure that the trial step distribution is symmetric (which is what makes this a Metropolis rather than a Metropolis-Hastings algorithm[13,14]). At each MCMC iteration, the type of trial step is chosen at random according to the probabilities listed in **Figure S18**. These probabilities were empirically chosen to roughly achieve an overall acceptance rate of ~0.1 to 0.6, as this range typically produces high efficiency for MCMC algorithms.[13] We stress, however, that the specific types of trial steps and their relative probabilities are *only* important for how long our MCMC runs take to converge. *Any* set of trial steps that allows the entire available space to be explored will produce the same final distribution, provided that the MCMC run includes sufficient steps to achieve convergence.

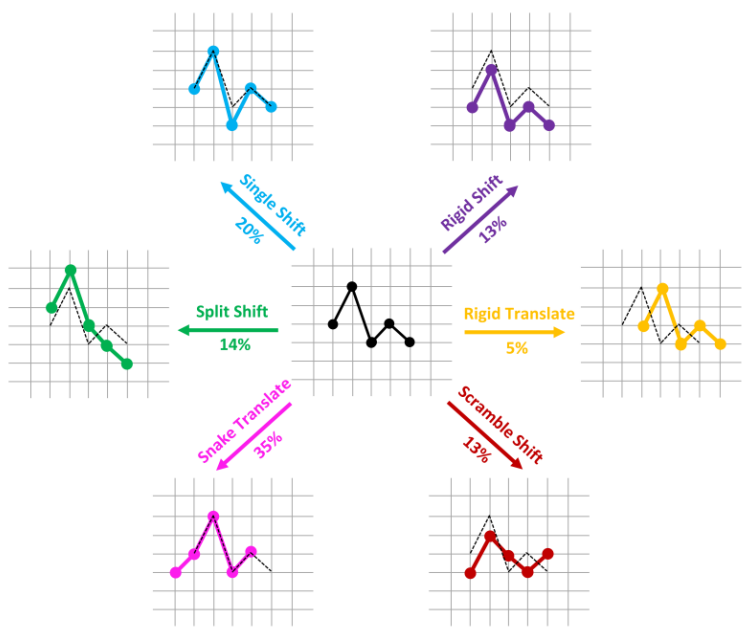

**Figure S18**. Examples of the six different types of trial steps being applied to a 5-node sequence. The middle grid shows the original 5-node sequence in black, and each of the surrounding grids shows one possible result of applying each type of trial step, in color, with the original sequence overlaid in dotted black. The percentages next to each arrow indicate the probability that that type of trial step is randomly

chosen at each MCMC iteration. Each type of trial step involves randomness in how it is applied (e.g., which node to shift and/or which direction to shift node(s) in).

As stated in the main text, we assign the probability of each *n*-node sequence *s* appearing in our desired distribution to be proportional to $\exp(\frac{SIG(s)}{T})$, where $T$ is an effective temperature (see section S.5.3). Because our trial steps are all fully symmetric, the Metropolis criterion states that a trial step will be accepted if and only if:[13]

$$x \leq \exp\left(\frac{SIG(s_{trial}) - SIG(s_{current})}{T}\right)$$

where x is a number between 0 and 1 generated uniformly at random, $s_{current}$ is the current *n*-node sequence and $s_{trial}$ is the *n*-node sequence produced by the trial step. If the trial *n*-node sequence includes one or more nodes that do not exist in the dataset being analyzed (see section S.4.2 above), or if the trial *n*-node sequence does not meet the user-specified criteria for the given MCMC run (see section S.5.2 below), $SIG(s_{trial})$ is set to infinity so that the trial step is always rejected.

*S.5.2 Including User-Specified Criteria*

As explained in the main text, because traces can be correlated with their own history in multiple different ways in the same dataset, it can be helpful to apply extra criteria to the *n*-node sequences produced by the MCMC feature-finder in order to focus attention on a particular type or category of rare event. In our algorithm, any trial *n*-node sequence that does not meet the user specified criteria is automatically rejected, resulting in a final distribution of only sequences that meet the criteria. Our algorithm is written to be extremely flexible, and so any imaginable set of criteria can be easily incorporated for a given run.

In this work, we apply what we call "range slope" criteria. We define the "range slope" of an *n*-node sequence to be the number of nodes spanned vertically by the sequence, divided by *n*. We then set a maximum allowable value for this range slope (a minimum allowable value could also be defined, but is not used in this work). This definition of slope was chosen because it works well for focusing on relatively flat features. Other types of slope definition work better for focusing on highly-slanted features, and are implemented in our publicly available code, but such highly-slanted features are outside the scope of the present work.

*S.5.3 The Role of Effective Temperature*

The significance of an *n*-node sequence (see section S.4.6 above) intuitively represents how "non-random-walk-like" the path represented by those nodes is. However, this significance value does not have any natural units associated with it, and the typical scale of significances for a set of *n*-node sequences can vary meaningfully depending on the dataset analyzed, the value of *n*, and the type of criteria the sequences must meet. Therefore, the effective temperature $T$ included in our MCMC algorithm (see section S.5.1 above) should be thought of as a way to set an appropriate relative scale for sequence significances during a given MCMC run.

Practically, $T$ has the effect of "tempering" the probability distribution simulated by the MCMC: small values of $T$ accentuate any peaks in this distribution so that the simulation spends more of its time there, while large values of $T$ flatten the distribution so that differences between high- and low-significance sequences are less pronounced (see, e.g., **Figure S25** and **Figure S26**). Because our MCMC feature finder is designed to find the most significant features, low $T$ values and highly-peaked probability distributions are desirable in our context.

As described above, different sets of node sequences can have meaningfully different scales on which their significance values fall. To correct for this, for each dataset we first perform an MCMC simulation with $T = \infty$—meaning that all possible node sequences are equally likely to appear in the final distribution—and then compute the standard deviation of the significance values from all node sequences produced by the simulation, which we call $\sigma_\infty$ (see **Figure S19**). Because $\sigma_\infty$ represents the typical significance difference between two potential node sequences, it can be used to put the significances from different sets of node sequences onto equivalent scales. We find empirically that effective temperatures in the range $T = \frac{\sigma_\infty}{3}$ to $T = \frac{2\sigma_\infty}{3}$ typically produce reasonable results—across a range of datasets, sequence lengths, and criteria types—in which the most significant feature is easy to identify (see, e.g., **Figure S25** and **Figure S26**). The results in this work are therefore taken from MCMC runs with $T = \frac{4\sigma_\infty}{9}$.

*S.5.4 Parallel Tempering*

As described in the previous section, we use relatively low effective temperatures in our MCMC feature finder in order to produce clear and highly-peaked features. However, these low temperatures can make it easy for the simulated node sequences to get "stuck" at a local maximum in the probability distribution and hence take a very long time to fully explore the available space. In the limit of infinite MCMC steps such a simulation will still converge to the correct final distribution, but in practice these "cold" simulations can produce misleading results and/or be extremely difficult to converge.

To address this issue, we employ a strategy known as parallel tempering.[13,15] This involves constructing a "super MCMC chain" composed of several MCMC chains run in parallel, each at a different effective temperature. In addition to the regular trial steps in which each sub-chain attempts to move its own node sequence independently of the other sub-chains, there is also a new type of trial step that attempts to swap the node sequences from two adjacent temperatures. This strategy helps the convergence of the low-temperature chains, because the high-temperature chains can efficiently explore the entire available space and then allow the low-temperature chains to "jump" between poorly-connect local probability maxima via the swap steps. An added benefit is that each parallel tempering MCMC simulation produces results for multiple effective temperatures, which helps us confirm that our conclusions are not overly sensitive to a particular value of $T$ (see section S.7.4).

For the results in this work, we employ parallel tempering with five different temperatures: $\frac{4\sigma_\infty}{9}, \frac{5\sigma_\infty}{9}, \frac{6\sigma_\infty}{9}, \sigma_\infty$, and $\infty$. A swap step is randomly attempted 20% of the time, and the other 80% of the time each sub-chain independently attempts a regular trial step according to the probabilities in **Figure S18**. Within a swap step, we randomly start with either the highest or lowest temperature sub-chain and then sequentially attempt to swap each node sequence with the one from the immediately lower or higher temperature sub-chain, respectively. This approach preserves the symmetry of the trial step distribution.

*S.5.5 Starting Point and Burn-In*

In order to generate an initial *n*-node sequence for each MCMC simulation, a random node is chosen in proportion to the number of traces passing through each node, then an *n*-node sequence is randomly generated centered on that node according to the exit probabilities. If this sequence does not meet the user-specified criteria, the process is repeated until a sequence that does meet criteria is obtained. A separate starting node sequence is independently chosen for each different temperature in the parallel tempering MCMC super-chain.

The above strategy for choosing starting points is chosen for convenience; the starting point for an MCMC simulation should have essentially no effect on the final results. The one caveat is that it is important to choose a starting point from a region of high probability,[16] because otherwise convergence of the MCMC chain will take a very long time. For example, if the starting point only has a probability of one billionth, it will take a minimum of one billion steps to converge to the correct final distribution. In our context it is difficult to even approximate the probability of any single node sequence *a priori*, and so we employ the strategy of "burn in": we discard the results from the first *N* MCMC steps so that the node sequence has an opportunity to move into a region of high probability before its locations start being included in the final simulated distribution. While some authors have criticized burn-in for being an inefficient way to start in a high-probability region, it has the advantage of being simple and easy, and should never cause any harm.[13] In this work, we use $N = 200{,}000$ unless otherwise stated (see **Figure S19**).

*S.5.6 Criteria for MCMC Convergence*

In order to assess when our MCMC simulations have run for sufficiently many steps that they will provide a good estimate of the desired distribution, we use a version of the Gelman-Rubin diagnostic, one of the most popular choices for this purpose.[16] This diagnostic relies on simulating multiple independent MCMC chains in parallel, each beginning at a different starting point. These parallel chains should not be confused with the sub-chains run at different temperatures for parallel tempering (which are not independent, because of the swap steps); rather, because we are employing parallel tempering, in order to use the Gelman-Rubin diagnostic we must run multiple independent copies of the entire parallel tempering super MCMC chain in parallel.

The intuition behind the Gelman-Rubin diagnostic is that, if the independent MCMC chains have suitably converged to the desired distribution, then the variance within each chain should be the same as the variance when data from all the chains are combined. This diagnostic test thus involves taking a ratio of between-chain to within-chain variance and requiring this ratio to be close to one.[13,14]

Formally, suppose that $\theta$ is any parameter that can be calculated at a single MCMC step, and that there are *m* independent chains run for *k* steps each. If $\theta_{ij}$ represents the value of $\theta$ at the $i^{\text{th}}$ step of the $j^{\text{th}}$ independent chain, then we can calculate the average within-chain variance as:[14]

$$W = \frac{1}{m}\sum_{j=1}^{m}\left(\frac{1}{n-1}\sum_{i=1}^{n}(\theta_{ij} - \bar{\theta}_j)^2\right)$$

where $\bar{\theta}_j$ is the average value of $\theta$ within chain *j*. Next we calculate the variance between the chain averages as:[14]

$$B = \frac{1}{m-1}\sum_{j=1}^{m}(\bar{\theta}_j - \bar{\bar{\theta}})^2$$

where $\bar{\bar{\theta}}$ is the average value of $\theta$ across all steps from all chains. We then take a ratio involving these two values, defined as:[14]

$$\hat{R} = \sqrt{\frac{\frac{n-1}{n}W + B}{W}}$$

This value of $\hat{R}$ will converge to 1 as $n \to \infty$,[14] and 1.1 is commonly used as a cutoff value, meaning that the MCMC simulation will be considered to be suitably converged when $\hat{R} < 1.1$.[16]

In this work, we apply convergence criteria in two different places. The first time is for our single-temperature, $T = \infty$, MCMC simulation used to determine $\sigma_\infty$ (see section S.5.3 above). No burn-in steps are used for this MCMC simulation since every possible node sequence is assigned equal probability at $T = \infty$. Following the advice of Brooks et al.,[13] we use three independent MCMC chains. Since $\sigma_\infty$ is calculated from the significances of each node sequence, the parameter for which we monitor convergence is the current node sequence's significance, $SIG(s_{current})$. Because our $\sigma_\infty$ will in turn affect our estimates from the main MCMC simulation, we apply very conservative convergence criteria of $\hat{R} < 1.005$. The three MCMC chains are run for a minimum of 150,000 steps each, after which $\hat{R}$ is calculated every 1000 steps until $\hat{R} < 1.005$ is achieved (see **Figure S19**).

The second place where we apply convergence criteria is to our parallel tempering MCMC simulations used to produce the node distributions that are the output of the MCMC feature-finder tool. We again use three independent parallel tempering MCMC super-chains, and we monitor convergence independently for each temperature. Within each temperature, we independently monitor convergence for each node, using whether or not a given node is included in the current node sequence—i.e., 0 or 1—as the parameter used to calculate $\hat{R}$. We require two criteria to be met for a parallel tempering MCMC simulation to achieve convergence. First, for every temperature, every node that is occupied in at least 0.1% of all steps at that temperature must have $\hat{R} < 1.1$. Second, in order to be extra-conservative, we also compute a single weighted average of $\hat{R}$ at each temperature by using the occupancy of each node, and this weighted average $\hat{R}$ must be less than 1.05. Each parallel tempering MCMC super-chain is run for a minimum of 200,000 steps, and then 200,000 additional steps are collected for each chain until the convergence criteria are met (see **Figure S19**).

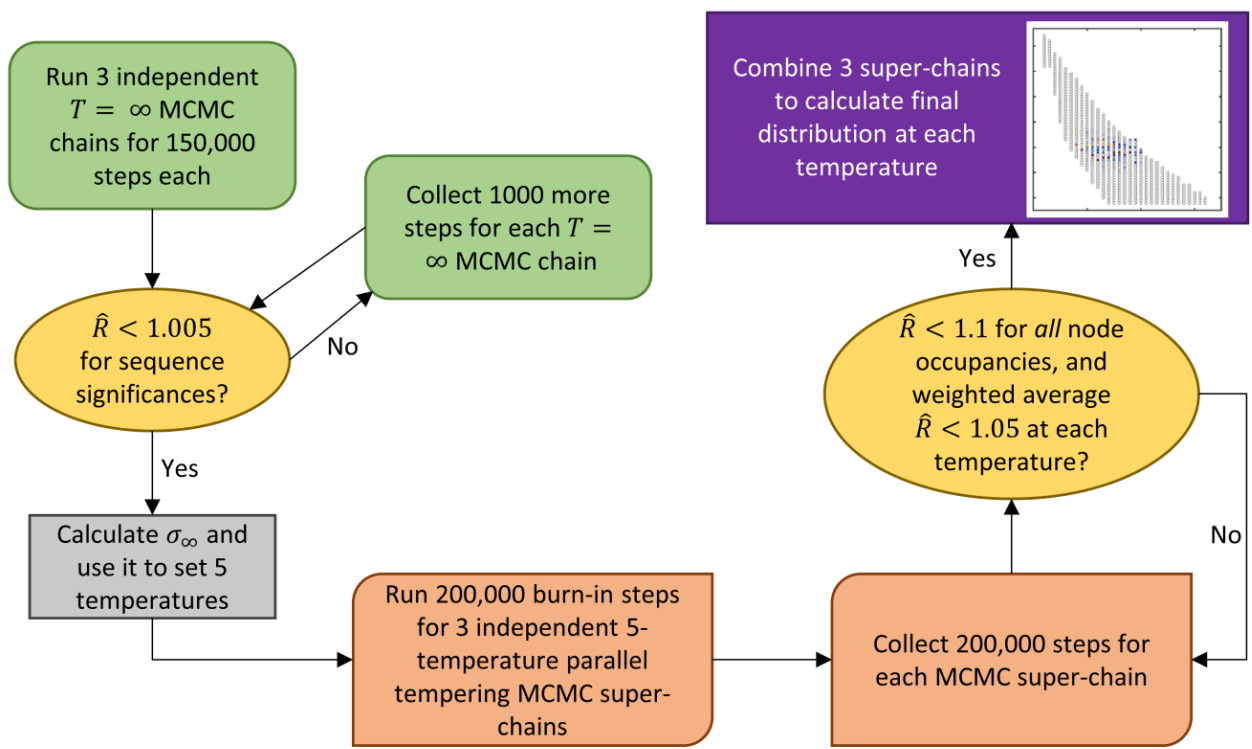

**Figure S19**. Flow chart for how the MCMC feature-finder is applied to each dataset. A single-temperature (at $T = \infty$) MCMC simulation is first performed (green) in order to determine $\sigma_\infty$, which is used to set the relative temperature scale for the 5-temperature parallel tempering MCMC simulation (red) used to obtain the final node probability distributions (purple). Convergence criteria (yellow) are used to ensure that the outputs of both the single-temperature and 5-temperature MCMC simulations are reliable.

S.6 Additional Data from MCMC Results

**Figure S20** shows the full two-dimensional node probability distribution MCMC results corresponding to the marginal distributions shown in Figure 4b of the main text. These results again show that the "seemingly empty" and "clearly molecular" datasets (panels b and c, respectively) produce essentially the same plateau feature, and that this feature is quite different form the one found in the "actually empty" dataset (panel a).

In **Figure S20**a, the 2D distribution for the actually empty dataset follows only a single path, meaning that the significance of this identified feature is primarily due to just a single trace from the original dataset. This supports our hypothesis, stated in the main text, that when our MCMC feature-finder is used to search for plateau-like features in a dataset that does not contain true molecular plateaus, it will likely identify whichever random trace behavior happens to appear plateau-like simply by chance.

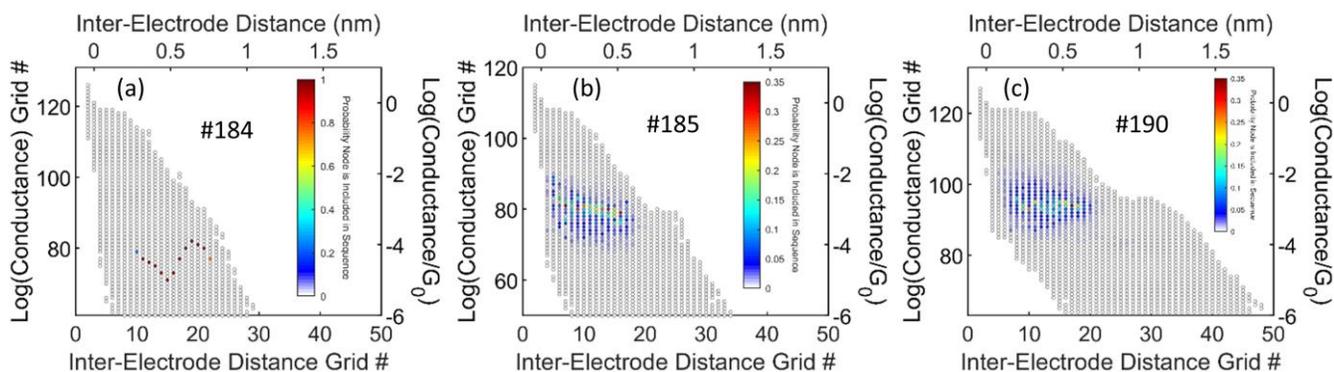

**Figure S20**. 2D node probability distributions from the MCMC results for the datasets from Figure 4 in the main text. Panels (a), (b), and (c) are for the "actually empty", "seemingly empty", and "clearly molecular" datasets, respectively, and the dataset ID# from **Table S2** for each dataset is overlaid. Each node visited by at least one trace from the given dataset is represented by a gray open circle. The color inside the circle corresponds to the probability that each node was included in the distribution of 12-node sequences produced by the MCMC feature-finder.

S.7 Robustness of MCMC Results

*S.7.1 Robustness to Different Gridding Sizes*

All of our grid-based correlation tools, including the MCMC feature-finder, rely on an initial step of coarse-gridding each trace, which requires choosing a gridding size. As explained in the main text, this gridding size has the potential to impact which correlations we are able to detect. Specifically, as the grid spacing becomes very coarse, different trace paths get lumped into the same nodes and so some correlations of these paths with their own histories may be lost. On the other hand, as the grid spacing becomes very fine, fewer traces pass through each node and so the power of our statistics declines, which can also lead to correlations being missed (or, equivalently, require a larger number of raw traces to detect the same correlation). In addition to these trade-offs in terms of correlation detection, more finely-spaced grids have the advantage of allowing us to more precisely distinguish between different distance and conductance values, whereas more coarsely-spaced grids are more computationally efficient.

While different gridding sizes may thus be appropriate for different applications—e.g., depending on the dataset size and/or the breadth of the feature under consideration—in practice we have found that a single gridding size can achieve a good balance between the trade-offs described above in a variety of contexts. In particular, throughout this work we use a grid with 25 nodes per nm of inter-electrode distance and 10 nodes per decade of log(conductance/$G_0$), a.k.a 25x10 gridding. To demonstrate that our final results are not overly sensitive to this gridding size, we re-ran our MCMC feature-finder on all of the datasets from MCBJ sample #114-5 using 10x4, 15x6, 30x12, and 35x14 gridding. We used node-sequence lengths of 5, 7, 14, and 17, respectively, for these new griddings in order to represent nearly the same actual length (~0.5 nm) as the 12-node sequences used in the main paper with the 25x10 gridding.

As shown in **Figure S21**, for each of the datasets collected in the presence of molecule—whether or not the raw data clearly displayed a molecular feature—the five different grid sizes have almost no impact on the distribution of the feature discovered by the MCMC, and do not qualitatively affect our main conclusion that the same feature is found in both the "seemingly empty" and "clearly molecular" datasets.

This demonstrates that the ability of our MCMC feature-finder to successfully identify rare plateau features is not dependent on a single choice of grid size. Interestingly, for the actually empty dataset, the different grid sizes do lead to meaningfully different features identified by the MCMC. This may be expected because, as shown in **Figure S20** above, the lack of true molecular plateaus in the "actually empty" dataset forces the MCMC feature-finder to focus on just the single plateau-like path that happens to be most significant, and single paths will of course be more sensitive to the exact gridding size. These results thus constitute additional evidence that the plateau-like features identified in the "seemingly empty" and "clearly molecular" datasets correspond to meaningful plateau behaviors in the experimental data, whereas the plateau-like features identified in the "actually empty" datasets do not.

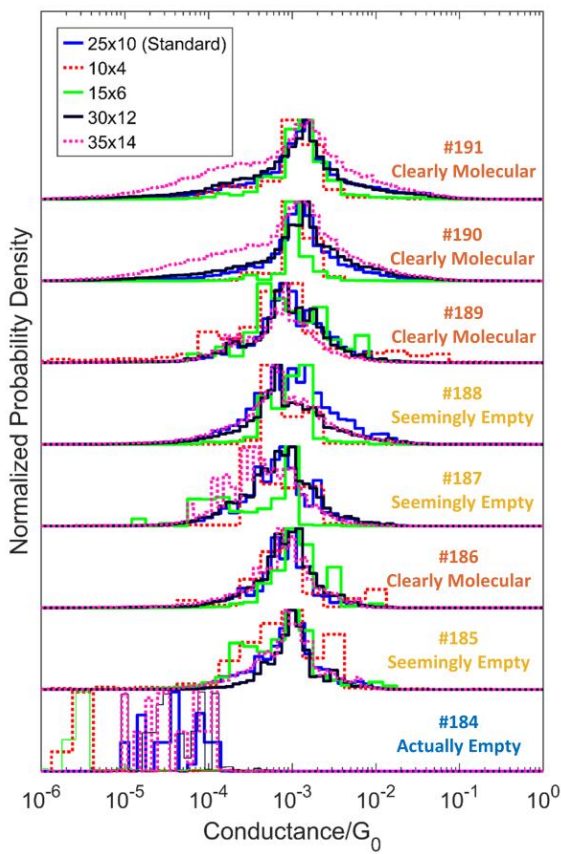

**Figure S21**. Demonstration of the robustness of the MCMC feature-finder results to the gridding size used. The final conductance distributions from the MCMC feature-finder are shown for the eight datasets collected with MCBJ sample #114-5 using the standard gridding size (25x10) as well as two coarser (10x4 and 15x6) and two finer (30x12 and 35x14) gridding sizes. Each dataset is labeled with its dataset ID# from **Table S2** as well as our categorization of either "actually empty", "seemingly empty", or "clearly molecular".

*S.7.2 Robustness to Different Section Lengths*

All of the three molecules investigated in this work have sulfur-to-sulfur lengths of ~1.3 nm, implying an apparent feature length of ~0.8 nm once snap-back is taken into account.[17,18] However, we also wanted include plateaus in which the molecule breaks off prematurely, because in the case of a *rare* plateau feature

S31

such "short" plateaus may constitute a significant fraction of all plateaus. We therefore use 12-node sequences, corresponding to 0.48 nm of inter-electrode distance, for the MCMC feature-finder results presented in the main text.

To demonstrate that our final results are not overly sensitive to this choice, we re-ran our MCMC feature-finder on all of the datasets from MCBJ sample #114-5 using either 8-node or 14-node sequences instead. As shown in **Figure S22**, these changes to the node-sequence length have essentially no effect on the locations of the discovered plateau features in conductance space. In the case of dataset #188 the 10-node sequences produce a somewhat different distribution, but this appears to simply be a consequence of our decision to set very loose slope criteria (see S.7.3 below); if we lower the maximum allowable range slope from 2.5 decades/nm to 1.25 decades/nm the 10-node sequence distribution for dataset #188 would move back in line with the other results. Similar to when the gridding was modified (see section S.7.1 above), there is more variation in the distributions for the "actually empty" dataset, consistent with our hypothesis that this dataset contains no meaningful plateau feature.

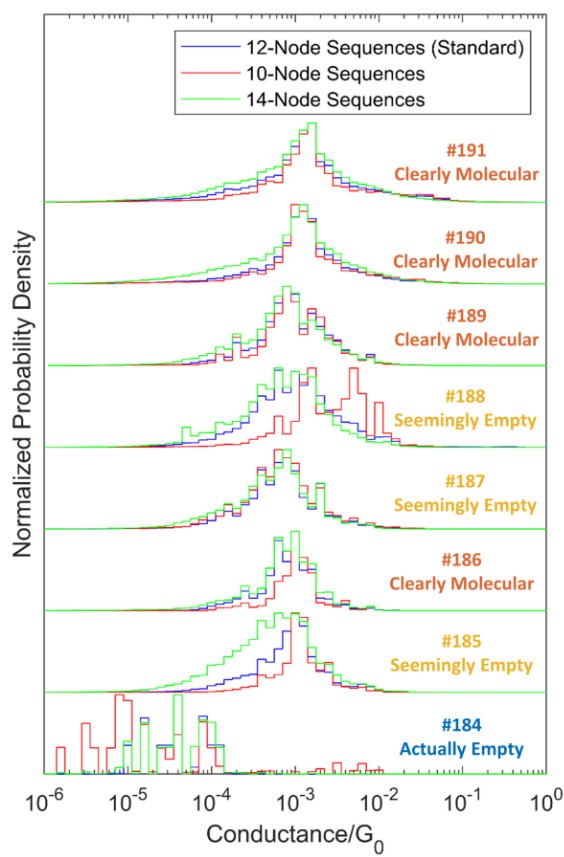

**Figure S22**. Demonstration of the robustness of the MCMC feature-finder results to the length of the node sequences being considered. The final conductance distributions are shown for the eight datasets collected with MCBJ sample #114-5 using the standard sequence length of 12, as well as a shorter length (8) and a longer length (14). Each dataset is labeled with its dataset ID# from **Table S2** as well as our categorization of either "actually empty", "seemingly empty", or "clearly molecular".

*S.7.3 Robustness to Different Slope Criteria*

As described in section S.5.2 above, we impose a maximum allowable "range slope" on our MCMC feature-finder in order to focus attention on plateau-like features. For the results in the main text, we set the maximum range slope to 1 *y*-grid/*x*-grid, which is equivalent to 2.5 decades/nm. This value was chosen to be extremely conservative in what can be considered "plateau-like"—plateaus identified in the literature typically have slopes *much* shallower than 2.5 decades/nm.[19] In the main text, we already demonstrated that using a slope restriction is *not* "forcing" the MCMC feature-finder results into plateau-like shapes which are unrepresentative of the experimental data. In order to also ensure that the specific value of the maximum does not have a strong effect on our results, we re-ran our MCMC feature-finder on all of the datasets from MCBJ sample #114-5 using either a 3.75 decades/nm or 1.25 decades/nm maximum range slope instead. As shown in **Figure S23**, these different slope criteria have essentially no effect on the locations of the discovered plateau-features in conductance space. Again (see sections S.7.1 and S.7.2 above), the main exception is the "actually empty" dataset, but this is consistent with our view that that dataset does not contain a meaningful plateau feature.

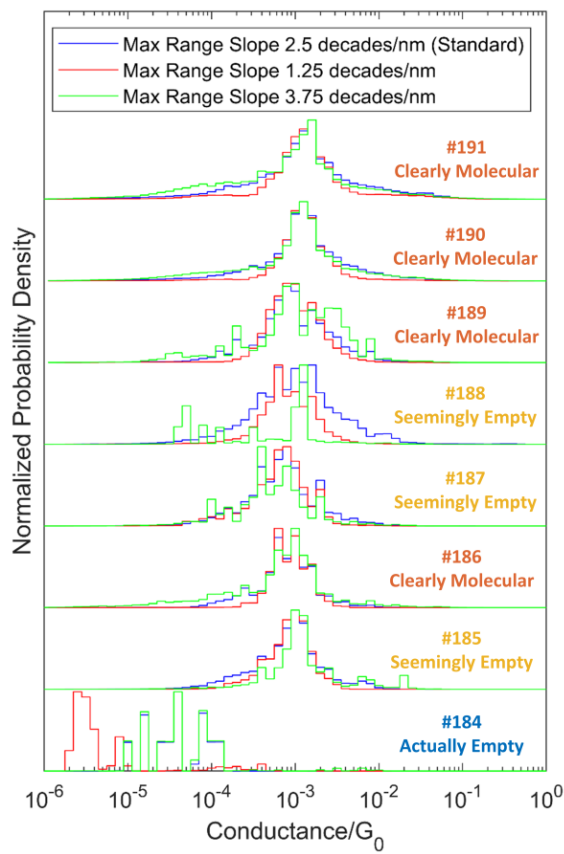

**Figure S23**. Demonstration of the robustness of the MCMC feature-finder results to the maximum allowable range slope imposed on the node sequences. The final conductance distributions are shown for the eight datasets collected with MCBJ sample #114-5 using the standard maximum range slope, as well as a more restrictive limit and a less restrictive limit. Each dataset is labeled with its dataset ID# from **Table S2** as well as our categorization of either "actually empty", "seemingly empty", or "clearly molecular".

The success of our MCMC feature-finder in identifying plateau features even with the extremely loose condition of a maximum range slope < 3.75 decades/nm raises the question of whether imposing slope criteria is necessary in the first place. The MCMC results for the eight datasets from MCBJ sample 114-5 using no slope restriction are thus shown in **Figure S24**, and indeed, the same plateau feature near $10^{-3}$ $G_0$ is still identified in several of them. In a couple others, however, a more-sloped feature is identified instead. This reinforces our original rationale for imposing slope restrictions: a single dataset will typically contain multiple types of rare behaviors, and the MCMC feature-finder is designed to primarily identify the single rare behavior that happens to be *least* random-walk-like. In a completely unrestricted MCMC simulation, a particular feature of interest may thus be identified in some datasets but not others—as seen in **Figure S24**—due to slight variations in the other behaviors present in each dataset. When an investigation is focused on a particular type of feature, such as molecular plateaus, it is therefore best to include that focus in the MCMC feature-finder by imposing appropriate criteria on the allowed node-sequences.

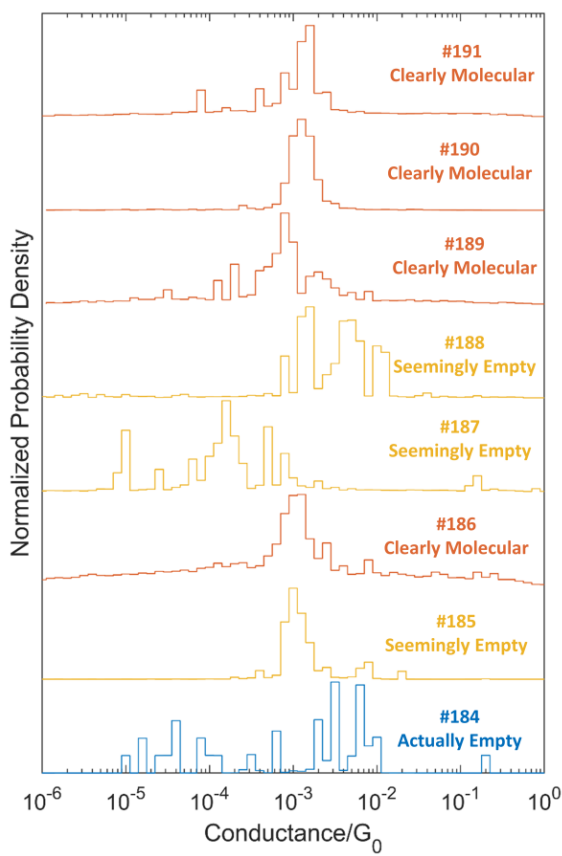

**Figure S24**. 1D conductance distributions from the MCMC feature-finder applied to the eight datasets collected with MCBJ sample #114-5 when no slope restrictions/criteria are used. Because these no-restriction MCMC simulations proved less efficient at mixing than our previous simulations, 600,000 rather than 200,000 burn-in steps were used (see section S.5.5 above). Some datasets still display a clear plateau feature, but others do not, because a single dataset can contain multiple types rare behaviors. Each dataset is labeled with its dataset ID# from **Table S2** as well as our categorization of either "actually empty", "seemingly empty", or "clearly molecular".

*S.7.4 Robustness to Different Effective Temperatures*

As explained above in section S.5.3, the effective temperature used by the MCMC feature-finder controls how peaked vs. flattened the final distribution will be. For the results presented in this work, we used an effective temperature of $4\sigma_\infty/9$, where $\sigma_\infty$ is determined separately for each dataset. However, since the MCMC feature-finder uses parallel tempering, results were simultaneously obtained for five total temperatures $\left(\frac{4\sigma_\infty}{9}, \frac{5\sigma_\infty}{9}, \frac{6\sigma_\infty}{9}, \sigma_\infty, \infty\right)$, and these results demonstrated that our main result—the location of the identified plateau feature in each dataset—is robust to the exact value of the effective temperature.

To illustrate this, **Figure S25** shows the overlaid 1D conductance distributions for dataset ID #185 (the "seemingly empty" dataset from Figure 4 in the main text) for six different effective temperatures. In addition to the five temperatures used for parallel tempering throughout this work, we added a sixth temperature of $\sigma_\infty/3$ in order to also test a value below our standard choice of $4\sigma_\infty/9$. The 2D distributions for these six temperatures are shown in **Figure S26**. These two figures show that effective temperatures in the range $\frac{\sigma_\infty}{3} \to \frac{2\sigma_\infty}{3}$ produce qualitatively equivalent results which clearly show the same location for the plateau feature. Temperatures much higher than $2\sigma_\infty/3$, on the other hand, produce such a flattened distribution that it becomes difficult to identify a single most-prominent feature. Our standard choice of $T = 4\sigma_\infty/9$ is thus justified.

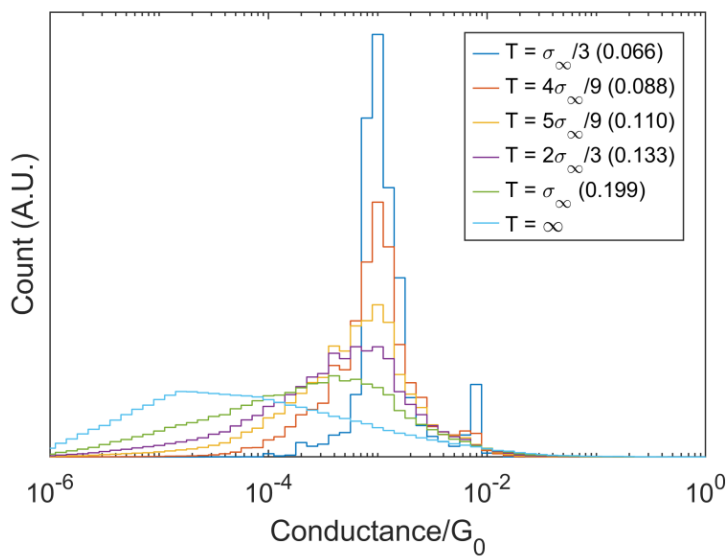

**Figure S25**. Example 1D conductance distributions produced by the MCMC feature-finder at six different temperatures, to illustrate the robustness of our results to the exact value of the effective temperature. These results are for dataset ID #185 from **Table S2**, and the six temperatures were run at the same time using parallel tempering. Qualitatively similar results are obtained for effective temperatures in the range $\frac{\sigma_\infty}{3} \to \frac{2\sigma_\infty}{3}$, justifying our standard choice of $T = 4\sigma_\infty/9$.

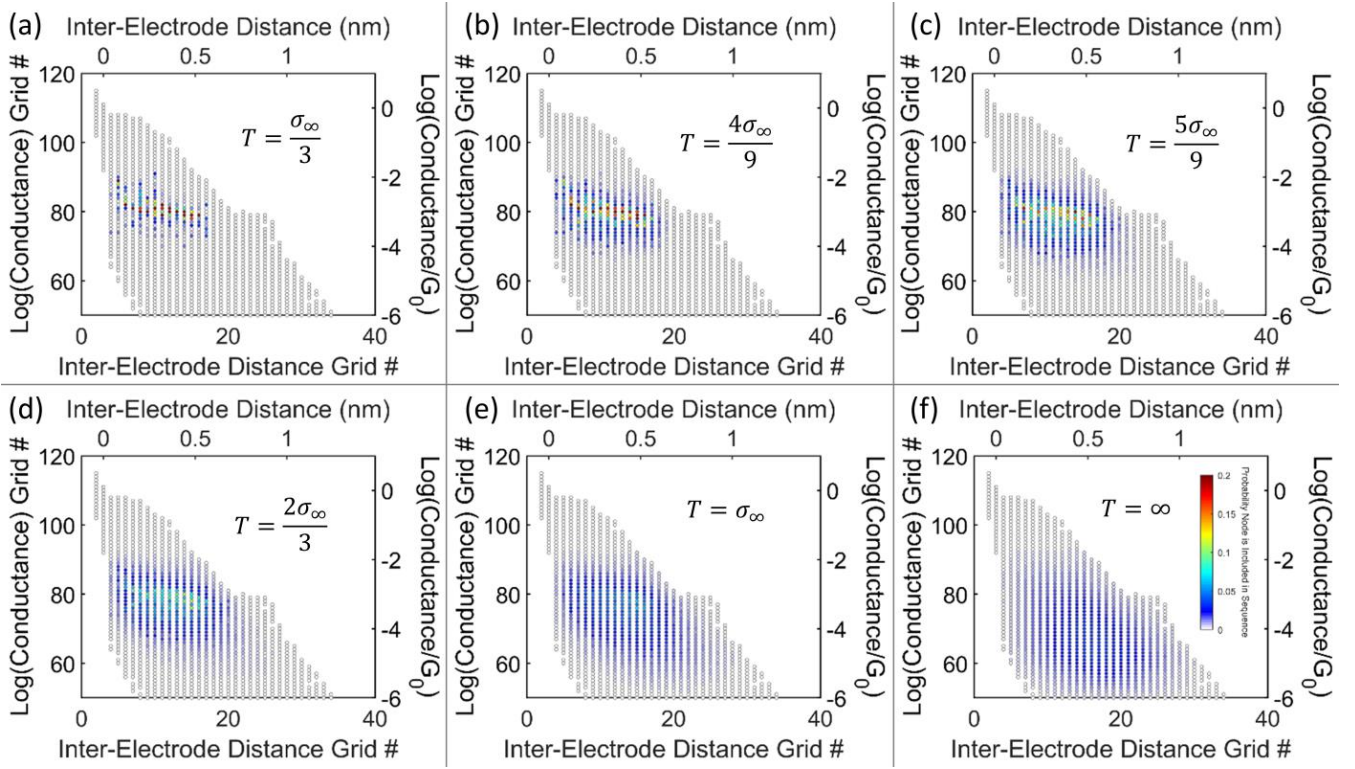

**Figure S26**. The 2D distributions corresponding to each 1D conductance distribution in **Figure S25**. These plots nicely illustrate how the effective temperature, $T$, controls how peaked vs. flattened the final distribution is, but does not meaningfully affect the location of the identified plateau feature until $T$ becomes high enough that the distribution approaches its fully flattened $T = \infty$ incarnation. Note that the $T = \infty$ distribution is "fully flattened" in that all allowed node sequences have equal probability, but the individual node probabilities shown in (f) are unequal because some nodes are included in more allowed sequences than others.

S.8 Details on Trace Scoring

As explained in the main text, we "score" a trace versus a particular node it passes through by taking the average connection strength between that selected node and all the other nodes the trace passes through. Mathematically, if a trace $t$ is composed of the nodes $t = (m_1, m_2, \cdots, m_L)$, and if $X \in \{m_i\}$ is one of those nodes, then the score of $t$ versus $X$ is defined as:

$$\frac{1}{L}\sum_{i=1}^{L} CS(m_i, X)$$

Two examples are shown in **Figure S27**: the pink trace passes through many nodes that have positive connection strengths with the selected (green-circled) node and so receives a fairly high score; conversely, the cyan trace mostly passes through nodes that have small or negative connection strengths with the selected node, and so receives a low score.

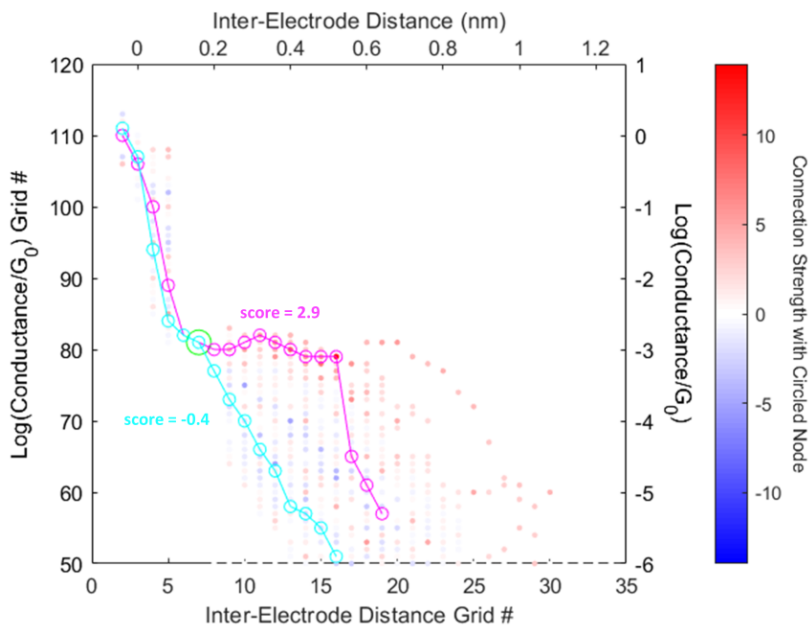

**Figure S27**. Examples of scoring traces versus particular nodes. The red and blue circles represent the connection strength distribution for dataset ID #185 versus the green-circled node. Overlaid in pink and cyan are two coarse traces, with the pink one receiving a high score versus the green-circled node since it passes through many high connection strength nodes, and the cyan one receiving a low score since it does not.

S.9 Robustness of Conclusions Drawn from Trace Scoring

    As explained above in section S.8, scoring of traces is always relative to a specific single node. The results in Figure 5 in the main text, therefore, are specific to the indicated node, which was selected because it appeared a high percentage of the time in the MCMC simulation output. To demonstrate that our conclusions drawn from Figure 5 are not overly sensitive to that specific choice, we consider the 16 nodes which, between them, account for 30% of the probability density in the MCMC simulation output for the dataset from Figure 5 (dataset ID #185). For each of these nodes, we scored every trace passing through that node, relative to that node. **Figure S28**a shows the distribution of scores for all of those traces (if a trace passed through more than one of the 16 nodes, we use its highest score). Then, just as in Figure 5 in the main text, we select the top-10% scoring traces from among this set (cut-off is shown by red dashed line in **Figure S28**a). As shown in **Figure S28**b,c, these more-robustly-chosen top-scoring traces support the same conclusions as Figure 5: the experimentally collected traces *do* contain plateau-like features, validating the plateau shape discovered by the MCMC feature-finder.

    The red curve in **Figure S28**b shows that a small molecular peak is observed even if *all* of the traces through the 16 selected nodes are considered, rather than just the top-10% scoring ones. This make sense, because those 16 nodes were selected because the MCMC feature-finder suggested that they were located on top of a plateau-like feature, and so they already help limit our focus to those traces most likely to contain such plateaus. However, we note that the molecular peak is much stronger in the top-10% scoring traces, demonstrating the utility of our scoring method for selectively extracting the traces corresponding to a given rare behavior.

    It is interesting to note that the extracted molecular traces display a sharper 1 $G_0$ peak than the full dataset (green curve in **Figure S28**b). This observation could indicate additional nanoscopic information

about the junction environment—e.g., perhaps nearby molecules stabilize the formation of gold atomic chains. Such hypotheses are intriguing directions for future work, but are outside the scope of the current manuscript.

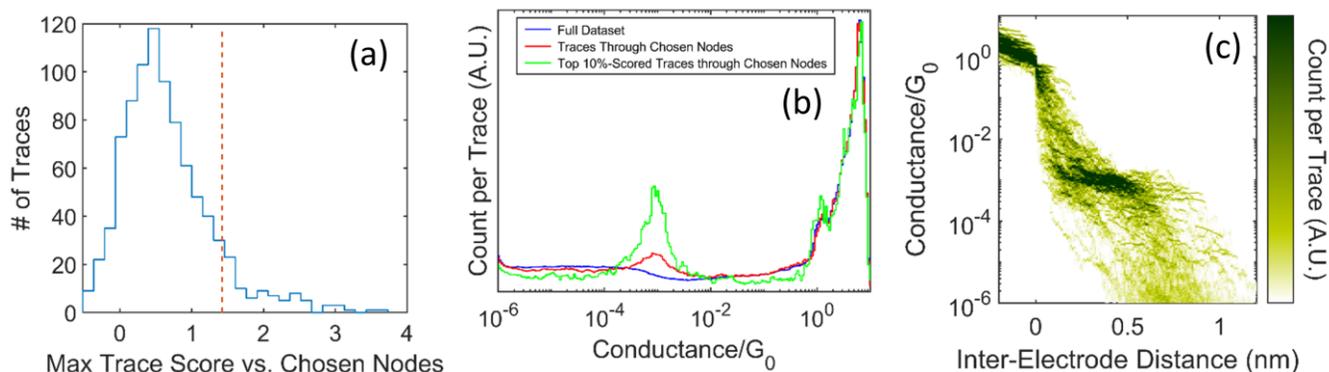

**Figure S28**. Robust extraction of the original experimental traces corresponding to the plateau feature discovered by the MCMC feature-finder for dataset ID #185. (a) Histogram of maximum trace scores for all traces passing through the 16 nodes selected using the MCMC results. The red dashed line indicates the cut-off for the top 10% of scores. (b) Overlaid conductance histograms for all traces in the dataset (blue), only the traces passing through the 16 selected nodes (red), and for the top 10%-scoring traces through the selected 16 nodes (green). (c) 2D histogram corresponding to the green curve from (b). Comparing the results in this plot to those in Figure 5 in the main text demonstrates that our scoring results are robust to which specific high-probability node is chosen to score against.

S.10 Comparison to Clustering Methods

In this section, we analyze one of our experimental datasets using clustering methods as a way to benchmark the results of our new grid-based correlation approach and demonstrate its advantages in rare event detection. We want to emphasize at the outset that the results of this type of comparison must be interpreted extremely cautiously. There are essentially an infinite number of different ways to implement clustering of breaking trace data: any number of different feature spaces can be defined for the data (parameterized traces,[20] trace histograms,[21] trace images,[22] histogram bins,[23] parameterized segments,[10] etc.); those features spaces can be further processed with different dimensionality-reduction techniques (principal component analysis (PCA),[21] auto-encoders,[24] uniform manifold approximation and projection (UMAP),[25] etc.); many different clustering algorithms can be employed (k-means,[22] OPTICS,[23] spectral clustering,[26] etc.); and, most importantly, even any single combination of these choices can be implemented in slightly different ways depending on data pre-processing details, parameter values, etc. Our goal in this section is thus decidedly *not* to demonstrate that our approach will always out-perform clustering, since such a claim is both impossible to prove and unlikely to be true. Instead, we use a few representative clustering approaches to demonstrate that these strategies, as they have been commonly implemented in the field so far, generally struggle with the identification of very rare behaviors. This helps motivate the need for a new approach specifically targeted at the challenge of rare event detection.

For our comparison, we consider three whole-trace clustering approaches: representing each trace with a 28 x 28 pixel image and then clustering with k-means (28x28 + k-means); representing each trace with a 28 x 28 pixel image, using UMAP to reduce to three dimensions, and clustering with Shi and Malik[27] spectral clustering (28x28 + UMAP + Spectral); and using PCA to reduce the raw traces to three dimensions then clustering with k-mediods (PCA + k-mediods). These approaches were chosen because they include many of the most common strategies employed to date in single-molecule transport data analysis, and because they are implemented in a publicly available software package.[25,28] Each of these clustering algorithms requires the number of clusters to be specified, so we calculated results for 2 to 9 clusters and used the Calinski-Harabasz (CH) index[29] to help determine an "optimal" number of clusters. Again, there are a plethora of cluster validation indices available, so our use of the CH-index is simply intended to be representative of the type of strategy often used in single-molecule transport analysis.[25,26] In addition to these whole-trace clustering approaches, we also compare to the results of segment clustering as described in Bamberger 2020 et al.[10] Because segment clustering is already designed to focus on linear segments, it is likely particularly well-suited to the task of identifying plateau features, even when they are rare, and thus constitutes an important benchmark for our new approach. In the interest of transparency, it is also important to note that segment clustering was developed by our research group, and so may also perform better in these comparisons because it could have been unconsciously "tuned" to the characteristics of data collected on our specific experimental set-up.

To perform these comparisons, we use the "seemingly empty" experimental dataset from Figure 4 in the main paper (dataset ID #185), since this is the primary dataset we used to demonstrate our new grid-based correlation approach. As a reminder, we know from other datasets collected in the presence of the same molecule that the rare plateau feature in this dataset should occur at ~$10^{-3}$ $G_0$ and extend to ~0.8 to 1.0 nm inter-electrode distance (e.g., see dataset ID #186 in **Figure S12**). As an additional "challenge case", we also created a synthetic experimental dataset by combining 600 consecutive traces from dataset ID #185 with 2400 consecutive traces from dataset ID #184, which was collected before any molecules were deposited. This "diluted" dataset should thus represent an example in which the plateau feature in the "seemingly empty" Figure 4 dataset has become even more rare.

The results of our comparisons are summarized in **Table S3**, with detailed results presented below in Sections S.10.1 and S.10.2. For the original rare-plateau dataset, the clustering approaches do have significant success extracting the plateau feature. However, in each case this extraction is somewhat of a struggle: it requires the CH-index to be ignored, it does not cleanly separate the plateau feature from the tunneling background, or it requires an approach (segment clustering) that is already focused on linear segments. When we move to the more challenging case of the diluted rare-plateau dataset, our MCMC feature-finder is still able to cleanly extract the molecular plateau, but all four clustering approaches now significantly struggle to unambiguously identify the plateau feature at all or cleanly separate it from the tunneling background. These results thus illustrate that while the typical clustering approaches used to date in the single-molecule transport field can locate rare plateau features in some cases, they are in general challenged by this task, especially as the "rareness" of the feature increases. Moreover, the results in Section S.10.2 for the diluted dataset provide insight into why this is the case: because the plateau feature is *so* rare in this dataset, the MCMC feature-finder only identified 11 traces that clearly belong to it; in contrast, the clustering approaches tend to produce clusters with far more traces in them, making it impossible for the plateau feature to be cleanly extracted. This is likely because these types of clustering approaches have primarily been designed and demonstrated with more-frequent features in mind, and so consequently they have been implicitly or explicitly "tuned" towards such major features rather than very rare ones. This helps motivate our claim that investigating rare behaviors requires an approach specifically targeted at this particular task.

**Table S3**. Summary of the success of our MCMC feature-finder as well as four representative clustering approaches in finding the known molecular feature in both our example rare plateau dataset (ID # 185) as well as a "dilution" of this dataset in which extra non-molecular traces were added in to make the plateau feature even more rare. Plots showing the detailed results that this table summarizes can be found in Sections S.10.1 and S.10.2.

|  | **Results for Rare Plateau Dataset** | **Results for "Diluted" Rare Plateau Dataset** |
|---|---|---|
| MCMC Feature-Finder | Plateau feature clearly and cleanly identified (**Figure S28** and/or Figure 5b,c in main paper) | Plateau feature clearly and cleanly identified (**Figure S33**) |
| 28x28 + k-means | Plateau feature cleanly extracted, but only if CH-index is ignored and many clustering solutions are searched through (**Figure S29**) | Weak plateau-like feature with poor separation from tunneling is identified, but only if the CH-index is ignored and many clustering solutions are searched through (**Figure S34**) |
| 28x28 + UMAP + Spectral | Plateau feature identified, but not completely separated from tunneling/other behaviors (**Figure S30**) | No clear plateau-like feature identified; weak feature possibly present but very poorly separated from tunneling background (**Figure S35**) |
| PCA + k-mediods | Plateau-like feature identified, but very poor separation from tunneling background (**Figure S31**) | Weak plateau-like feature with very poor separation from tunneling is identified (**Figure S36**) |
| Segment Clustering | Plateau feature clearly identified and mostly well-separated from the background tunneling/other behaviors (**Figure S32**) | Three potential plateau features are extracted, not all of which match the expected molecular feature (**Figure S37**) |

*S.10.1 Detailed Comparison Results for Example Dataset*

When we apply 28x28 + k-means clustering to our example rare plateau dataset, the CH-index indicates that the optimal number of clusters is 2 (**Figure S29**a), but this clustering solution shows no identification of the molecular feature (**Figure S29**b). If we ignore the CH-index and search through other clustering solutions, with 9 clusters (**Figure S29**c) we find a cluster that nicely extracts the plateau feature (**Figure S29**d).

Moving on to 28x28 + UMAP + Spectral clustering, the CH-index again indicates 2 as the optimal number of clusters (**Figure S30**a), but there is also a local maximum at 7 clusters. Focusing on this 7-cluster solution (**Figure S30**b), there is a cluster that extracts the plateau feature (**Figure S30**c). However, the feature is not extracted as cleanly from the background as the result of 28x28 + k-means or of our MCMC feature-finder.

The results of PCA + k-mediods clustering are broadly similar to 28x28 + UMAP + Spectral clustering, but in this case the extraction of the plateau feature is even less clean (**Figure S31**).

When we apply segment clustering to the example rare plateau dataset and consider the results following the procedure described in Bamberger et al. 2020,[10] a single "main plateau cluster" is unambiguously identified. This plateau cluster is for the most part well-separated from the background, as can be seen by looking at both the segments assigned to this cluster (**Figure S32**a) and the 2D and 1D histograms of the traces which these segments originated from (**Figure S32**b,c). It is important to note that segment clustering is perhaps one of the best-suited clustering approaches for identifying rare plateau features, since it is already focused on roughly linear trace segments such as plateaus.

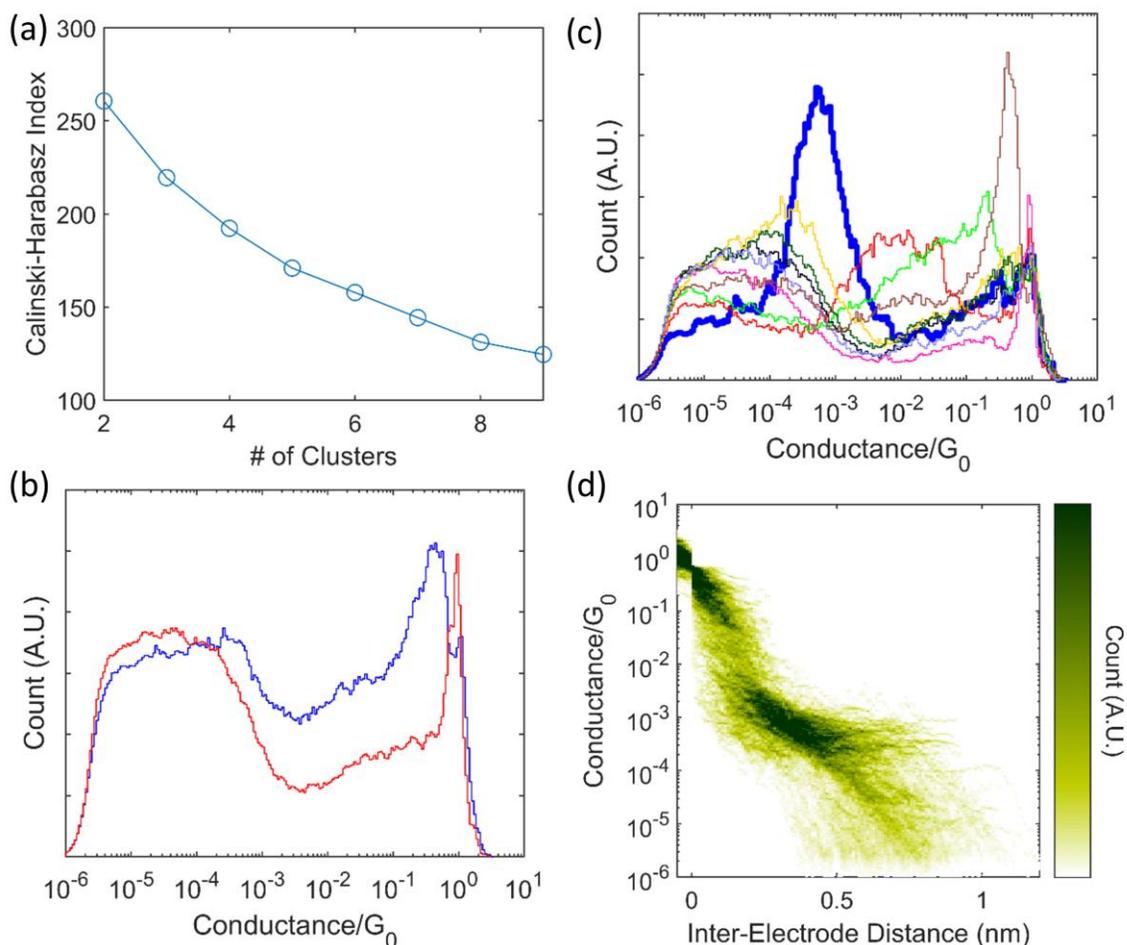

**Figure S29**. Clustering results when 28x28 + K-means is applied to our example rare plateau dataset. (a) CH-index for solutions with 2 through 9 clusters, showing a maximum value at 2. (b) Overlaid 1D conductance histograms for the 2-cluster solution, showing essentially no extraction of the molecular feature. (c) Overlaid 1D conductance histograms for the 9-cluster solution. One of these clusters (thick blue) cleanly extracts the plateau feature, as can be seen in the 2D histogram for the 318 traces in this cluster in (d).

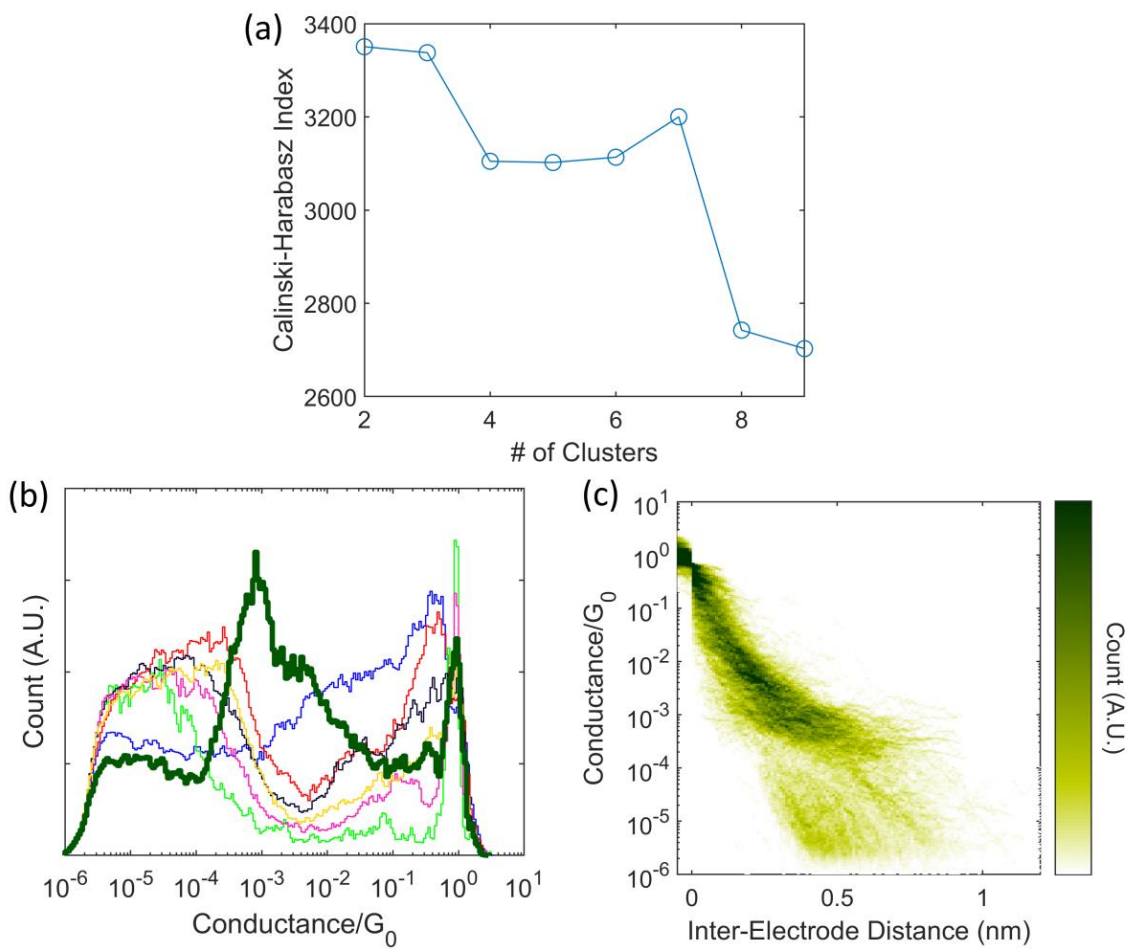

**Figure S30**. Clustering results when 28x28 + UMAP + Spectral is applied to our example rare plateau dataset. (a) CH-index for solutions with 2 through 9 clusters, showing a maximum value at 2 but also a local maximum at 7. (b) Overlaid 1D conductance histograms for the 7-cluster solution. (c) 2D histogram for the 477 traces in the cluster shown in thick dark green in (b). This cluster certainly contains the plateau feature, but the separation from the tunneling and other background behaviors is not completely clean.

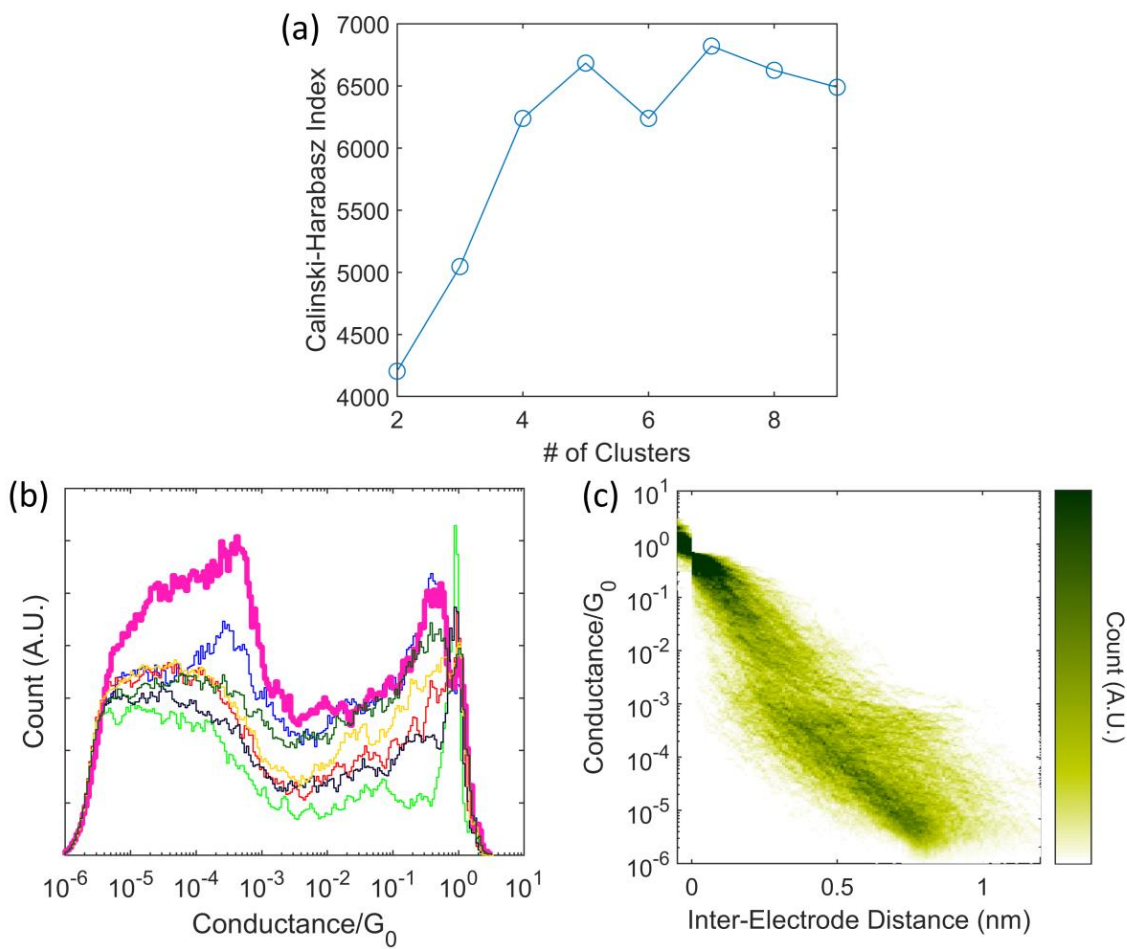

**Figure S31**. Clustering results when PCA + k-mediods is applied to our example rare plateau dataset. (a) CH-index for solutions with 2 through 9 clusters, showing a maximum value at 7. (b) Overlaid 1D conductance histograms for the 7-cluster solution. (c) 2D histogram for the 487 traces assigned to the cluster in (b) that most resembles the known molecular feature (thick pink). The separation of the plateau feature from the tunneling is quite incomplete.

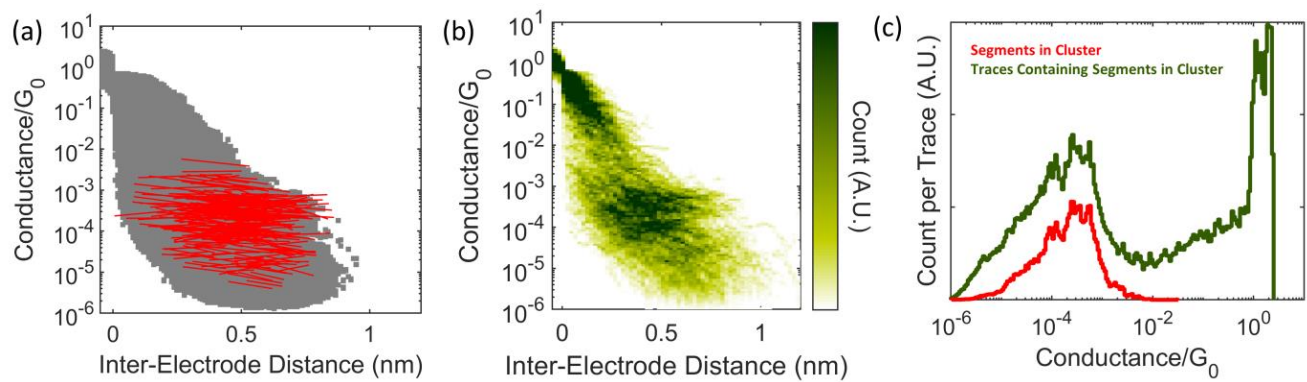

**Figure S32**. Results when segment clustering is applied to our example rare plateau dataset. (a) Linear fits to the segments in what was unambiguously identified as the main plateau cluster (red), overlaid on the full dataset distribution in gray. (b) 2D histogram of the 259 traces which contain segments assigned

to the cluster in (a). (c) Overlaid 1D conductance histograms for both the data from all fitted segments in (a) (red) and for all traces in (b) (dark green).

*S.10.2 Detailed Comparison Results for "Diluted" Example Dataset*

The 2D node distribution produced by applying our MCMC feature-finder to the "diluted" rare plateau feature dataset (using the same parameters as in Figure 4 in the main paper; 12-node sequences restricted to have a slope of no more than 2.5 decades/nm) is shown in **Figure S33**a, along with a projection of this distribution onto the conductance axis. The known plateau feature is clearly identified; to confirm this, in **Figure S33**b we show the 2D and 1D histograms for the top-10% scoring traces through the nodes identified by the MCMC (using the method described in Section S.9). This demonstrates that the MCMC feature-finder works well even when an already rare plateau feature is "diluted" by five times.

We next apply the 28x28 + k-means clustering to this diluted rare plateau dataset. The CH-index again indicates an optimal number of clusters of 2 (**Figure S34**a), and this 2-cluster solution again shows no separation of the plateau feature (**Figure S34**b). If we consider the other clustering solutions as well, the most promising result occurs with 7 clusters (**Figure S34**c). However, the specific cluster which appears to contain plateau-like traces is not at all well-separated from the tunneling background (**Figure S34**d).

With 28x28 + UMAP + Spectral clustering of the diluted rare plateau dataset, the CH-index indicates an optimal number of clusters of 2, but a local maximum is also found at 7 (**Figure S35**a). This 7-cluster solution contains a cluster that perhaps includes the plateau feature (**Figure S34**b), but if so the separation from the tunneling background is extremely poor.

Moving on to PCA + k-mediods clustering, the CH-index indicates that 8 is the optimal number of clusters (**Figure S36**a). One of the clusters in this 8-cluster solution does appear to contain a plateau feature (**Figure S36**b), but again, examination of this cluster shows a large contribution from tunneling that makes the molecular feature difficult to identify without already knowing that is there (**Figure S36**c).

Lastly, we apply segment clustering to the diluted rare plateau dataset (**Figure S37**). Following the procedure described in Bamberger et al. 2020,[10] three possible "main plateau clusters" are identified (**Figure S37**a-c), each with different peak conductances (**Figure S37**d). This ambiguity means that a single molecular plateau feature cannot be extracted with confidence.

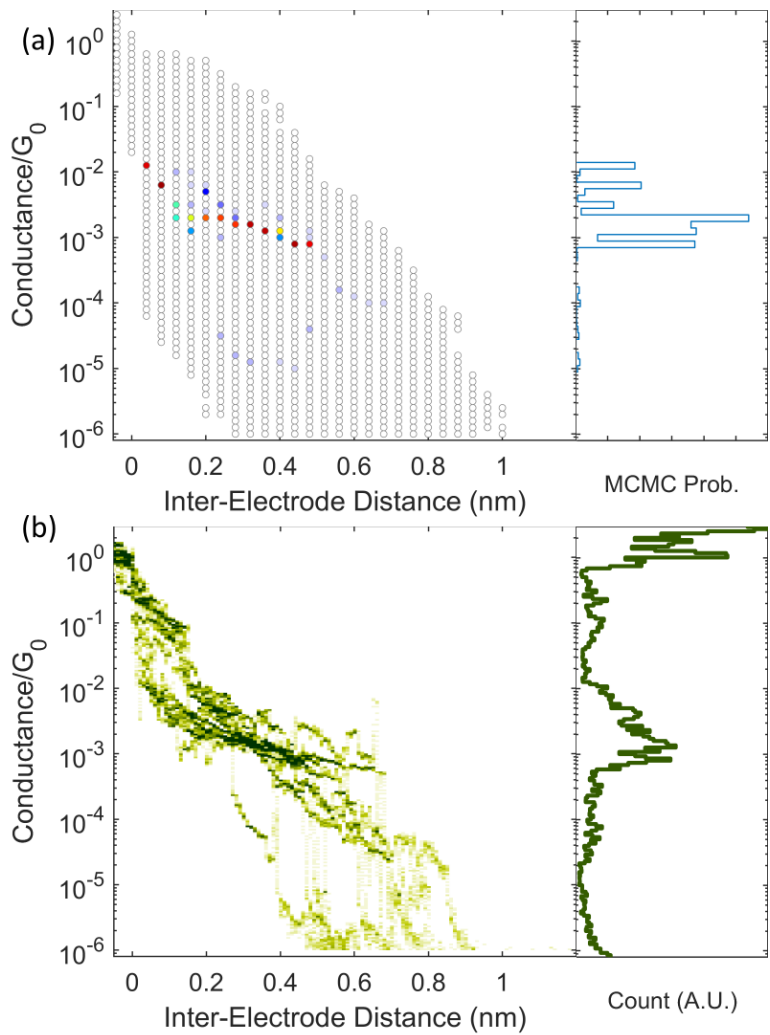

**Figure S33**. Results when our MCMC feature-finder is applied to the diluted rare plateau dataset. (a) 2D node distribution produced by the MCMC, along with its projection onto the conductance axis. (b) 2D and 1D histograms of the 11 traces selected using the distribution in (a) and the method described in Section S.9.

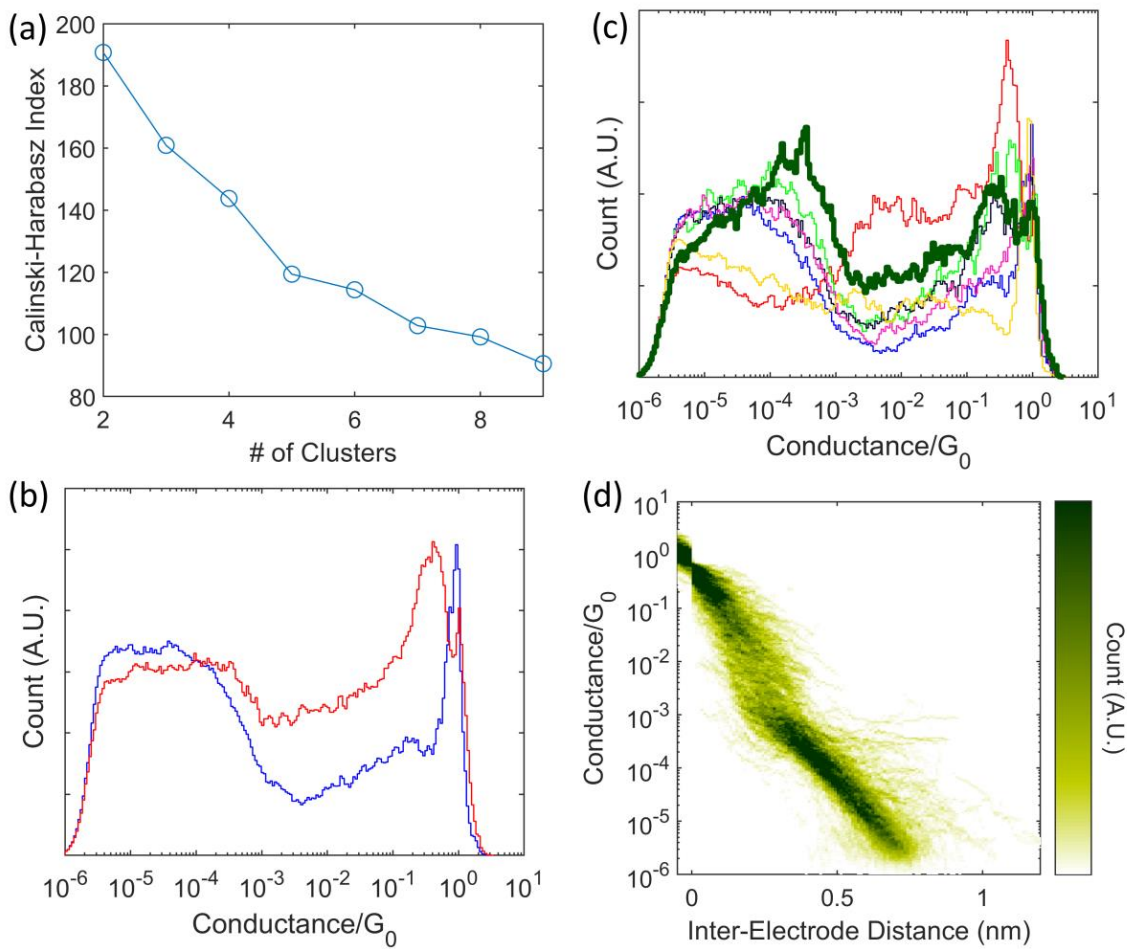

**Figure S34**. Clustering results when 28x28 + k-means is applied to the diluted rare plateau dataset. (a) CH-index for solutions with 2 through 9 clusters, showing a maximum value at 2. (b) Overlaid 1D conductance histograms for the 2-cluster solution, showing no extraction of the molecular feature. (c) Overlaid 1D conductance histograms for the 7-cluster solution, which came the closest to extracting a molecular plateau feature (the histogram in thick dark green). (d) 2D histogram for 332 traces in the most plateau-like cluster from (c), showing a large tunneling component and only a weak plateau-like contribution.

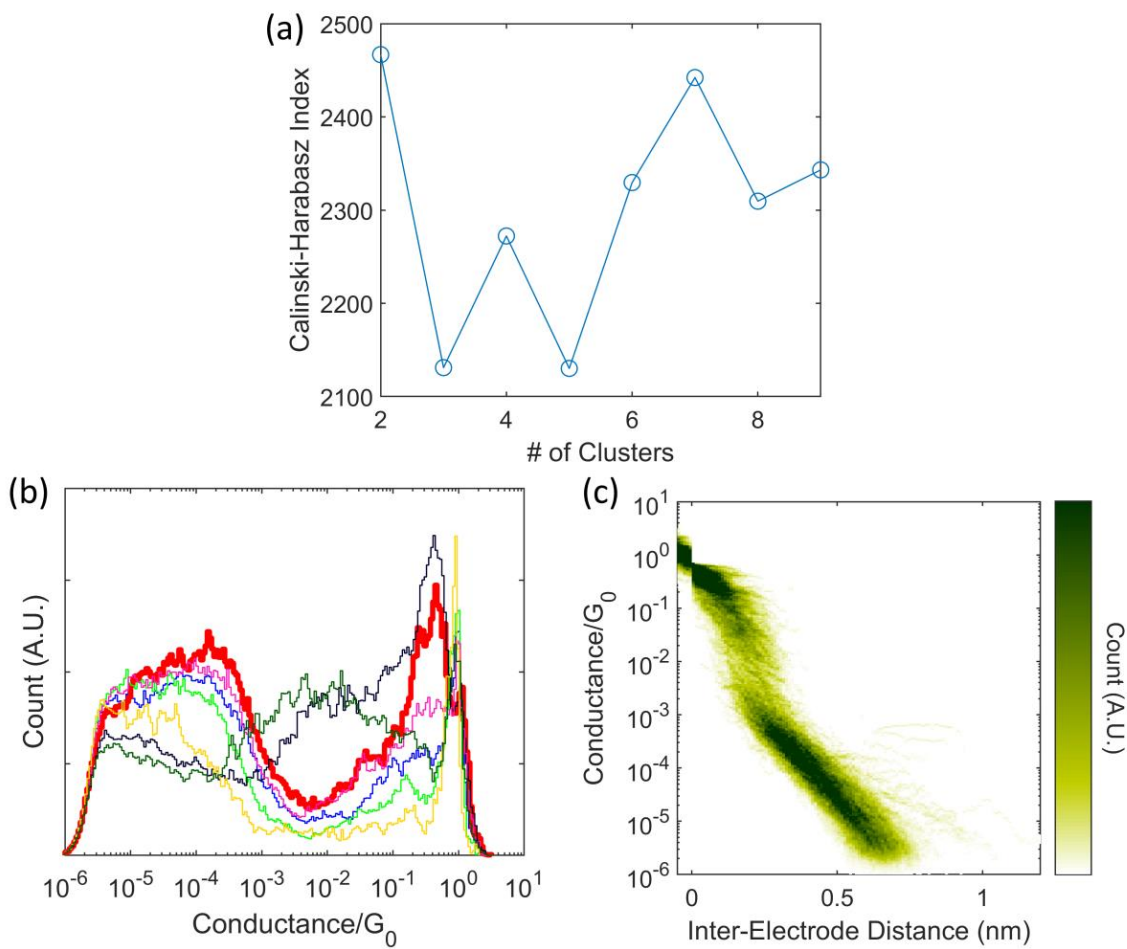

**Figure S35**. Clustering results when 28x28 + UMAP + Spectral is applied to the diluted rare plateau dataset. (a) CH-index for solutions with 2 through 9 clusters, showing a maximum value at 2, but also a local maximum at 7. (b) Overlaid 1D conductance histograms for the 7-cluster solution. (c) 2D histogram of the 458 traces from the most plateau-like cluster in (b) (thick red). This cluster is dominated by a tunneling component, indicating only very poor separation of the molecular feature.

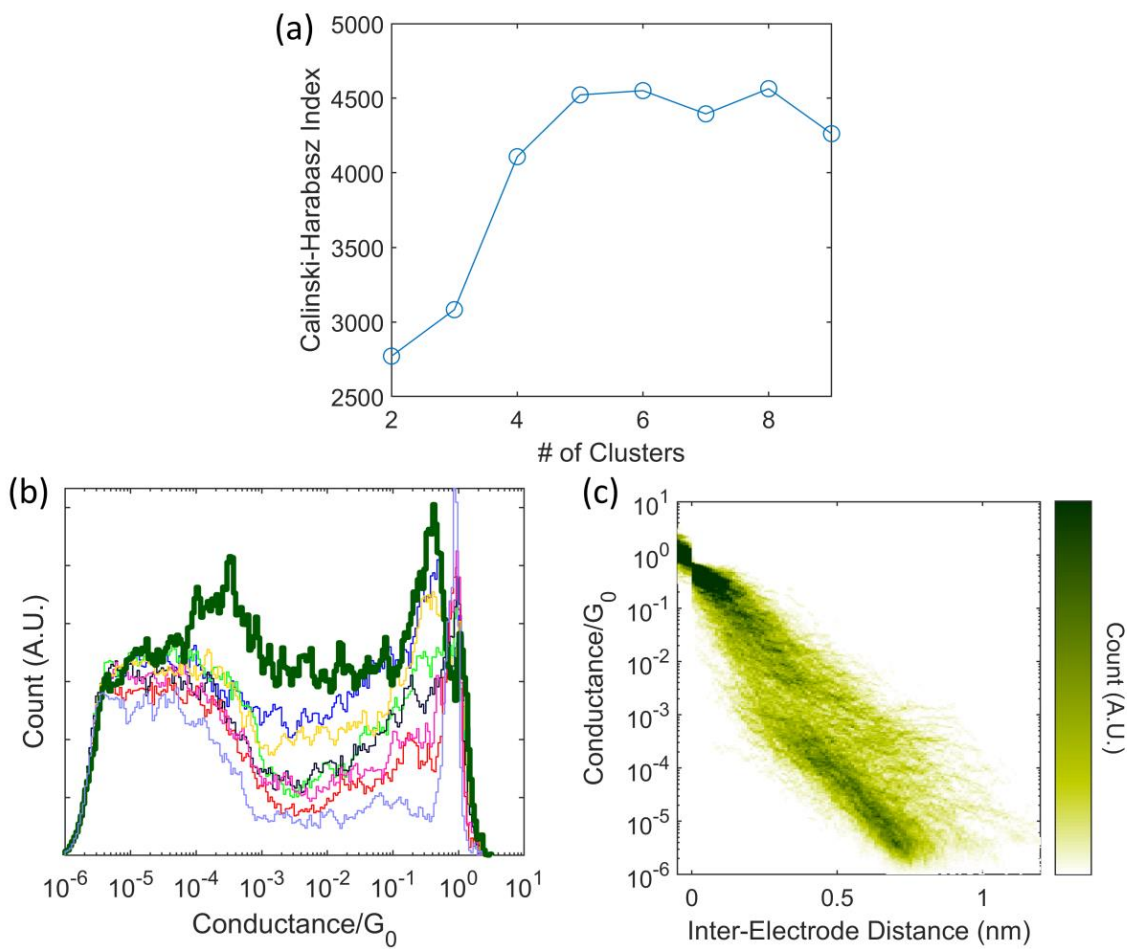

**Figure S36**. Clustering results when PCA + k-mediods is applied to the diluted rare plateau dataset. (a) CH-index for solutions with 2 through 9 clusters, showing a maximum value at 8. (b) Overlaid 1D conductance histograms for the 8-cluster solution, which came the closest to extracting a plateau-like cluster (the histogram in thick dark green). (c) 2D histogram of the 364 traces from the thick dark green cluster in (b). The very weak separation from the tunneling background makes the molecular feature difficult to identify.

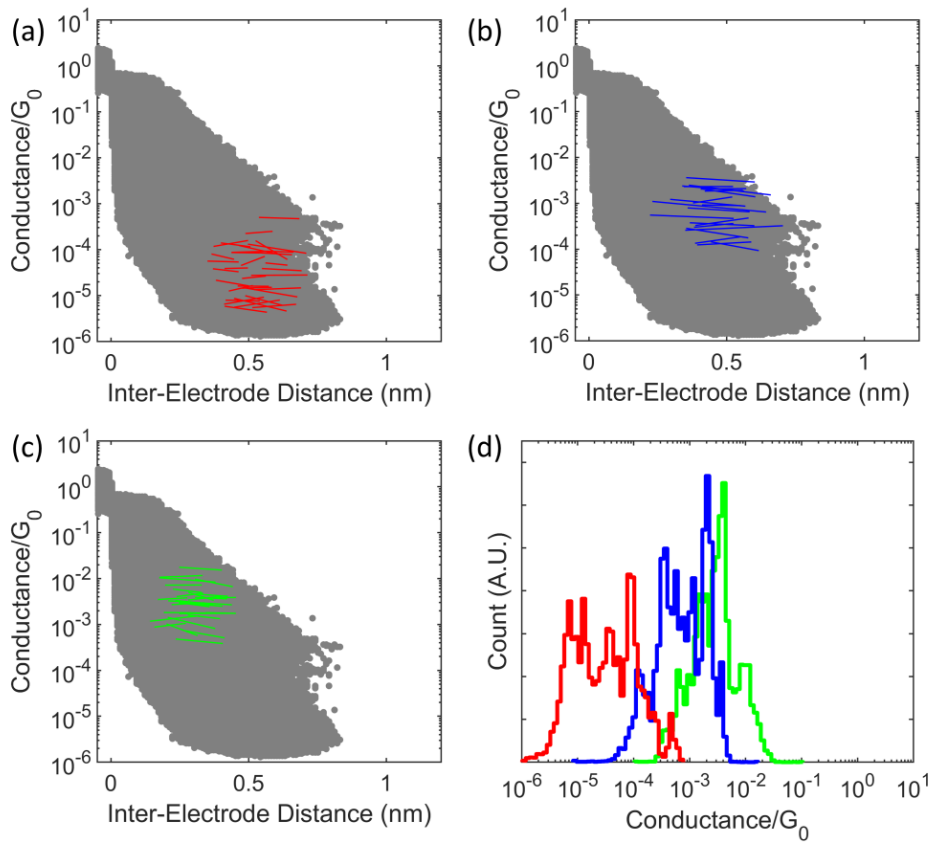

**Figure S37**. Results when segment clustering is applied to the diluted rare plateau dataset. (a-c) Linear fits to the segments in the three potential "main plateau clusters" (red, blue, and green, respectively), each overlaid onto the full dataset distribution in gray. (d) Overlaid 1D conductance histograms for the data from all the fitted segments in (a-c).

S.11 Extensions and Limitations

In this work we have focused on rare molecular plateaus in particular, but as explained in the main text our approach is designed to be compatible with rare events more generally. In this section, we briefly present some examples using simulated data to illustrate how our work can be extended, but also the limitations to that extension.

*S.11.1 Multiple Plateau Features*

The first example we consider is a dataset that has two rare plateau features at different conductances (e.g., if a molecule can bind in two different conformations). To simulate a dataset for this case, we generated 1950 tunneling traces and 50 molecular traces using the same methodology and parameters described in Section S.2, except that for half of the molecular traces we let $G_{plateau} = 10^{-3.5}$ $G_0$ and for the other half we let $G_{plateau} = 10^{-4.5}$ $G_0$. We then combined these 2000 simulated traces to create a dataset with a rare double-plateau feature (**Figure S38**a,b).

The results of applying our MCMC feature-finder to this dataset (using 8-node sequences required to have slopes no more than 2.5 decades/nm) are shown in **Figure S38**c,d. Both features are identified, showing that the feature-finder is capable in principle of locating multiple features at the same time. However, despite both plateau features consisting of 25 traces, the MCMC results are heavily weighted towards just one of them. The reason is that the MCMC feature-finder can be quite sensitive to small variations in how much the traces in each rare feature overlap (which is the price we pay for being able to detect such features in the first place). This phenomenon is not limited to multiple plateaus, but rather represents a general limitation of our approach: it is best-suited to datasets in which only a single rare feature matches the user-specified criteria. This is in fact the same issue that arose in Figure 3a in the main text, where because two highly-correlated paths were present in the data (a tunneling path and a plateau path), the MCMC feature-finder mostly focus on just the former until we added criteria that excluded it.

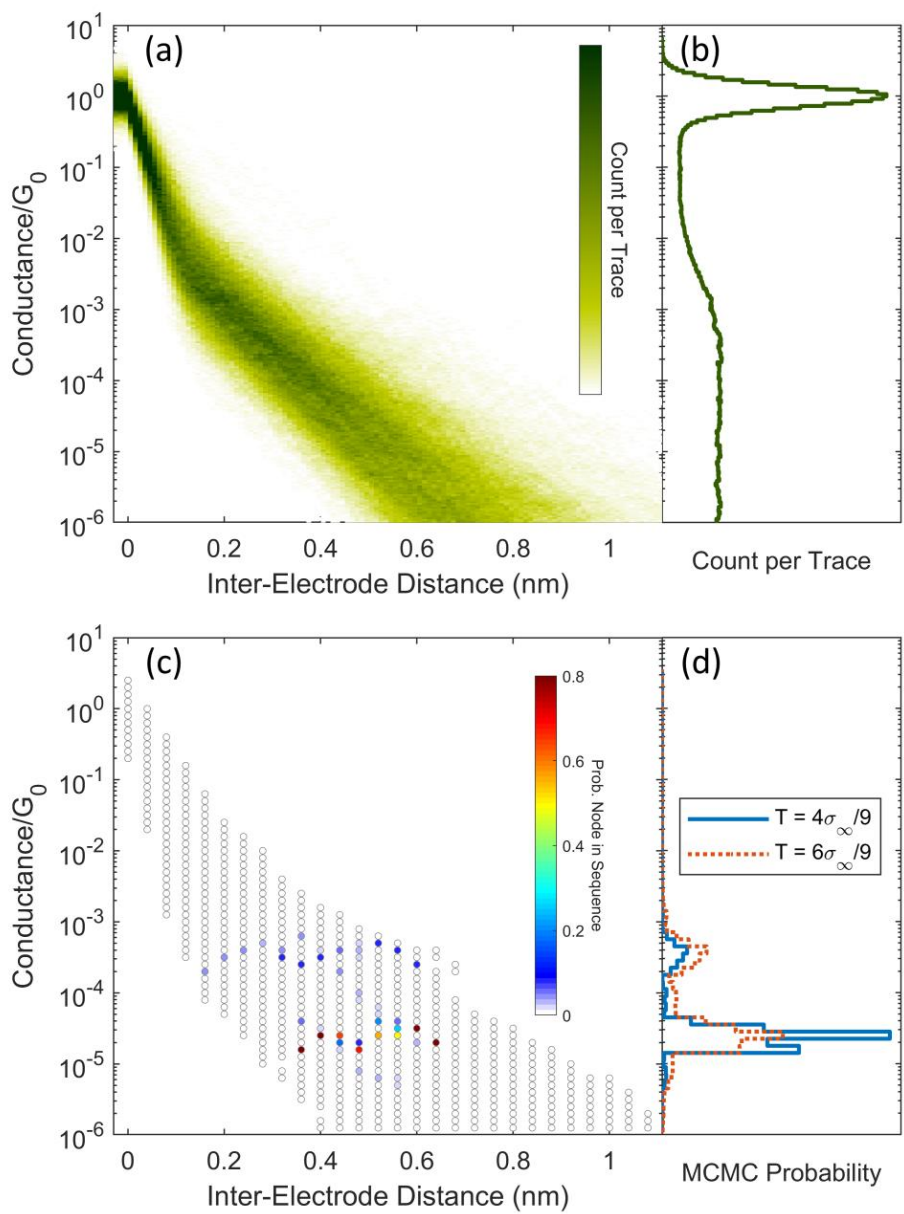

**Figure S38**. Example of applying the MCMC feature-finder to a dataset containing two rare plateau features. (a,b) A simulated dataset created by combining 1950 simulated tunneling traces with 50

simulated molecular traces divided evenly between two plateau features shows no sign of either feature in the raw data. (c,d) The distribution produced by the MCMC feature-finder clearly picks out both plateaus. However, the distribution is heavily weighted towards just one of the features, which is a common result in practice and a limitation of our approach. Raising the effective temperature (red dashed line in (d)) make the distribution somewhat more balanced (the distribution in (c) is for $T = 4\sigma_\infty/9$; the two-dimensional distribution for $T = 6\sigma_\infty/9$ is not shown; see Section S.5.3 for definition of $\sigma_\infty$).

While this limitation is intrinsic to the design of our approach, it can be partially addressed. The primary strategy, as suggested by the example in Figure 3 mentioned just above, is to add criteria to the node sequences generated by the MCMC which are specific to a single feature of interest. For example, in the hypothetical case of a molecule with two plateau features, if the features differed in their length and/or slope, they could be independently identified using the MCMC feature-finder twice with different criteria settings. Alternatively, if one plateau feature has already been identified and the user suspects that another may be present, they could add criteria restricting the plateau conductance (e.g., in the example in **Figure S38**, requiring the MCMC to only consider node-sequences with conductances $> 10^{-4}$ $G_0$ will easily identify the upper plateau feature). Our public MATLAB code is written so that users can very easily add their own custom criteria as options for the MCMC feature-finder. Finally, we also note that in the case of two rare features, raising the effective temperature used by the MCMC feature-finder will in general decrease its preference for just one of the two features (see red dotted line in **Figure S38**d). This is indeed the role that effective temperature is designed to play (see SI sections S.5.3 and S.7.4); the trade-off is that very rare features are harder to identify at higher effective temperatures.

*S.11.2 Switching Behavior*

The next example we consider is a molecule that stochastically (and rarely) switches between two different conductance states. The limitation of our approach that is relevant to address in this context is that the MCMC feature-finder is designed to locate behaviors that are localized in both conductance *and* distance (and thus likely to repeatedly pass through the same set of nodes). Therefore, if rare conductance switching preferentially occurs at a narrow range of inter-electrode distance—which could be the case, for instance, if the switching mechanism depends on stretching-induced conformational change—then our tools should be able to identify this behavior. On the other hand, if the rare switching is broadly distributed in distance—which could be the case, for instance, if the switching mechanism is light-induced or something else independent of stretching—then our approach would not be suitable.

To demonstrate these points, we constructed two types of simulated datasets containing rare switching features: "preferential switching" datasets and "random switching" datasets. For both dataset types, we start with 1900 simulated molecular traces evenly split between a high and a low conductance feature (**Figure S39**a). These molecular traces were created using the same parameters described in Section S.2, except that the average *PlateauLength* was increased to 1.2 nm and the average $G_{plateau}$ was set to $10^{-2.8}$ $G_0$ for half the traces and $10^{-4.3}$ $G_0$ for the other half. We then simulated switching traces in the same way as these molecular traces, except that the average $G_{plateau}$ value abruptly changes from $10^{-2.8}$ $G_0$ to $10^{-4.3}$ $G_0$ (or vice versa) partway through. To create "random switching" traces, we chose the distance at which this change occurs uniformly at random between 0.5 and 1.0 nm (**Figure S39**b), and to create "preferential switching" traces we chose this distance uniformly at random between 0.7 and 0.75 nm (**Figure S39**c).

Next, we created a preferential switching *dataset* by combining the 1900 molecular traces with 50 random switching traces and 50 preferential switching traces (both types of switching traces were evenly split between low-to-high and high-to-low). This dataset represents a molecule that rarely switches between two conductance features, with the switching events occurring over a broad range but with a moderate preference for one specific distance. For comparison, we created a random switching dataset by combining the 1900 molecular traces with 100 random switching traces (evenly split between low-to-high and high-to-low). To demonstrate the effects of random variation, we created two separate preferential switching datasets and two separate random switching datasets by independently generating the switching traces for the two copies.

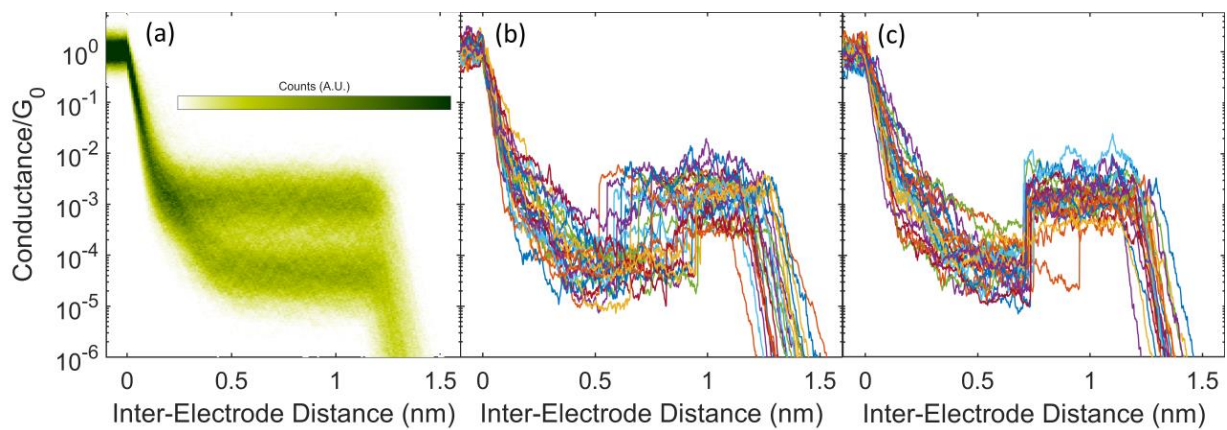

**Figure S39**. Example of simulated traces used to construct different types of rare conductance-switching datasets. (a) 2D histogram of 1900 simulated traces with unbroken plateaus evenly split between a high-conductance feature and a low-conductance feature. (b) 25 simulated low-to-high switching traces with the switching distance chosen uniformly at random between 0.5 and 1.0 nm ("random switching" traces; analogous high-to-low random switching traces were also created). (c) Same as (b), but with the switching distance chosen uniformly at random between 0.70 and 0.75 nm ("preferential switching" traces; again, analogous high-to-low preferential switching traces were also created).

The feature of interest in these simulated datasets is a sharp change in conductance, so we set the MCMC feature-finder to consider 3-node sequences with a slope greater than 10 decades/nm (i.e., an increase of at least 1.2 decades in the 0.12 nm spanned by the three nodes). In addition to applying this minimum slope to the "range slope" described in section S.5.2, which is always positive, in this case we also applied it to the "fit slope" (i.e., the slope of a linear regression to the node sequence). This allowed us to only focus on low-to-high conductance switching and thus avoid contributions from the end-of-plateau break-offs.

The final 2D distributions calculated by the MCMC feature-finder for the two preferential switching datasets and the two random switching datasets are shown in **Figure S40**a,b,d,e, and these distributions have been projected onto the distance axis in **Figure S40**c,f. As expected, the MCMC feature-finder clearly identifies the switching feature centered at 0.725 nm in both preferential switching datasets (**Figure S40**a-c), despite the rareness of this feature (only 25 out of 2000 traces) and the presence of a "background" of broadly distributed random switching. In contrast, when the switching feature is not localized at a particular distance, the MCMC produces a broader and inconsistent distribution because it is not well-suited to identifying this type of rare behavior (**Figure S40**d-f). These examples thus illustrate the points made in the first paragraph of this section. The fact that peaks are still visible in **Figure S40**f provides another example of a limitation of the MCMC feature-finder already mentioned in the main text: this tool will always find *something*, so users must check that a feature is found *consistently* in order to ensure that it is meaningful.

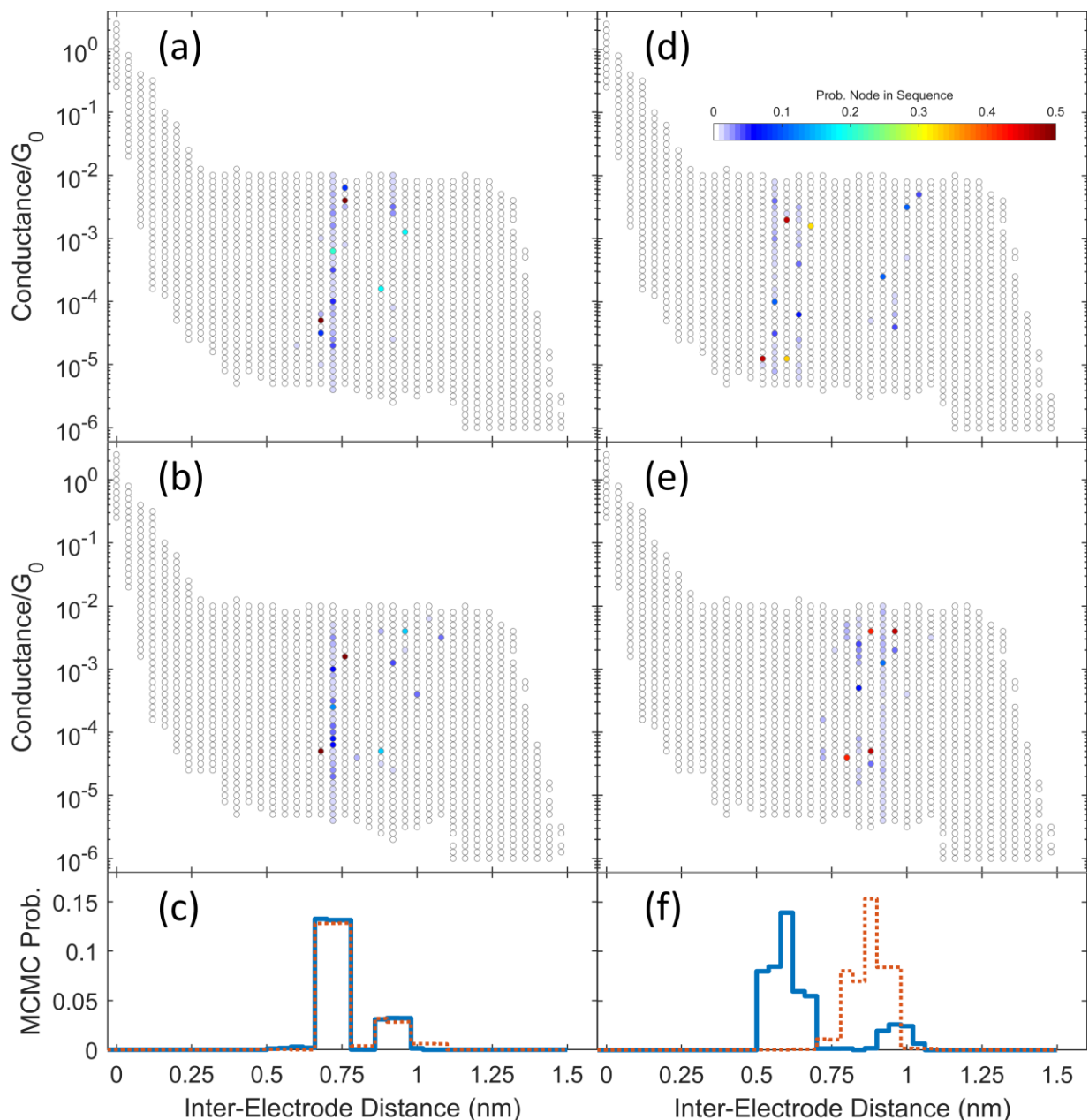

**Figure S40**. (a,b) 2D node distributions produced by the MCMC feature-finder for two separate datasets containing a rare switching feature with a preferential switching distance. (c) Projection of the distributions from (a) and (b) (blue and dotted red, respectively) onto the inter-electrode distance axis. (d-f) Same as (a-c), but for two separate datasets containing a rare switching feature with no preferred switching distance.

# References

(1) Talipov, M. R.; Navale, T. S.; Hossain, M. M.; Shukla, R.; Ivanov, M. V.; Rathore, R. Dihedral-Angle-Controlled Crossover from Static Hole Delocalization to Dynamic Hopping in Biaryl Cation Radicals. *Angewandte Chemie International Edition* **2017**, *56* (1), 266–269. https://doi.org/10.1002/anie.201609695.

(2) Ivie, J. A.; Bamberger, N. D.; Parida, K. N.; Shepard, S.; Dyer, D.; Saraiva-Souza, A.; Himmelhuber, R.; McGrath, D. V.; Smeu, M.; Monti, O. L. A. Correlated Energy-Level Alignment Effects Determine Substituent-Tuned Single-Molecule Conductance. *ACS Appl. Mater. Interfaces* **2021**, *13* (3), 4267–4277. https://doi.org/10.1021/acsami.0c19404.

(3) Cao, X.-Y.; Yang, J.; Dai, F.; Ding, D.-J.; Kang, Y.-F.; Wang, F.; Li, X.-Z.; Liu, G.-Y.; Yu, S.-S.; Jin, X.-L.; Zhou, B. Extraordinary Radical Scavengers: 4-Mercaptostilbenes. *Chemistry – A European Journal* **2012**, *18* (19), 5898–5905. https://doi.org/10.1002/chem.201103897.

(4) Zhou, C.; Chia, G. W. N.; Ho, J. C. S.; Seviour, T.; Sailov, T.; Liedberg, B.; Kjelleberg, S.; Hinks, J.; Bazan, G. C. Informed Molecular Design of Conjugated Oligoelectrolytes To Increase Cell Affinity and Antimicrobial Activity. *Angewandte Chemie International Edition* **2018**, *57* (27), 8069–8072. https://doi.org/10.1002/anie.201803103.

(5) Pedersen, S. K.; Ulfkjær, A.; Newman, M. N.; Yogarasa, S.; Petersen, A. U.; Sølling, T. I.; Pittelkow, M. Inverting the Selectivity of the Newman–Kwart Rearrangement via One Electron Oxidation at Room Temperature. *J. Org. Chem.* **2018**, *83* (19), 12000–12006. https://doi.org/10.1021/acs.joc.8b01800.

(6) Błaszczyk, A.; Elbing, M.; Mayor, M. Bromine Catalyzed Conversion of S-Tert-Butyl Groups into Versatile and, for Self-Assembly Processes Accessible, Acetyl-Protected Thiols. *Org. Biomol. Chem.* **2004**, *2* (19), 2722–2724. https://doi.org/10.1039/B408677E.

(7) Xie, H.; Ng, D.; Savinov, S. N.; Dey, B.; Kwong, P. D.; Wyatt, R.; Smith, A. B.; Hendrickson, W. A. Structure−Activity Relationships in the Binding of Chemically Derivatized CD4 to Gp120 from Human Immunodeficiency Virus. *J. Med. Chem.* **2007**, *50* (20), 4898–4908. https://doi.org/10.1021/jm070564e.

(8) Wang, C.; Batsanov, A. S.; Bryce, M. R.; Sage, I. An Improved Synthesis and Structural Characterisation of 2-(4-Acetylthiophenylethynyl)-4-Nitro-5-Phenylethynylaniline: The Molecule Showing High Negative Differential Resistance (NDR). *Synthesis* **2003**, *2003* (13), 2089–2095. https://doi.org/10.1055/s-2003-41451.

(9) Meisner, J. S.; Ahn, S.; Aradhya, S. V.; Krikorian, M.; Parameswaran, R.; Steigerwald, M.; Venkataraman, L.; Nuckolls, C. Importance of Direct Metal−π Coupling in Electronic Transport Through Conjugated Single-Molecule Junctions. *J. Am. Chem. Soc.* **2012**, *134* (50), 20440–20445. https://doi.org/10.1021/ja308626m.

(10) Bamberger, N. D.; Ivie, J. A.; Parida, K.; McGrath, D. V.; Monti, O. L. A. Unsupervised Segmentation-Based Machine Learning as an Advanced Analysis Tool for Single Molecule Break Junction Data. *J. Phys. Chem. C* **2020**, *124* (33), 18302–18315. https://doi.org/10.1021/acs.jpcc.0c03612.

(11) Johnson, T. K.; Ivie, J. A.; Jaruvang, J.; Monti, O. L. A. Fast Sensitive Amplifier for Two-Probe Conductance Measurements in Single Molecule Break Junctions. *Review of Scientific Instruments* **2017**, *88* (3), 033904. https://doi.org/10.1063/1.4978962.

(12) Sericola, B. *Markov Chains: Theory and Applications*; John Wiley & Sons, Incorporated: Somerset, UNITED STATES, 2013.

(13) Brooks, S.; Gelman, A.; Jones, G.; Meng, X.-L. *Handbook of Markov Chain Monte Carlo*; CRC Press LLC: London, UNITED KINGDOM, 2011.
(14) Gelman, A.; Carlin, J. B.; Stern, H. S.; Dunson, D. B.; Vehtari, A.; Rubin, D. B. *Bayesian Data Analysis*, 3rd ed.; Chapman and Hall/CRC Texts in Statistical Science Ser.; CRC Press LLC, 2013.
(15) Earl, D. J.; Deem, M. W. Parallel Tempering: Theory, Applications, and New Perspectives. *Phys. Chem. Chem. Phys.* **2005**, *7* (23), 3910–3916. https://doi.org/10.1039/B509983H.
(16) Roy, V. Convergence Diagnostics for Markov Chain Monte Carlo. *Annu. Rev. Stat. Appl.* **2020**, *7* (1), 387–412. https://doi.org/10.1146/annurev-statistics-031219-041300.
(17) Hong, W.; Manrique, D. Z.; Moreno-García, P.; Gulcur, M.; Mishchenko, A.; Lambert, C. J.; Bryce, M. R.; Wandlowski, T. Single Molecular Conductance of Tolanes: Experimental and Theoretical Study on the Junction Evolution Dependent on the Anchoring Group. *J. Am. Chem. Soc.* **2012**, *134* (4), 2292–2304. https://doi.org/10.1021/ja209844r.
(18) Yanson, A. I.; Bollinger, G. R.; van den Brom, H. E.; Agraït, N.; van Ruitenbeek, J. M. Formation and Manipulation of a Metallic Wire of Single Gold Atoms. *Nature* **1998**, *395* (6704), 783–785. https://doi.org/10.1038/27405.
(19) Quek, S. Y.; Venkataraman, L.; Choi, H. J.; Louie, S. G.; Hybertsen, M. S.; Neaton, J. B. Amine−Gold Linked Single-Molecule Circuits:  Experiment and Theory. *Nano Lett.* **2007**, *7* (11), 3477–3482. https://doi.org/10.1021/nl072058i.
(20) Lemmer, M.; Inkpen, M. S.; Kornysheva, K.; Long, N. J.; Albrecht, T. Unsupervised Vector-Based Classification of Single-Molecule Charge Transport Data. *Nature Communications* **2016**, *7*, 12922. https://doi.org/10.1038/ncomms12922.
(21) Hamill, J. M.; Zhao, X. T.; Mészáros, G.; Bryce, M. R.; Arenz, M. Fast Data Sorting with Modified Principal Component Analysis to Distinguish Unique Single Molecular Break Junction Trajectories. *Phys. Rev. Lett.* **2018**, *120* (1), 016601. https://doi.org/10.1103/PhysRevLett.120.016601.
(22) Cabosart, D.; El Abbassi, M.; Stefani, D.; Frisenda, R.; Calame, M.; van der Zant, H. S. J.; Perrin, M. L. A Reference-Free Clustering Method for the Analysis of Molecular Break-Junction Measurements. *Appl. Phys. Lett.* **2019**, *114* (14), 143102. https://doi.org/10.1063/1.5089198.
(23) Wu, B. H.; Ivie, J. A.; Johnson, T. K.; Monti, O. L. A. Uncovering Hierarchical Data Structure in Single Molecule Transport. *The Journal of Chemical Physics* **2017**, *146* (9), 092321. https://doi.org/10.1063/1.4974937.
(24) Huang, F.; Li, R.; Wang, G.; Zheng, J.; Tang, Y.; Liu, J.; Yang, Y.; Yao, Y.; Shi, J.; Hong, W. Automatic Classification of Single-Molecule Charge Transport Data with an Unsupervised Machine-Learning Algorithm. *Physical Chemistry Chemical Physics* **2019**, *22*, 1674–1681. https://doi.org/10.1039/C9CP04496E.
(25) El Abbassi, M.; Overbeck, J.; Braun, O.; Calame, M.; van der Zant, H. S. J.; Perrin, M. L. Benchmark and Application of Unsupervised Classification Approaches for Univariate Data. *Communications Physics* **2021**, *4* (1), 1–9. https://doi.org/10.1038/s42005-021-00549-9.
(26) Lin, L.; Tang, C.; Dong, G.; Chen, Z.; Pan, Z.; Liu, J.; Yang, Y.; Shi, J.; Ji, R.; Hong, W. Spectral Clustering to Analyze the Hidden Events in Single-Molecule Break Junctions. *J. Phys. Chem. C* **2021**, *125* (6), 3623–3630. https://doi.org/10.1021/acs.jpcc.0c11473.
(27) Shi, J.; Malik, J. Normalized Cuts and Image Segmentation. *IEEE Transactions on Pattern Analysis and Machine Intelligence* **2000**, *22* (8), 888–905. https://doi.org/10.1109/34.868688.
(28) Perrin, M. *Data Clustering Tool*; 2021.
(29) Caliński, T.; Harabasz, J. A Dendrite Method for Cluster Analysis. *Communications in Statistics* **1974**, *3* (1), 1–27. https://doi.org/10.1080/03610927408827101.